\documentclass[a4paper,11pt]{article}
\pdfoutput=1

\usepackage{amssymb,amsmath,bm}
\usepackage{a4wide}
\usepackage{color}
\usepackage{slashed}
\usepackage{graphicx}
\usepackage{amsfonts}
\usepackage{lscape}
\usepackage{hyperref}
\usepackage{comment}
\usepackage{amsthm}
\hypersetup{
    colorlinks=true, 
    linktoc=all,     
    linkcolor=blue,  
    citecolor= red
}
\usepackage{booktabs}
\usepackage{array}
\usepackage{rotating}
\usepackage[numbers, sort&compress]{natbib}
\usepackage{float}
\usepackage[utf8]{inputenc}
\usepackage[T1]{fontenc}
\newcommand{\GeV}{{\, \rm GeV}}

\newcommand{\eps}{\epsilon}

\newcommand{\be}{\begin{equation}}
\newcommand{\ee}{\end{equation}}

\newcommand{\ssb} {s_\beta}
\newcommand{\ccb} {c_\beta}

\newcommand{\ssa} {s_\alpha}
\newcommand{\cca} {c_\alpha}

\newcommand{\cref}[1]{Chapter~\ref{ch:.#1}}

\newcommand{\beq}{\begin{equation}} 
\newcommand{\eeq}{\end{equation}} 
\newcommand{\ba}{\begin{array}}  
\newcommand{\ea}{\end{array}} 
\newcommand{\bea}{\begin{eqnarray}}  
\newcommand{\eea}{\end{eqnarray} }  
\newcommand{\bal}{\begin{align}}
\newcommand{\eal}{\end{align}}   
\newcommand{\bi}{\begin{itemize}}  
\newcommand{\ei}{\end{itemize}}  
\newcommand{\ben}{\begin{enumerate}}  
\newcommand{\een}{\end{enumerate}}  
\newcommand{\bc}{\begin{center}}
\newcommand{\ec}{\end{center}} 
\newcommand{\bt}{\begin{table}}
\newcommand{\et}{\end{table}}  
\newcommand{\btb}{\begin{tabular}}
\newcommand{\etb}{\end{tabular}}

\definecolor{sanddune}{rgb}{0.59, 0.44, 0.09}

\definecolor{mypink}{RGB}{219, 48, 122}

\begin{document}

\vspace{1cm}
\begin{titlepage}
\vspace*{-1.0truecm}
\begin{flushright}
TTP21-025  \\
P3H-21-051 \\
 \vspace*{2mm}
 \end{flushright}
\vspace{0.8truecm}

\vspace{1mm}
\begin{center}
\huge\bf
Flavor-Violating Higgs Decays and \\  Stellar Cooling Anomalies in Axion Models
\end{center}

\vspace{1mm}
\begin{center}
\bf 
Marcin Badziak$^a$, Giovanni Grilli di Cortona$^{b}$,  Mustafa Tabet$^c$, Robert Ziegler$^{c}$\\
\vspace{5mm} 
{\small\sl $^a${\sl Institute of Theoretical Physics, Faculty of Physics, University of Warsaw, ul. Pasteura 5, PL-02-093 Warsaw, Poland  \vspace{0.2truecm}}

$^b${\sl Istituto Nazionale di Fisica Nucleare, Laboratori Nazionali di Frascati, C.P. 13, 00044 Frascati, Italy  \vspace{0.2truecm}}

$^c${\sl Institut  f\"ur  Theoretische  Teilchenphysik,  Karlsruhe  Institute  of  Technology,  Karlsruhe,  Germany \vspace{0.2truecm}}}
\end{center}

\begin{abstract}
\noindent We study a class of DFSZ-like models for the QCD axion that can address observed anomalies in stellar cooling. Stringent constraints from SN1987A and neutron stars are avoided by suppressed couplings to nucleons, while axion couplings to electrons and photons are sizable. All axion couplings depend on few parameters that also control the extended Higgs sector, in particular lepton flavor-violating couplings of the Standard Model-like Higgs boson $h$. This allows us to correlate axion and Higgs phenomenology, and we find that that ${\rm BR}(h \to \tau e)$ can be as large as the current experimental bound of 0.22\%, while ${\rm BR} (h \to \mu \mu)$ can be larger than in the Standard Model by up to 70\%. Large parts of the parameter space will be tested by the next generation of axion helioscopes such as the IAXO experiment. 
\end{abstract}

\end{titlepage}

\newpage

\renewcommand{\theequation}{\arabic{section}.\arabic{equation}}


\tableofcontents

\section{Introduction and Motivation}

Arguably the QCD axion is one of the best candidates for New Physics beyond the Standard Model (SM), being motivated not only by the Peccei-Quinn solution to the strong CP Problem~\cite{PQ1,PQ2,WW1,WW2}, but also by the observed Dark Matter abundance~\cite{AxionDM1,AxionDM2,AxionDM3}. Interestingly, the axion can also account for excessive energy losses observed in several stellar environments~\cite{Raffelt:2011ft,Ringwald:2015lqa, Giannotti:2015dwa, Giannotti:2015kwo, Giannotti:2016hnk, CoolingAnom3}, which hints to new cooling channels such as a light axion with large couplings to electrons. This kind of scenarios are however constrained by other astrophysical observations that strongly constrain axion couplings to nucleons, such as the observation of the neutrino burst in SN1987A~\cite{SNbound2} or neutron star cooling~\cite{Keller:2012yr,Sedrakian:2015krq,Hamaguchi:2018oqw,Beznogov:2018fda,Sedrakian:2018kdm}. 

Taking the stellar cooling hints seriously therefore points to a rather special structure of axion couplings, which definitely prefers the class of DFSZ-like axion models~\cite{DFSZ1,DFSZ2}, in which the axion has large couplings to SM fermions~\cite{CoolingAnom3}. Since all couplings are proportional to the axion mass, the required size of electron couplings puts a lower bound on the axion mass of the order of a few meV, corresponding to axion decay constants of the order of $10^9 \GeV$. While the standard DFSZ benchmark models discussed in Ref.~\cite{CoolingAnom3} have some tension with perturbativity and SN1987A constraints for such heavy axions, several modifications of DFSZ models (or ``axion variant models''~\cite{Peccei:1986pn,Krauss:1986wx}) have been proposed that can ease this tension as a result of suppressed couplings to nucleons~\cite{DiLuzio:2017ogq, Bjorkeroth:2019jtx,Saikawa:2019lng}. Moreover, these models can feature a trivial domain wall number that elegantly avoids the cosmological domain wall problem~\cite{Vilenkin:1982ks}.

In this article we revisit these ``nucleophobic'' axion models which are simple Two-Higgs-Doublet models (2HDMs) with a global Peccei-Quinn (PQ) symmetry, that are constructed analogous to the standard DFSZ axion models, but with \emph{flavor-dependent} PQ charges. In general the axion coupling to electrons depends on free parameters that also control lepton flavor-violating (LFV) effects, which are mediated by LFV couplings of both the axion and the physical Higgs scalars. In contrast to earlier analyses, which have focussed on implications for axion searches, here we are interested also in the Higgs phenomenology, which is correlated to the axion phenomenology to a large extent. Thus, we have in mind a DFSZ models with a \emph{light} second Higgs doublet, whose scale is subject only to present experimental constraints. This setup allows us to predict various flavor-violating Higgs decays probed at the LHC in terms of the same parameters that control the axion couplings to nucleons, electrons and photons, which are constrained by astrophysical observations and probed by dedicated axion searches such as IAXO~\cite{IAXO2}. For similar studies of the phenomenological implications of DFSZ models with light Higgs doublets see e.g. Refs.~\cite{Chiang:2015cba,Chiang:2017fjr,Chiang:2018bnu}.  
We find, in particular, that in the region explaining the stellar cooling anomaly, the branching ratio for the $h\to \tau e$ decay can be as large as the current LHC upper bound, while the couplings of the Higgs to taus and muons may deviate significantly from the SM prediction. 

This article is structured as follows: In Section~\ref{sec1} we define our basic setup and provide the expressions for axion and Higgs couplings to the SM in terms of the model parameters. In  Section~\ref{sec:axion_pheno} we provide the constraints on these parameter from axion physics, and identify the region preferred by stellar cooling anomalies. In Section~\ref{sec3} we discuss the phenomenology of the extended Higgs sector, focussing on modifications of the SM-like Higgs couplings to leptons, in particular LFV couplings. We combine all constraints in Section~\ref{sec4} and study the implications for future searches at the LHC and axion helioscopes in the cooling hint region, before we conclude in Section~\ref{sec5}.


\section{Setup}
\label{sec1}

In this section we first discuss the general effective Lagrangian for the QCD axion, before we define the special DFSZ-like UV completions that we want to consider. These models fix not only the couplings of the QCD axion, but also the couplings in the extended Higgs sector in terms of a few parameters. This structure gives rise to a correlated axion-Higgs phenomenology that we will analyze in the following sections.

\subsection{Axion Effective Lagrangian}

At energies much below the PQ breaking scale, the effective axion couplings to gauge fields and fermions are given by
\begin{equation}
{\cal L}  =  \frac{a}{f_a} \frac{\alpha_s}{8 \pi}  G \tilde{G} +  \frac{E}{N}  \frac{a}{f_a} \frac{\alpha_{\rm em}}{8 \pi} F \tilde{F} + \frac{\partial_\mu a}{2 f_a} \overline{f}_i \gamma^\mu \left( C^V_{ij} + C^A_{ij} \gamma_5 \right) f_j \, ,
\label{LaPQ}
\end{equation}
where $f_a$ is the axion decay constant, $F \tilde{F} \equiv \frac{1}{2} \eps_{\mu \nu \rho \sigma} F^{\mu \nu} F^{\rho \sigma}$ with the electromagnetic (EM) field strengths and similar in the gluon sector, $E/N$ is the ratio of EM and color anomaly coefficients and we use the convention $\eps^{0123} = -1$ (for more details see Appendix~\ref{genDFSZ}). 

The first term in Eq.~\eqref{LaPQ} gives rise to the axion mass, which can be conveniently calculated in chiral perturbation theory, giving~\cite{Villadoro2} 
\begin{align}
m_a & = 5.691(51) \, \mu {\rm eV} \left( \frac{10^{12} \GeV}{f_a} \right)  \, .
\end{align}
Below the QCD scale the relevant couplings are those to photons, nucleons $n,p$ and electrons, 
\begin{equation}
{\cal L}  =    C_\gamma \frac{a}{f_a} \frac{\alpha_{\rm em}}{8 \pi} F \tilde{F} + \frac{\partial_\mu a}{2 f_a} \left( C_n \overline{n} \gamma^\mu \gamma_5  n  + C_p \overline{p} \gamma^\mu \gamma_5  p  + C_e \overline{e} \gamma^\mu \gamma_5  e \right) \, ,
\label{LaQCD}
\end{equation}
where the matching to the UV coefficients in the Lagrangian of Eq.~\eqref{LaPQ} is given by
$C_{\gamma} = |E/N - 1.92(4)|$ and
\begin{align} 
\label{CppCn}
C_p +C_n &= 
0.50(5)\left( C_u  + C_d  -1\right) - 2 \delta  \, ,  \\
C_p -C_n &= 1.273(2)\left( C_u  - C_d  - {\frac{1-z}{1+z}} \right) \, ,
\label{CpmCn}
\end{align}
where $C_q \equiv C^A_{qq} (\mu = f_a)$, $z = m_u/m_d = 0.48(3)$ and $\delta \equiv 0.038(5) C_s + 0.012(5) C_c +  0.009(2)  C_b + 0.0035(4) C_t$ arises from QCD running effects~\cite{diCortona:2015ldu}.  

For the purpose of addressing the stellar cooling anomalies with axions it is helpful to have small couplings to nucleons in order to avoid the stringent constraints from SN1987A and neutron star cooling, cf. Section~\ref{anconstraints}. From Eqs.~\eqref{CppCn} and \eqref{CpmCn} it is clear that axion couplings to nucleons can be suppressed if the UV quark couplings satisfy the approximate relations
\begin{align}
C_u &  \simeq \frac{1}{1+z} \approx 2/3 \, , & 
C_u + C_d &  \simeq  1 \, . 
\end{align}
As analyzed in detail in Ref.~\cite{astrophobic} (see also Refs.~\cite{Krauss:1987ud, Hindmarsh:1997ac}), one can realize  these conditions in the context of ``nucleophobic'' DFSZ models that  we will now discuss in more details.
\subsection{UV Lagrangian}

We add two Higgs doublets $h_i$ with hypercharge $Y=-1/2$ and a SM singlet $\phi$ to the SM. The Lagrangian admits a $U(1)_{\rm PQ}$ symmetry, with fermion charges that are consistent with a $2+1$ flavor structure. This symmetry is broken twice: explicitly by the QCD anomaly and spontaneously by Higgs and singlet vacuum expectation values (VEVs). The QCD axion is the pseudo-Nambu-Goldstone boson of the PQ symmetry, and thus a linear combination of the CP-odd components of all PQ-charged fields with VEVs, with each coefficient given by the respective PQ charge and VEV (up to a normalization). In Appendix~\ref{genDFSZ} we summarize the general structure of these models, which are defined by the Yukawa couplings and the Higgs-singlet interactions needed to break additional $U(1)$ factors. These Lagrangian parameters govern the axion couplings to matter, apart from fermion flavor mixing. As discussed above, here we are only interested in models that have potentially suppressed couplings to nucleons, which requires approximately $C_u \approx 2/3, C_d \approx 1/3$. The axion coupling to electrons is set by a free rotation angle that also controls lepton flavor-violating axion and Higgs couplings. Moreover, all models have a trivial domain wall number (see also~\cite{Saikawa:2019lng}), which evades the cosmological domain wall problem in scenarios when PQ symmetry is broken after inflation~\cite{Vilenkin:1982ks}.

 The Lagrangian is given by (apart from kinetic terms):
\begin{align}
{\cal L} = {\cal L}_{\rm yuk} (h_1, h_2,f_i) - V (|h_1|, |h_2|, |\phi|) +  \left(a h_1^\dagger h_2 \phi + {\rm h.c.}  \right) \, , 
\label{Lag}
\end{align}
where the first term comprises Yukawa couplings, the second term is that part of the scalar potential which only depends on the modulus of the two Higgs fields and the singlet, and the last term is needed to ensure that $U(1)_{\rm PQ}$ is the only global symmetry. The Yukawa Lagrangian reads
\begin{align}
{\cal L}_{\rm yuk} & = - y^{u}_{33} \overline{q}_{L3} u_{R3} h_{A_1} - y^{u}_{3a}  \overline{q}_{L3} u_{Ra} h_{A_2} - y^{u}_{a3} \overline{q}_{La} u_{R3} h_{A_3} - y^{u}_{ab}  \overline{q}_{La} u_{Rb} h_{A_4} \nonumber \\
&  + y^{d}_{33} \overline{q}_{L3} d_{R3} \tilde{h}_{A_5} + y^{d}_{3a}  \overline{q}_{L3} d_{Ra} \tilde{h}_{A_6} + y^{d}_{a3} \overline{q}_{La} d_{R3} \tilde{h}_{A_7} + y^{d}_{ab} \overline{q}_{La} d_{Rb} \tilde{h}_{A_8} \nonumber \\ 
& + y^{e}_{33} \overline{l}_{L3} e_{R3} \tilde{h}_{A_9} + y^{e}_{3a} \overline{l}_{L3} e_{Ra} \tilde{h}_{A_{10}} + y^{e}_{a3} \overline{l}_{La} e_{R3} \tilde{h}_{A_{11}} + y^{e}_{ab} \overline{l}_{La} e_{Rb} \tilde{h}_{A_{12}} 
+ {\rm h.c.}   
\label{L1}
\end{align}
where $\tilde{h}_i = i \sigma^2 h_i^*$, $a,b = 1,2$ runs over the first two fermion generations, and $A_{1 \hdots 12} \in \{ 1,2 \}$ are parameters that define the Higgs field to which a given fermion bilinear structure couples to. 

We consider four different structures in the quark sector, Q1-Q4, which are defined by the choice of $(A_{1 \hdots 4})(A_{5 \hdots 8})$:
\begin{align}
{\rm Q1}&: (2222)(1212) \, , &
y_u & \sim \begin{pmatrix} H_2 & H_2 \\ H_2 & H_2 \end{pmatrix} \, , &
y_d & \sim \begin{pmatrix} H_2 & H_1 \\ H_2 & H_1 \end{pmatrix} \, ,  \nonumber \\
 {\rm Q2}&: (1122)(2211) \, , &
 y_u & \sim \begin{pmatrix} H_2 & H_2 \\ H_1 & H_1 \end{pmatrix} \, , &
y_d & \sim \begin{pmatrix} H_1 & H_1 \\ H_2 & H_2 \end{pmatrix} \, ,  \nonumber \\
  {\rm Q3}&: (1212)(2121) \, , &
  y_u & \sim \begin{pmatrix} H_2 & H_1 \\ H_2 & H_1 \end{pmatrix} \, , &
y_d & \sim \begin{pmatrix} H_1 & H_2 \\ H_1 & H_2 \end{pmatrix} \, ,  \nonumber \\
  {\rm Q4}&: (2121)(1111) \, , & 
  y_u & \sim \begin{pmatrix} H_1 & H_2 \\ H_1 & H_2 \end{pmatrix} \, , &
y_d & \sim \begin{pmatrix} H_1 & H_1 \\ H_1 & H_1 \end{pmatrix} \, ,
\label{eq:Qmodels}
\end{align}
where we also indicate in 2+1 flavor space notation to which Higgs field the quark bilinears couple to. Together with the last term in Eq.~\eqref{Lag} this choice fixes the PQ color anomaly coefficient $N$, the quark contribution to the electromagnetic anomaly coefficient $E_Q$, and all couplings of the QCD axion to quarks $C^{A,V}_{q_i q_j}$ in terms of the parameters $\tan \beta$ and $\xi_{ij}^{q_P}$ (to be defined below), which we summarize in Table~\ref{Qmodels}.  

The quark Yukawa Lagrangians of each model is combined with one out of the four following structures in the charged lepton sector, defined by $(A_{9 \hdots 12})$, i.e. the Higgs to which a lepton bilinear couples in $2 + 1$ flavor space
\begin{align}
{\rm E1L}&: (1122)  \, , &
y_e & \sim \begin{pmatrix} H_2 & H_2 \\ H_1 & H_1 \end{pmatrix} \, ,  \nonumber \\
 {\rm E1R}&: (1212)  \, ,&
 y_e & \sim \begin{pmatrix} H_2 & H_1 \\ H_2 & H_1 \end{pmatrix} \, ,  \nonumber \\
 {\rm E2L}&: (2211) \, , & 
 y_e & \sim \begin{pmatrix} H_1 & H_1 \\ H_2 & H_2 \end{pmatrix} \, ,
\nonumber \\
 {\rm E2R}&: (2121)  \, , & 
 y_e & \sim \begin{pmatrix} H_1 & H_2 \\ H_1 & H_2 \end{pmatrix} \, .  \label{eq:Lmodels}
\end{align}
This choice fixes the charged lepton contribution to the  electromagnetic anomaly coefficient $E_L$ and the axion couplings to leptons $C^{V,A}_{\ell_i \ell_j}$, which we summarize in Table~\ref{Lmodels}.
\begin{table}[t]
\centering
\begin{tabular}{|c||c||c||c|c|c||c|c|c||}
\hline
Model & $E_Q/N$  & $C^A_{u_i u_i}$ & $C^A_{d_i d_i}$ & $C^{V,A}_{u_i \ne u_j} $ & $C^{V,A}_{d_i \ne d_j} $   \\
\hline
Q1 &  $2/3 + 6 c_\beta^2$ & $c_\beta^2 $ & $\xi^{d_R}_{ii} - c_\beta^2$ & 0 & $\xi^{d_R}_{ij}$ \\
Q2 &$-4/3 + 6 c_\beta^2$ & $c_\beta^2 - \xi^{u_L}_{ii}  $ & $-\xi^{d_L}_{ii} + s_\beta^2$ & $\pm \xi^{u_L}_{ij}$ & $ \pm\xi^{d_L}_{ij}$  \\
Q3 &  $- 4/3 + 6 c_\beta^2$ & $c_\beta^2 - \xi^{u_R}_{ii}$ & $-\xi^{d_R}_{ii} + s_\beta^2 $ & $- \xi^{u_R}_{ij}$ & $ - \xi^{d_R}_{ij}$  \\
Q4 & $-10/3 + 6 c_\beta^2$ & $- s_\beta^2 + \xi^{u_R}_{ii} $ & $s_\beta^2$ & $ \xi^{u_R}_{ij}$ & 0  \\
\hline
\end{tabular}
\caption{Axion couplings in the four ``nucleophobic'' models Q1-Q4, see Eq.~\eqref{eq:Qmodels}, as a function of the parameters $\xi^{q_P}_{ij}$ and $c_\beta \equiv \cos \beta, s_\beta \equiv \sin \beta$. Here $E_Q$ denotes the contribution of the quark sector to the electromagnetic anomaly coefficient $E$, to be added to the contribution from the charged lepton sector. In all models the domain wall number is trivial, $2N = 1$.  \label{Qmodels} }
\end{table}
\begin{table}[t]
\centering
\begin{tabular}{|c||c||c||c|c|c||c|c|c||}
\hline
Model & $E_L/N$ & $C^A_{e_i e_i}$  & $C^{V,A}_{e_i \ne e_j}$  \\
\hline
E1L & $ 2  - 6 c_\beta^2 $ & $- c_\beta^2 + \xi^{e_L}_{ii}$ & $\mp \xi^{e_L}_{ij}$\\
E1R & $2 - 6 c_\beta^2 $ & $- c_\beta^2 + \xi^{e_R}_{ii}$ & $ \xi^{e_R}_{ij}$\\
E2L & $ 4  - 6 c_\beta^2 $ & $ s_\beta^2 - \xi^{e_L}_{ii}$ & $\pm \xi^{e_L}_{ij}$ \\
E2R & $4 - 6 c_\beta^2 $ & $ s_\beta^2 - \xi^{e_R}_{ii}$ & $ - \xi^{e_R}_{ij}$\\
\hline
\end{tabular}
\caption{\label{Lmodels} Axion couplings in the four models EL1, E1R, E2L, E2R, see Eq.~\eqref{eq:Lmodels}, as a function of the parameters $\xi^{e_P}_{ij}$ and $c_\beta \equiv \cos \beta, s_\beta \equiv \sin \beta$. Here $E_L$ denotes the contribution of the charged lepton sector to the electromagnetic anomaly coefficient $E$, to be added to the contribution from the quark sector. }
\end{table}

 Since each quark sector model can be combined with any charged lepton sector model, we have in total 16 different models, which we denote by e.g. ``Q1E1L'', which has the Higgs structure (2222)(1212)(1122), axion couplings to quarks and charged leptons as in Tables~\ref{Qmodels} and \ref{Lmodels}, and an electromagnetic anomaly coefficient $E/N$ that is the sum of both sectors, $E/N = E_Q/N + E_L/N = 8/3$ in this example. 
 
The couplings to quarks and leptons depend on the Higgs vacuum angle $\tan \beta$ and the parameters $\xi^{f_P}_{ij}$, with $f = u,d,e$ and $P= L,R$, which are defined by
\begin{align}
\tan \beta & \equiv \langle H_2 \rangle/ \langle H_1 \rangle \, , & 
\xi^{f_{P}}_{ij} & \equiv (V_{fP})^*_{3i}  (V_{fP})_{3j} \, , 
\end{align}
where $V_{fP}$ are the unitary matrices which diagonalize quark and charged lepton masses according to $V_{fL}^\dagger M_f V_{fR}  = m_f^{\rm diag} $. Unitarity implies the relations
  \begin{align}
  |\xi^{f_P}_{ij}| & = \sqrt{\xi^{f_P}_{ii} \xi^{f_P}_{jj}} \, , & 0 & \le \xi^{f_P}_{ii} \le 1 \, , & \sum_i \xi^{f_P}_{ii}  & = 1 \, , 
  \label{xidef}
 \end{align}
so apart from complex phases 
there are only two parameters $\xi^{f_P}_{ii}$ in each chiral fermion sector, which depend on the structure of quark and charged lepton masses. 

By construction in all models the axion couplings to nucleons can be suppressed by choosing $\cos^2 \beta \approx 2/3$ and special values for the parameters $\xi^{q_P}_{11}$, which are $\xi^{u_L}_{11} = \xi^{d_L}_{11} =0 $  (Q2), $\xi^{u_R}_{11} = \xi^{d_R}_{11} =0 $ (Q3), $\xi^{d_R}_{11} =1 $ (Q1) and  $\xi^{u_R}_{11} =1 $ (Q4). The coupling to electrons is then controlled mainly by the parameter $\xi_{11}^{e_L}$ (E1L,E2L) and $\xi_{11}^{e_R}$ (E1R,E2R), and addressing the stellar cooling anomalies will generically correspond to $\xi_{11}^{e_{L/R}} \ne 0,1$. Therefore lepton flavor-violating axion couplings, which are controlled by $\xi_{i \ne j}^{e_{L/R}}$, are a generic consequence of the cooling anomalies in these models (since two different $\xi_{ii}^e$ are non-zero), while it is always possible to avoid quark flavor violation (i.e. having only one non-zero $\xi_{ii}^q$). In the following we focus for simplicity on scenarios without quark flavor violation, obtained from choosing quark Yukawa matrices such that to very good approximation (neglecting small corrections necessary to reproduce the CKM matrix)
\begin{align}
{\rm Q1:} \quad &  \xi^{d_R}_{11}  =  \xi^{u_R}_{33} =  \xi^{u_L}_{33} =  \xi^{d_L}_{33}  = 1\, , \nonumber \\
{\rm Q2:} \quad & \xi^{u_R}_{33} =     \xi^{d_R}_{33}  =   \xi^{u_L}_{33} =  \xi^{d_L}_{33}  = 1\, , \nonumber \\
{\rm Q3:} \quad &  \xi^{u_R}_{33} =   \xi^{d_R}_{33}  =   \xi^{u_L}_{33} =  \xi^{d_L}_{33}  = 1\, , \nonumber \\
{\rm Q4:} \quad & \xi^{u_R}_{11} =   \xi^{d_R}_{33}  =   \xi^{u_L}_{33} =  \xi^{d_L}_{33}  = 1\, ,
\label{epsQ}
\end{align}
which implies that all other $\xi_{ii}^{q_P} $ are vanishing (cf.~Eq.~\eqref{xidef}). This choice implies that models Q2 and Q3 give identical predictions for all  axion couplings,  while models Q1 and Q4 only give the same contribution for axion couplings to quarks in the 1st generation.  In all models axion couplings to nucleons can be suppressed by taking  $c_\beta \approx \sqrt{2/3}$ or equivalently $t_\beta \approx \sqrt{1/2} \approx 0.7$, which fixes the 1st generation quark couplings to identical values in all four models, $C_u \approx 2/3, C_d \approx 1/3$, while the axion couplings to heavy quarks are either close to $2/3$ or $1/3$. 
Axion couplings to leptons  are  controlled by $\tan \beta$ and the free parameters  $\xi_{ii}^{e_L}$  [E1L,E2L] or $\xi_{ii}^{e_R}$  [E1R,E2R]. 

In the following we study the consequences of the above choices for $\xi^{f_P}_{ii}$ for the structure of the quark and charged lepton Yukawa sectors.

\subsection{Quark Yukawa Sector}
 The quark Yukawa Lagrangian is given by
\begin{align}\label{eq:Yukawa couplings}
{\cal L}_{\rm Q}& = - \overline{q}_{L,i} u_{R,j} \left[ Y^u_{1,ij} h_{1} + Y^u_{2,ij} h_2 \right] 
                     + \overline{q}_{L,i} d_{R,j} \left[ Y^d_{1,ij} \tilde{h}_1 + Y^d_{2,ij} \tilde{h}_2 \right]
                 + \mathrm{h.c.} \,, 
 \end{align}
where the structure of the couplings $Y^u_{1,2}$ and $Y^d_{1,2}$ depends on the model under consideration. These couplings have to be chosen appropriately, such that the diagonalization of the quark mass matrices
\begin{align}
M_u & = v c_\beta Y^u_1 + v s_\beta Y^u_2 \, , & 
M_d & = v c_\beta Y^d_1 + v s_\beta Y^d_2 \, , 
\end{align}
reproduces i) the quark masses as singular values, ii) suitable left-handed rotations in order to obtain the correct CKM matrix, and iii) third rows of mixing matrices that match the parameter choice in Eq.~\eqref{epsQ}, up to correction of small CKM angles. This gives the following parametric structure of Yukawa matrices, where $\lambda \approx 0.2$ is of the order of the Cabibbo angle, and a numerical coefficient of order unity is understood in front of the $\lambda, \lambda^2, \lambda^3$ entries in order to reproduce the exact values of the CKM matrix:
 \paragraph{Model Q1:} 

\begin{align}
Y^u_1 & = 0 \, , &
Y^u_2 & = \frac{1}{s_\beta v} \begin{pmatrix} m_u   &  \lambda m_c   & \lambda^3 m_t    \\ 0  & m_c    & \lambda^2 m_t   \\  0   & 0   & m_t \\ \end{pmatrix} \, , \nonumber \\
Y^d_1 & = \frac{1}{c_\beta v}\begin{pmatrix} 0 & 0 &  m_d    \\ 0 & 0 & 0   \\  0 & 0 & 0  \\ \end{pmatrix} \, , & 
Y^d_2 & = \frac{1}{s_\beta v} \begin{pmatrix} 0 &  0   & 0  \\ 0  & m_s  & 0  \\  m_b &0 & 0 \\ \end{pmatrix}  \, .
\label{YukQ1}
\end{align}

\paragraph{Model Q2:} 

\begin{align}
Y^u_1 & = \frac{1}{c_\beta v} \begin{pmatrix} 0 & 0 &  0   \\ 0 & 0 & 0 \\  0 & 0 & m_t    \\ \end{pmatrix} \, , &
Y^u_2 & = \frac{1}{s_\beta v} \begin{pmatrix} m_u   &  \lambda m_c   & \lambda^3 m_t   \\ 0   & m_c    & \lambda^2 m_t   \\  0   & 0   & 0 \\ \end{pmatrix} \, , \nonumber \\
Y^d_1 & = \frac{1}{c_\beta v} \begin{pmatrix} m_d&  0   & 0  \\ 0  & m_s  & 0  \\  0 &0 & 0 \\ \end{pmatrix} \, , & 
Y^d_2 & = \frac{1}{s_\beta v}\begin{pmatrix} 0 & 0 &  0   \\ 0 & 0 & 0   \\  0 & 0 & m_b  \\ \end{pmatrix} \, .
\end{align}

\paragraph{Model Q3:} 

\begin{align}
Y^u_1 & = \frac{1}{c_\beta v} \begin{pmatrix} 0 & 0 &  \lambda^3 m_t    \\ 0 & 0 & \lambda^2 m_t  \\  0 & 0 & m_t    \\ \end{pmatrix} \, , &
Y^u_2 & = \frac{1}{s_\beta v} \begin{pmatrix} m_u   &  \lambda m_c   & 0  \\ 0   & m_c    & 0  \\  0   & 0   & 0 \\ \end{pmatrix} \, , \nonumber \\
Y^d_1 & = \frac{1}{c_\beta v} \begin{pmatrix} m_d&  0   & 0  \\ 0  & m_s  & 0  \\  0 &0 & 0 \\ \end{pmatrix} \, , & 
Y^d_2 & = \frac{1}{s_\beta v}\begin{pmatrix} 0 & 0 &  0   \\ 0 & 0 & 0   \\  0 & 0 & m_b  \\ \end{pmatrix} \, .
\end{align}

\paragraph{Model Q4:} 

\begin{align}
Y^u_1 & = \frac{1}{c_\beta v} \begin{pmatrix} 0 & 0 & 0   \\ 0 & m_c & 0  \\  m_t & 0 & 0    \\ \end{pmatrix} \, , &
Y^u_2 & = \frac{1}{s_\beta v} \begin{pmatrix} 0   &  0   & m_u  \\ 0   & 0    & 0  \\  0   & 0   & 0 \\ \end{pmatrix} \, , \nonumber \\
Y^d_1 & = \frac{1}{c_\beta v} \begin{pmatrix} m_d&  \lambda m_s   & \lambda^3 m_b  \\ 0  & m_s  & \lambda^2 m_b  \\  0 &0 &  m_b  \\ \end{pmatrix} \, , & 
Y^d_2 & = 0 \, .
\label{YukQ4}
\end{align}

\subsection{Charged Lepton Yukawa Structure}
The charged lepton Yukawa Lagrangian is defined as
\begin{align}
{\cal L}_{\rm L3}& =   \overline{\ell}_{L,i} e_{R,j} \left[ Y^e_{1,ij} \tilde{h}_1 + Y^e_{2,ij} \tilde{h}_2 \right] \, , 
 \end{align}
where the structure of the couplings $Y^e_{1}$ and $Y^e_{2}$ depends on the model under consideration. We conveniently parametrize these couplings as follows: one matrix can be chosen to  have only a 33-entry without loss of generality (for E1L and E1R $Y^e_{1}$ and for E2L and E2R $Y^e_{2}$), while the other matrix can be implicitly defined through  charged lepton masses and rotations upon the relation
\begin{align}
v c_\beta Y^e_1 + v s_\beta Y^e_2 = M_e = V_{EL} m_e^{\rm diag} V_{ER}^\dagger \, .
\end{align}
Thus we take for models E1L and E1R 
\begin{align}
Y^e_1 & = \frac{1}{c_\beta v}  \begin{pmatrix} 0 & 0 &  0   \\ 0 & 0 & 0   \\  0 & 0 &   m_\tau  \\ \end{pmatrix}  \, , & 
Y^e_2 & =  \frac{1}{s_\beta v} V_{EL} m_e^{\rm diag} V_{ER}^\dagger -  \frac{1}{s_\beta v}  \begin{pmatrix} 0 & 0 &  0   \\ 0 & 0 & 0   \\  0 & 0 &   m_\tau  \end{pmatrix}  \, , 
\label{YukE1}
\end{align}
and for models E2L and E2R 
\begin{align} 
Y^e_1 & =  \frac{1}{c_\beta v} V_{EL} m_e^{\rm diag} V_{ER}^\dagger -  \frac{1}{c_\beta v}  \begin{pmatrix} 0 & 0 &  0   \\ 0 & 0 & 0   \\  0 & 0 &   m_\tau  \end{pmatrix}  \, , 
  & Y^e_2 & = \frac{1}{s_\beta v}  \begin{pmatrix} 0 & 0 &  0   \\ 0 & 0 & 0   \\  0 & 0 &   m_\tau  \\ \end{pmatrix} \, . 
  \label{YukE2}
\end{align}
The six rotation angles in $V_{EL}$ and $V_{ER}$ are taken as free parameters, apart from the constraints imposed by the choice of $\xi_{ii}^{e_L}$ and  $\xi_{ii}^{e_R}$ in the specific scenario. 

\subsection{Lagrangian in the Mass Basis}\label{sec:L_mass_basis}
The Lagrangian in the mass basis (where the mass terms of fermions and Higgs fields are diagonal) is given  by
\begin{align}
    \mathcal{L}_\mathrm{H} &= 
            - \overline{u}_{L,i} H^0_k \left[ \frac{m_{u_i} }{v \ssb} \delta_{ij} \alpha_k^u + \eps^u_{ij} \beta_k^u \right] u_{R,j} 
            - \overline{d}_{L,i} H^0_k \left[ \frac{m_{d_i} }{v \ccb} \delta_{ij} \alpha_k^d + \eps^d_{ij} \beta_k^d \right] d_{R,j} \nonumber \\ 
          & - \overline{e}_{L,i} H^0_k \left[ \frac{m_{e_i} }{v \ccb} \delta_{ij} \alpha_k^e + \eps^e_{ij} \beta_k^e \right] e_{R,j}
            + \overline{d}_{L,i} H^- V_{ki}^*    \sqrt2 \left[ \frac{ m_{u_j}}{v \ssb}\delta_{kj} \ccb - \frac{\eps^u_{kj} }{\ssb} \right] u_{R,j} \nonumber \\
 &   + \overline{u}_{L,i} H^+ V_{ik}\sqrt2 \left[ \frac{ m_{d_j}}{v \ccb} \delta_{kj} \ssb - \frac{\eps^d_{kj}}{\ccb}  \right] d_{R,j}
 + \overline{\nu}_{L,i} H^+ U_{ki}^*\sqrt2 \left[ \frac{ m_{e_j}}{v \ccb} \delta_{kj} \ssb - \frac{\eps^e_{kj}}{\ccb}  \right] e_{R,j} +  {\rm h.c.} \,, 
 \label{LH}
\end{align}
where the index $k=1,2,3$ runs over the  three neutral physical Higgs fields $H^0_k = (h, H, A)$ while $i,j$ are fermion flavor indices. The couplings depend  on rotation angles in the scalar sector through $\alpha^f_k, \beta^f_k$ and on rotation matrices in the fermion sector through $V_{ij}, U_{ij}, \eps^f_{ij}$.

The couplings from the scalar rotations are given by
\begin{align}
    \alpha^u  &= ( \cca \,, \ssa \,, -i \ccb ) \,, 
    & \alpha^d  &= \alpha^e = ( -\ssa \,, \cca \,, -i \ssb ) \,,   \\
     \beta^u &= \frac{1}{ s_\beta} \left( -c_{\alpha - \beta} \,, -s_{\alpha - \beta} \,, i \right) \,, 
      & \beta^d &= \beta^e =  \frac{1}{ c_\beta} \left( c_{\alpha - \beta} \,, s_{\alpha - \beta} \,, i \right)   \,, 
\end{align}
and only depend on the two Higgs sector angles $\alpha$ and $\beta$ which we treat as free parameters (obtained by a suitable scalar potential). The relation between the above Higgs mass eigenstates $H^0_k$ and the Higgs doublet fields $h_{1,2}$ in Eq.~\eqref{eq:Yukawa couplings} is spelled out in Appendix~\ref{AppB}. 

The couplings from the fermion rotations read 
\begin{align}
    \eps^u & = V_{UL}^\dagger Y^u_1 V_{UR} \,,
    & \eps^d & = V_{DL}^\dagger Y^d_2 V_{DR} \,, 
    & \eps^e & = V_{EL}^\dagger Y^e_2 V_{ER} \,,  
\end{align}
and only depend on the unitary rotations $V_{fP}$ that diagonalize quark and  lepton masses, and the Yukawa couplings $Y^u_1, Y^d_2, Y^e_2$. The unitary rotations also fix the physical CKM and PMNS matrices $V$ and $U$
 \begin{align}
       V &\equiv V_\mathrm{CKM} = V_{UL}^\dagger V_{DL}   \,,
     & U &\equiv U_\mathrm{PMNS}  = V_{EL}^\dagger V_{\nu L}   \,,
\end{align}
The explicit structure of the quark Yukawa couplings in Eqs.~\eqref{YukQ1}-\eqref{YukQ4} determines the quark unitary rotations for a given model Q1-Q4, and gives for the final couplings  from the fermion rotations the following matrices:
\paragraph{Model Q1:} 

\begin{align}
\eps^u &= 0,\label{epsQ1} \\ 
\eps^d &= \frac{1}{s_\beta v}  \begin{pmatrix} 0 & 0 &  0   \\ 0 & m_s & 0   \\  0 & 0 &   m_b  \\ \end{pmatrix} \, ,
\end{align}

\paragraph{Model Q2:} 
\begin{align}
\eps^u &\sim \frac{1}{c_\beta v}
    \begin{pmatrix}
     \lambda^6 m_u & \lambda^5 m_c & \lambda^3 m_t \\
     \lambda^5 m_u & \lambda^4 m_c & \lambda^2 m_t   \\ 
     \lambda^3 m_u & \lambda^2 m_c &   m_t 
     \end{pmatrix} + \mathrm{higher\ orders},\\
\eps^d &= \frac{1}{s_\beta v}  \begin{pmatrix} 0 & 0 &  0   \\ 0 & 0 & 0   \\  0 & 0 &   m_b  \\ \end{pmatrix}  \, , 
\end{align}

\paragraph{Model Q3:} 
\begin{align}
\eps^u &\sim \frac{1}{c_\beta v}
\begin{pmatrix}
\lambda^2 \frac{m_c^2 m_u}{m_t^2} & \lambda\frac{m_c^3}{m_t^2} &  \frac{1}{\lambda} \frac{m_c^2}{m_t} \\
\lambda^9 m_u & \lambda^2 \frac{m_c  (m_t^2 \lambda^6 - m_c^2)}{m_t^2} & \frac{m_c^2 - m_t^2 \lambda^6}{m_t}  \\
\lambda^3 m_u & \lambda^2 m_c &   m_t 
\end{pmatrix} + \mathrm{higher\ orders},\\
\eps^d &= \frac{1}{s_\beta v}  \begin{pmatrix} 0 & 0 &  0   \\ 0 & 0 & 0   \\  0 & 0 &   m_b  \\ \end{pmatrix}  \, , 
\end{align}

\paragraph{Model Q4:} 
\begin{align}
\eps^u &\simeq \frac{1}{c_\beta v}  \begin{pmatrix} 0 & 0 &  0   \\ 0 & m_c & 0   \\  0 & 0 &   m_t  \\ \end{pmatrix} + \mathrm{higher\,\,orders},\\
\eps^d &= 0  \, ,  \label{epsQ4} \\
\nonumber 
\end{align}
where we have neglected sub-leading corrections suppressed by 
small mass ratios and/or by higher power of the Cabibbo angle.

Similarly the explicit structure of the lepton Yukawa couplings in Eqs.~\eqref{YukE1}-\eqref{YukE2} gives for the final couplings  from the lepton rotations the following matrices:
\paragraph{Model  E1L and E1R:} 
 \begin{align}
(\eps^e_{E1})_{ij} =  \frac{m_{e_i} \delta_{ij} }{s_\beta v}   -  \frac{m_\tau}{s_\beta v} (V^*_{EL})_{3i}   (V_{ER})_{3j}   \, , 
 \label{epsE1}
\end{align}
\paragraph{Model  E2L and E2R:} 
\begin{align}
(\eps^e_{E2})_{ij} =  \frac{m_\tau}{s_\beta v} (V^*_{EL})_{3i}   (V_{ER})_{3j}   \, , 
 \label{epsE2}
\end{align}
which only depends on the third row of the rotation matrices in the left- and right-handed sectors.  These can be expressed through the  four real parameters $\xi^{e_L}_{11}, \xi^{e_L}_{22}, \xi^{e_R}_{11} , \xi^{e_R}_{22}$, up to complex phases, which we set to zero for simplicity: they do not enter $h \to \tau \ell$ and would only have a minor impact on the Higgs couplings to $\tau$'s. Thus we finally have for models E1L and E1R 
\begin{align}
(\eps^e_{E1})_{ij} =  \frac{m_{e_i} \delta_{ij} }{s_\beta v}   -  \frac{m_\tau}{s_\beta v} \sqrt{\xi^{e_L}_{ii} \xi^{e_R}_{jj}}   \, , 
 \label{epsE1_xi}
\end{align}
and  for models E2L and E2R 
\begin{align}
(\eps^e_{E2})_{ij} =  \frac{m_\tau}{s_\beta v}\sqrt{\xi^{e_L}_{ii} \xi^{e_R}_{jj}}   \, , 
 \label{epsE2_xi}
\end{align}
where we remind the reader that $0 \le \xi^{e_{L}}_{ii} ,\xi^{e_{R}}_{ii} \le 1 $. 

To summarize, we consider generalized DFSZ-like models that are defined by the Yukawa structure in Eq.~\eqref{L1}. In total we have 4 different models in the quark sector (Q1-Q4) and 4 different models in the charged lepton sector (E1L, E1R, E2L, E2R). These scenarios depend on 7 parameters; the 4 angular parameters $\xi_{11}^{e_L}, \xi_{22}^{e_L}, \xi_{11}^{e_R}, \xi_{22}^{e_R}$ controlling lepton flavor-violation, the vacuum angle $\beta$ entering both axion and Higgs couplings, and two parameters that control overall decoupling: the Higgs mixing angle $\alpha$   for Higgs phenomenology and the axion decay constant $f_a$ for axion phenomenology. The corresponding predictions for the axion couplings to the SM are summarized in Tables \ref{Qmodels} and \ref{Lmodels}, while the structure of neutral and charged Higgs couplings to fermions are given in Eq.~\eqref{LH}, along with the predictions in Eqs.~\eqref{epsQ1}-\eqref{epsQ4} and \eqref{epsE1_xi}-\eqref{epsE2_xi}. In the following we study first the phenomenology of axion couplings, focussing on the possibility to address stellar cooling anomalies, before we turn to Higgs phenomenology, paying particular attention to flavor-violating decays of the SM-like Higgs. 


\section{Axion Phenomenology}
\label{sec:axion_pheno}
In this section we provide the expressions for the low-energy axion couplings to nucleons, leptons and photons and use the present constraints in order to find the allowed regions in the parameter space. Finally we discuss the stellar cooling anomalies which hint to non-vanishing axion couplings to electrons and select a preferred region of model parameters, in particular yielding an upper bound on the axion decay constant.

\subsection{Predictions for Axion Couplings}
The axion has no relevant flavor-violating couplings to quarks, and the most important observables in the quark sector arise from axion couplings to nucleons constrained by SN1987A and neutron star cooling. In the different quark models the predictions for the low-energy axion couplings to protons $C_p$ and neutrons $C_n$ read~\cite{diCortona:2015ldu}
\begin{align}
C_p &= 0.44 - 1.30  s_\beta^2 \, , &  C_n & = -0.38 + 1.24 s_\beta^2 \, , & ( & {\rm Q1}) \\
C_p &= 0.41 - 1.30 s_\beta^2 \, , &  C_n & = -0.41 + 1.24 s_\beta^2 \, , &  ( & {\rm Q2,Q3})\\
C_p &=  0.41 - 1.30 s_\beta^2 \, , & \,\,\, C_n & = -0.41 +1.24 s_\beta^2 \, . &  ( & {\rm Q4})
\end{align}
To derive the above expressions we have neglected Yukawa running effects (see  Ref.~\cite{Choi:2017gpf,  MartinCamalich:2020dfe, Heiles:2020plj, Chala:2020wvs, Bauer:2020jbp, Choi:2021kuy}), which in DFSZ models are relevant only below the scale where the heavy Higgs doublet of the 2HDM decouples~\cite{Choi:2021kuy}. Since in our setup this mass scale is rather low (close to the TeV scale) we do not expect large corrections from top Yukawa running.

In the lepton sector instead there are flavor-violating couplings controlled by the parameters $\xi_{ii}^e$, and the most important observable in the lepton sector are $\mu \to e a$ decays arising from the flavor-violating axion coupling $C_{\mu e} = \sqrt{|C_{\mu e}^A|^2 + |C_{\mu e}^V|^2}$, and axion couplings to muons $C_\mu$ and electrons $C_e$, which are constrained by SN1987A and white dwarf cooling, respectively. The predictions for these couplings are 
\begin{align}
C_e &= -c_\beta^2 + \xi^{e_L}_{11} \, , & C_\mu &= -c_\beta^2 + \xi^{e_L}_{22} \, , & C_{\mu e} & =  \sqrt{2 \xi^{e_L}_{11} \xi^{e_L}_{22}} \, , & (& {\rm E1L}) \\
C_e &= -c_\beta^2 + \xi^{e_R}_{11} \, ,& C_\mu &= -c_\beta^2 + \xi^{e_R}_{22} \, ,& C_{\mu e} & = \sqrt{2 \xi^{e_R}_{11} \xi^{e_R}_{22}} \, , & (& {\rm E1R})  \\
C_e &= s_\beta^2 -  \xi^{e_L}_{11}\, , & C_\mu &= s_\beta^2 - \xi^{e_L}_{22} \, ,& C_{\mu e} & = \sqrt{2 \xi^{e_L}_{11} \xi^{e_L}_{22} }\, , & (& {\rm E2L})   \\
C_e &= s_\beta^2 -  \xi^{e_R}_{11}\, , & C_\mu &= s_\beta^2 - \xi^{e_R}_{22} \, ,& C_{\mu e} & = \sqrt{2 \xi^{e_R}_{11} \xi^{e_R}_{22}} \, . & (& {\rm E2R})  
\end{align}
Finally there is the coupling to photons, which is determined by the ratio of the electromagnetic and color anomaly coefficient that is fixed in each model, see  Tables~\ref{Qmodels} and \ref{Lmodels}. There are only four distinct values $E/N = \{ -4/3, 2/3, 8/3, 14/3 \}$, which determine the axion couplings to photons (cf. Eq.~\eqref{LaQCD})
\begin{align}
C_{ \gamma} & = E/N - 1.92 \, . 
\end{align}
 The axion phenomenology thus depends on four  parameters: the Higgs vacuum angle $\beta$, the relevant rotations in the charged lepton sector $\xi_{11}^{e_L}, \xi_{22}^{e_L}$ (E1L, E2L) or  $\xi_{11}^{e_R}, \xi_{22}^{e_R}$ (E1R, E2R) and the axion decay constant $f_a$. In the following we will simplify the notation and use the parameters $\xi_{11}, \xi_{22}$ for all four models, where the chirality is left understood from the model under consideration. Thus we will distinguish only between E1 and E2 models, which represent (E1L,E1R) and (E2L,E2R) models upon the proper identification of  $\xi_{11}, \xi_{22}$. The phenomenology is indeed essentially independent of the chirality of $\xi_{ii}$, except bounds from LFV decays as we are going to see below.  
\subsection{Constraints on Axion Couplings to Photons}
Axion couplings to photons are constrained mainly by the CAST experiment~\cite{Anastassopoulos:2017ftl} and the evolution of horizontal branch stars in globular clusters~\cite{Ayala:2014pea}, which at 95\% CL require 
\begin{align}
\frac{\alpha}{2 \pi f_a}C_{\gamma} & \le 6.6 \times 10^{-11} \GeV^{-1} \, ,
\end{align}
and exclude the region where
\begin{align}
f_a \le 1.8 \times 10^7 \GeV \left| E/N - 1.92 \right| \, .
\end{align}
The most stringent constraint arise for models with $E/N = -4/3$, and exclude $f_a \le 5.7 \times 10^7 \GeV$, which is typically weaker than other constraints. However, the bound on photon couplings will be improved by helioscopes of the next generation, in particular the IAXO experiment, by about an order of magnitude~\cite{Armengaud:2014gea}.
\subsection{Constraints on Axion Couplings to Nucleons}
\label{anconstraints}
The couplings to neutron and protons can be bounded by the burst duration of the neutrinos observed in the SN1987A and reads~\cite{SNbound2}
\begin{equation}
0.61 g_{ap}^2+g_{an}^2 + 0.53 g_{an} g_{ap} < 8.26\times 10^{-19} \, , 
\end{equation}
where $g_{ai}=C_{i} m_{i}/f_a$ with $i=n,p$.  This excludes the region where\footnote{Taking into account also axion production from thermal pions besides nucleon bremsstrahlung would strengthen the bound roughly by  a factor 2~\cite{Carenza:2020cis}.}
\begin{align}
f_a &\lesssim 4.3 \times 10^{8} \GeV \frac{\sqrt{1-4.2 \, t_\beta^2+4.5 \, t_\beta^4}}{1+t_\beta^2} \, , & ( & {\rm Q1})  \nonumber\\
f_a &\lesssim 4.4 \times 10^{8} \GeV \frac{\sqrt{1-4.1 \, t_\beta^2+4.2 \, t_\beta^4}}{1+t_\beta^2} \, . & ( & {\rm Q2, Q3,Q4}) 
\label{eq:SN}
 \end{align}
Since the nature of the axion emission is not completely understood and present simulations do not take into account all the relevant physics (for example the feedback from axion emission  presumably modifies the bound when included in simulations~\cite{SNbound2}), these constraints should not be considered as a robust bound but rather as an indication.
 
However, also the cooling of neutron stars provides information about the axion nucleon coupling~\cite{Keller:2012yr,Sedrakian:2015krq,Hamaguchi:2018oqw,Beznogov:2018fda,Sedrakian:2018kdm} and yields upper bounds that are at least of the same order as the SN1987A bound. For example the observation of the neutron star in HESS J1731-347~\cite{Beznogov:2018fda} sets a strong bound on the axion neutron coupling as
\begin{equation}
g_{an} \lesssim 2.8\times 10^{-10},
\end{equation}
which translates to the excluded region
\begin{align}
f_a &\lesssim 1.3 \times 10^9  \GeV  \frac{| 1- 2.3\,  t_\beta^2 |}{1+t_\beta^2} \, , & ( & {\rm Q1}) \nonumber\\
f_a &\lesssim 1.4 \times 10^9  \GeV \frac{| 1- 2.0 \, t_\beta^2 |}{1+t_\beta^2} \, . & ( & {\rm Q2, Q3,Q4})
\label{eq:NS}
\end{align}
Still, the limited understanding of the cooling of neutron stars and the lack of observational data suggest that these constraints should be taken with some grain of salt.

\subsection{Constraints on Axion Couplings to Electrons}
The axion coupling to electrons can be constrained by the shape of the white dwarf luminosity function, giving the 95\% CL bound~\cite{WDbound} 
\begin{equation}
|g_{ae}|\lesssim2.2 \times 10^{-13}\, , 
\end{equation}
where $|g_{ae}|=C_e m_e / f_a$, which excludes the region
 \begin{align}
f_a & \lesssim 2.3 \times 10^{9} \GeV \frac{|1- \xi_{11} (1+t_\beta^2) |}{1+t_\beta^2} \, ,  & ( & {\rm E1L, E1R}) \\
f_a & \lesssim 2.3 \times 10^{9} \GeV \frac{|t_\beta^2- \xi_{11} (1+t_\beta^2) |}{1+t_\beta^2} \, .  & ( & {\rm E2L,E2R})
\end{align}

 \subsection{Constraints on Axion Couplings to Muons}
It has been shown recently~\cite{Bollig:2020xdr,Croon:2020lrf}
 that SN1987A also constraints the axion couplings to muons 
 \begin{align}
\frac{2 f_a} {|C_{\mu}|} \ge 1.3 \times 10^{8} \GeV \, ,
\end{align}
excluding the region where
\begin{align}
f_a & \lesssim 6.5 \times 10^{7} \GeV \frac{|1- \xi_{22} (1+t_\beta^2) |}{1+t_\beta^2} \, ,  & ( & {\rm E1L, E1R}) \\
f_a & \lesssim 6.5 \times 10^{7} \GeV \frac{|t_\beta^2- \xi_{22} (1+t_\beta^2) |}{1+t_\beta^2} \, .  & ( & {\rm E2L,E2R})
\end{align}

 \subsection{Constraints on LFV Axion Couplings}
Finally, limits on the LFV coupling $C^{V,A}_{\mu e}$ arise from constraints on the two-body lepton decay $\mu^+\to e^+ a$, which depends on the chiral structure. For an isotropic decay  the most stringent bound was provided by an experiment at TRIUMF, which sets
the limit ${\rm BR}(\mu^+ \to e^+ a) < 2.6 \times 10^{-6}$ (at $90\%$ CL.)~\cite{muea}. If the decay has the same angular distribution as the SM, the weaker bound from the TWIST experiment~\cite{Bayes:2014lxz} applies, ${\rm BR}(\mu^+ \to e^+ a) < 5.8 \times 10^{-5}$ (at $90\%$ C.L.). The resulting bounds on the axion couplings (including a recast of the TRIUMF experiment) have been given in Ref.~\cite{Calibbi:2020jvd} and read for purely left-handed or right-handed  couplings (at  $95\%$ C.L.)
\begin{align}
\frac{2 f_a}{|C_{\mu e}|} & \ge 1.0 \times 10^9 \GeV \, , & [C_{\mu e}^V & = - C_{\mu e}^A] & \\
\frac{2 f_a}{|C_{\mu e}|} & \ge 4.9 \times 10^9 \GeV \, , &  [C_{\mu e}^V & = C_{\mu e}^A]
\end{align}
which excludes the region
\begin{align}
f_a & \lesssim 7.1 \times 10^8 \GeV \, \sqrt{\xi_{11} \xi_{22}} \, , & ( & {\rm E1L,E2L}) \\
f_a & \lesssim 3.5 \times 10^9 \GeV \, \sqrt{\xi_{11} \xi_{22}}\, . & ( & {\rm E1R,E2R}) 
\end{align}
Note that this constraint is different for left-handed  or right-handed  lepton couplings, because the SM background on $\mu \to e + {\rm invis.}$ is purely left-handed and thus the LH models have weaker constraints. This is the only feature that allows to distinguish LH and RH models with axion physics (when $\xi_{ii}^{e_L} = \xi_{ii}^{e_R}$). 
\subsection{Summary of Constraints from Axion Physics}
\begin{figure}[t!]
\begin{center}
	\includegraphics[width=0.49\textwidth]{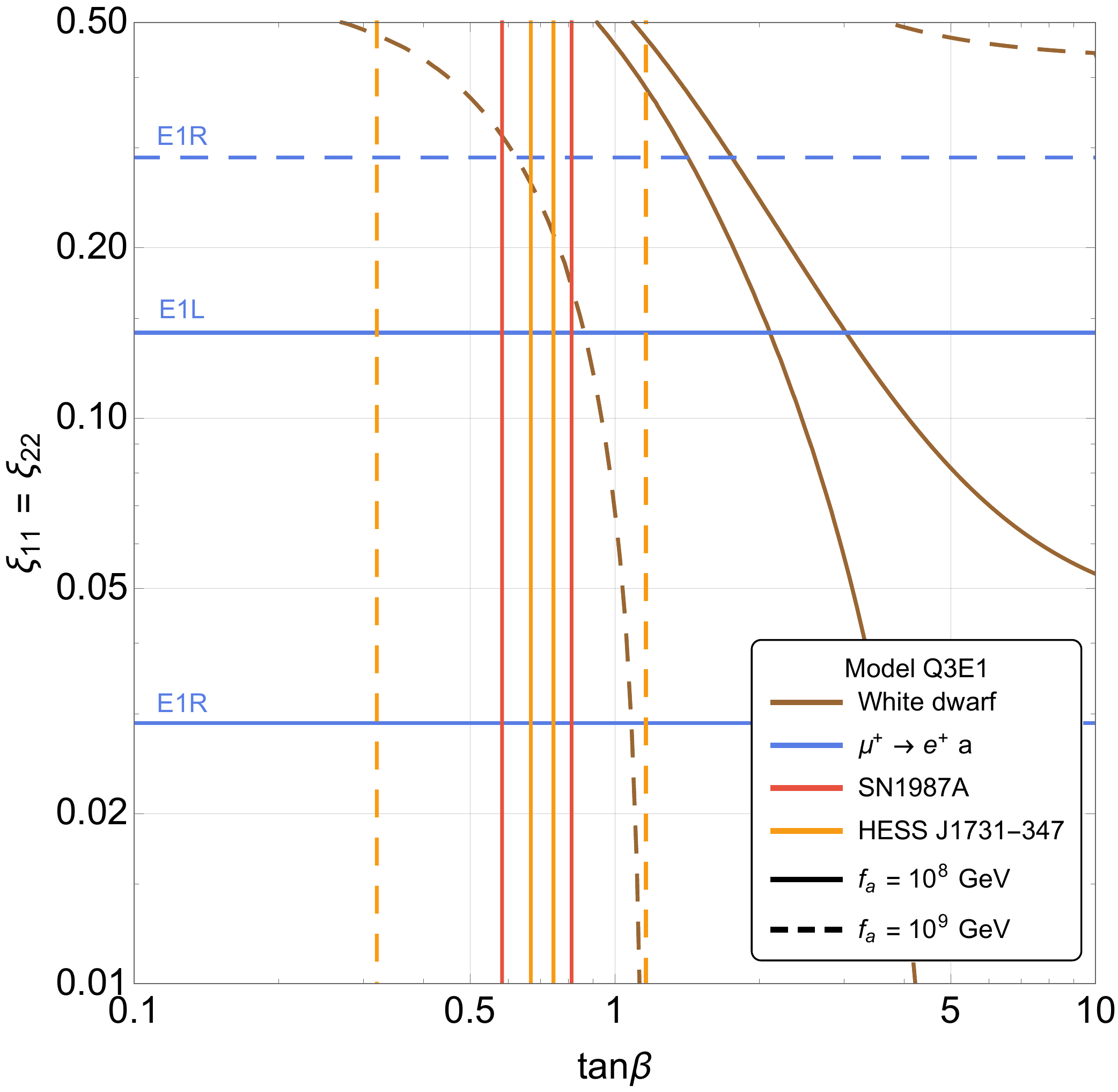}
	\includegraphics[width=0.49\textwidth]{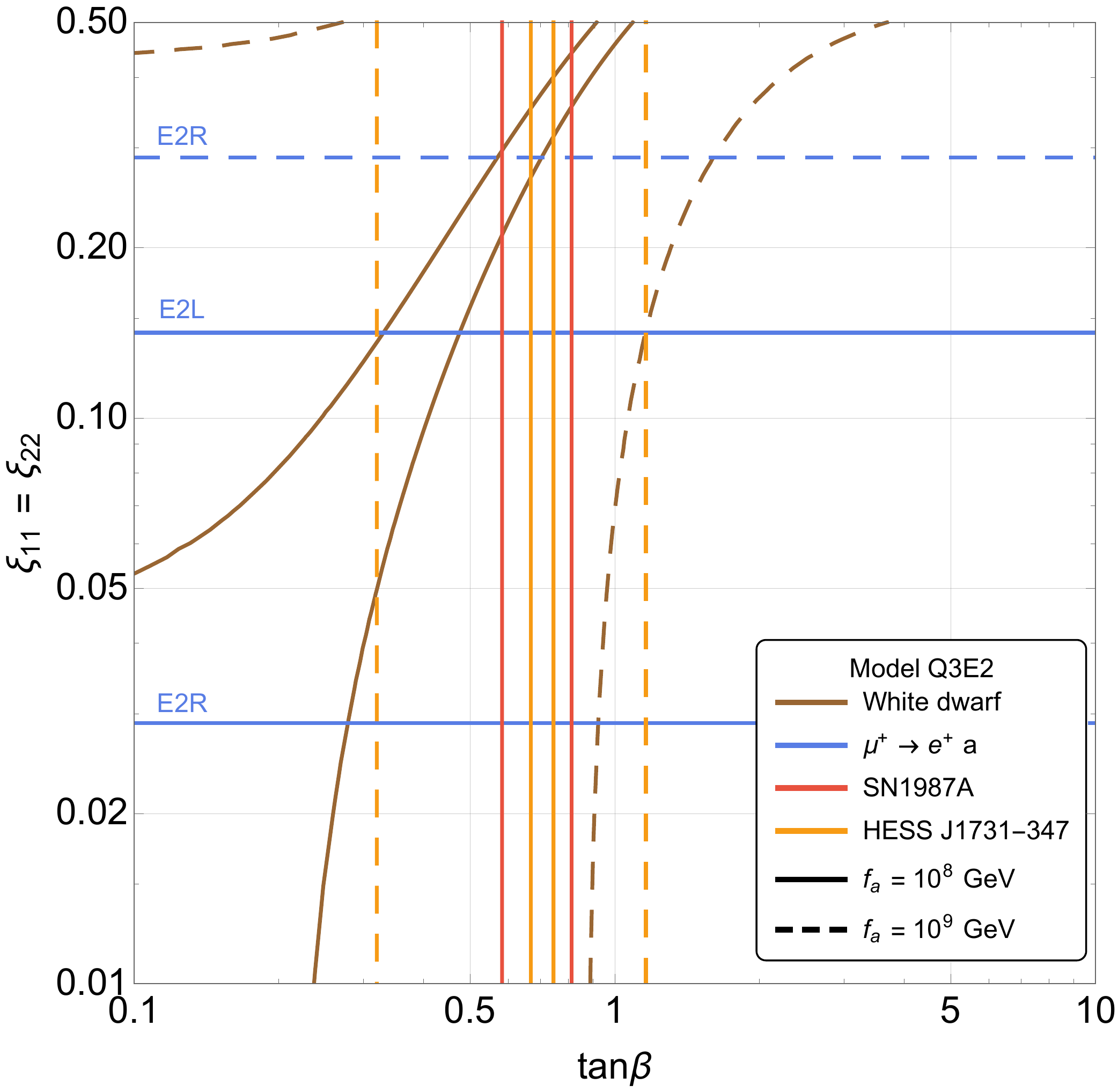}
\caption{Allowed regions of the Q3E1L/Q3E1R (left panel) and Q3E2L/Q3E2R (right panel) models  in the $\tan\beta-(\xi_{11}=\xi_{22})$ plane.  The constraints from white dwarf cooling (brown), neutron star cooling (orange) and SN1978A (red) allow only the region between  the two solid curves (for $f_a = 10^8 \GeV$) and dashed curves  (for $f_a = 10^9 \GeV$), while constraints from $\mu \to e a$ exclude the regions above  the horizontal blue lines, for $f_a = 10^8 \GeV$ (solid) and $f_a = 10^9 \GeV$ (dashed). Note that choosing $\xi_{22}= 0$ gives the same plot with the constraint from $\mu \to e a$ removed. 
\label{fig:fa}}
\end{center}
\end{figure}

The constraints on $f_a$ are summarized in Fig.~\ref{fig:fa} for the models Q3E1L/Q3E1R (left panel) and Q3E2L/Q3E2R (right panel) for $\xi_{11}=\xi_{22}$. The figures show the allowed regions in the $\tan\beta$ and $\xi_{11} = \xi_{22}$ parameters space for a given $f_a$. The constraints from white dwarf cooling (brown), neutron star cooling (orange) and SN1978A (red) allow only the region between  the two solid curves (for $f_a = 10^8 \GeV$) and dashed curves  (for $f_a = 10^9 \GeV$), while the constraints from $\mu \to e a$ exclude the regions above  the horizontal blue lines, for $f_a = 10^8 \GeV$ (solid) and $f_a = 10^9 \GeV$ (dashed), distinguishing between E1L/E1R and E2L/E2R models (for  $f_a = 10^9 \GeV$ the entire region of E1L and E2L models is allowed by $\mu \to e a$). We show only the Q3 model as a representative, because choosing a different quark model would only slightly change the bounds from supernovae and neutron stars, cf.~Eqs.~\eqref{eq:SN} and \eqref{eq:NS}.  Note that the bounds from white dwarfs are the same for the models E1L, E1R and for the models E2L, E2R, while the bounds from $\mu \to e a$ are equal for the models E1L, E2L and the models E1R, E2R.

Fig.~\ref{fig:fa} shows that it is not possible to have models with $f_a\lesssim$ a few $10^8$ GeV and $\xi_{11} = \xi_{22}$ that satisfy all  constraints simultaneously. This is due to the fact that the bounds from $\mu \to e a$ and white dwarfs select opposite regions in $\xi_{ii}$. However, when $\xi_{11}\neq \xi_{22}\to0$ the bound from $\mu \to e a$ does not apply anymore, and it is possible to have (mild) cancellations in nucleon and electron couplings near $\tan \beta \approx 0.7$ and $\xi_{11} \approx 2/3$ (E1L,E1R) or $\xi_{11} \approx 1/3$ (E2L,E2R) such that $f_a\lesssim 10^8$ GeV is allowed. For  $f_a\lesssim 10^9$ GeV the parameter space opens up, but still sizable regions of parameter space are excluded even in the limit $\xi_{22} \to 0$. While clearly all bounds can be relaxed by further increasing the axion decay constant, an upper bound on $f_a$ arises when the axion is responsible for explaining the stellar cooling anomalies.
\subsection{Stellar Cooling Anomalies}
Several hints for excessive cooling in stellar objects have been observed in the past years. These observations include (see e.g. Ref.~\cite{Giannotti:2015dwa, Giannotti:2015kwo, Giannotti:2016hnk, CoolingAnom3}: i) the cooling efficiency of pulsating white dwarfs (WDs) extracted by the rate of period change; ii) the WD  luminosity function, relating the WD distribution to their brightness; iii)  the luminosity of the tip of the red giant branch (RGB) in globular clusters;  iv) the ratio of the number of horizontal branch (HB) stars over RGB stars in globular clusters (R-parameter); v) the ratio of blue and red supergiants in open clusters. 

The interpretation of these data in terms of BSM physics (such as millicharged particles, hidden photons and axions) was discussed in Ref.~\cite{CoolingAnom2}. This analysis revealed that axions coupled to electrons are perfectly suited to address all anomalies, in contrast to the other candidates. A combined fit of the WD luminosity function (WDLF), WD  pulsation and RGB stars is driven mainly by the WDLF and favors\footnote{A more conservative assessment on the systematic uncertainties   reduces the significance only slightly~\cite{Saikawa:2019lng}.} a non-zero axion coupling to electrons~\cite{CoolingAnom3} 
\begin{equation}
g_{ae} = 1.6^{+0.29}_{-0.34} \times 10^{-13} \, , 
\end{equation}
which translates to the region
\begin{align}
f_a & = 3.2^{+0.86}_{-0.49} \times 10^{9} \GeV \frac{|1- \xi_{11} (1+t_\beta^2) |}{1+t_\beta^2} \, ,  & ( & {\rm E1L, E1R}) \\
f_a & = 3.2^{+0.86}_{-0.49} \times 10^{9} \GeV \frac{|t_\beta^2- \xi_{11} (1+t_\beta^2) |}{1+t_\beta^2} \, .  & ( & {\rm E2L,E2R})
\end{align}
Since $0 \le \xi_{11} \le 1$, the fraction is always smaller than unity giving an upper bound on the axion decay constant $f_a \lesssim 4 \times 10^9 \GeV$ at $1\sigma$, corresponding to an axion heavier than at least $1$~meV.    

\begin{figure}[thbp]
\begin{center}
	\includegraphics[width=0.4\textwidth]{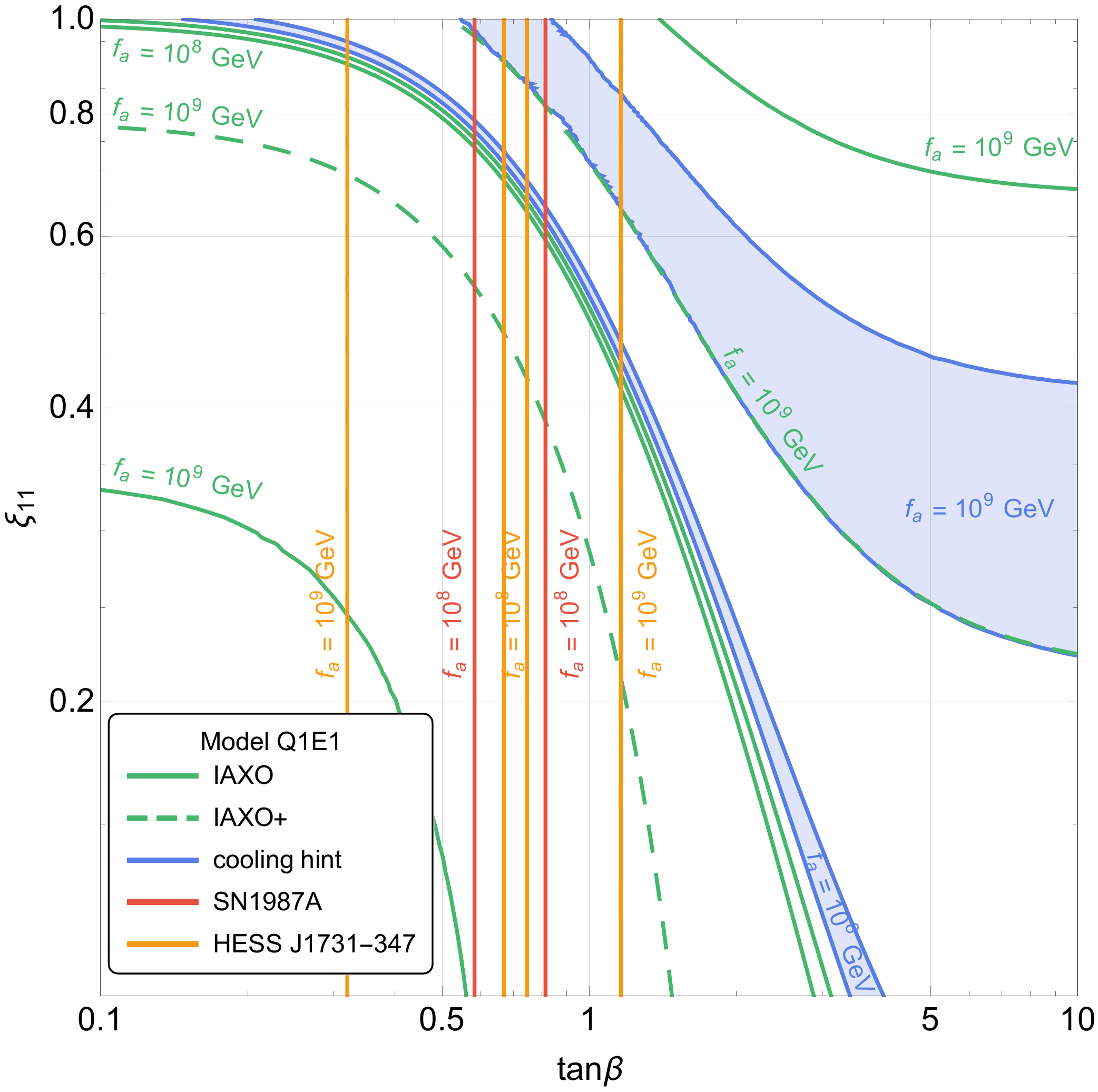}
	\includegraphics[width=0.4\textwidth]{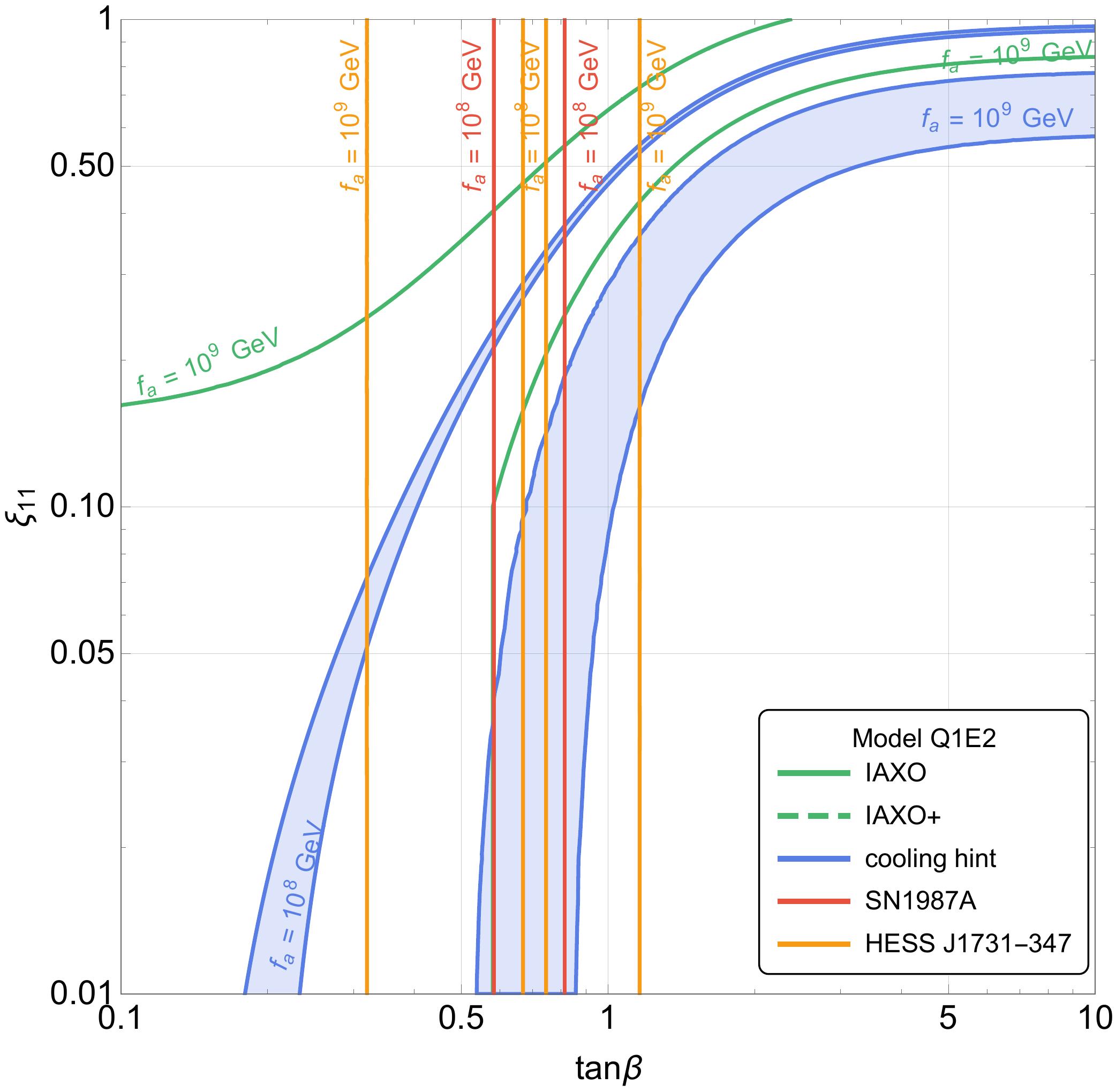}
	\includegraphics[width=0.4\textwidth]{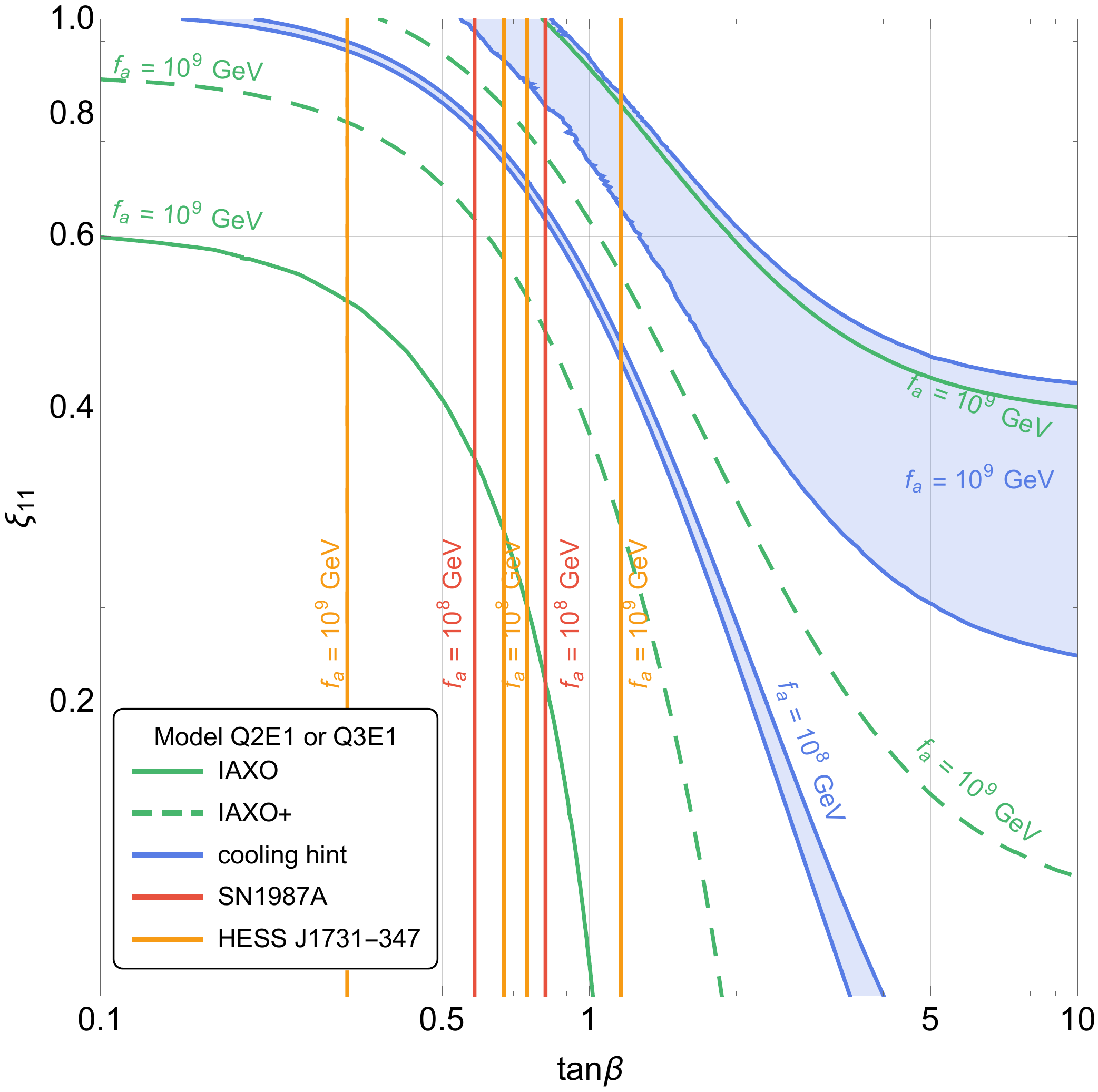}
	\includegraphics[width=0.4\textwidth]{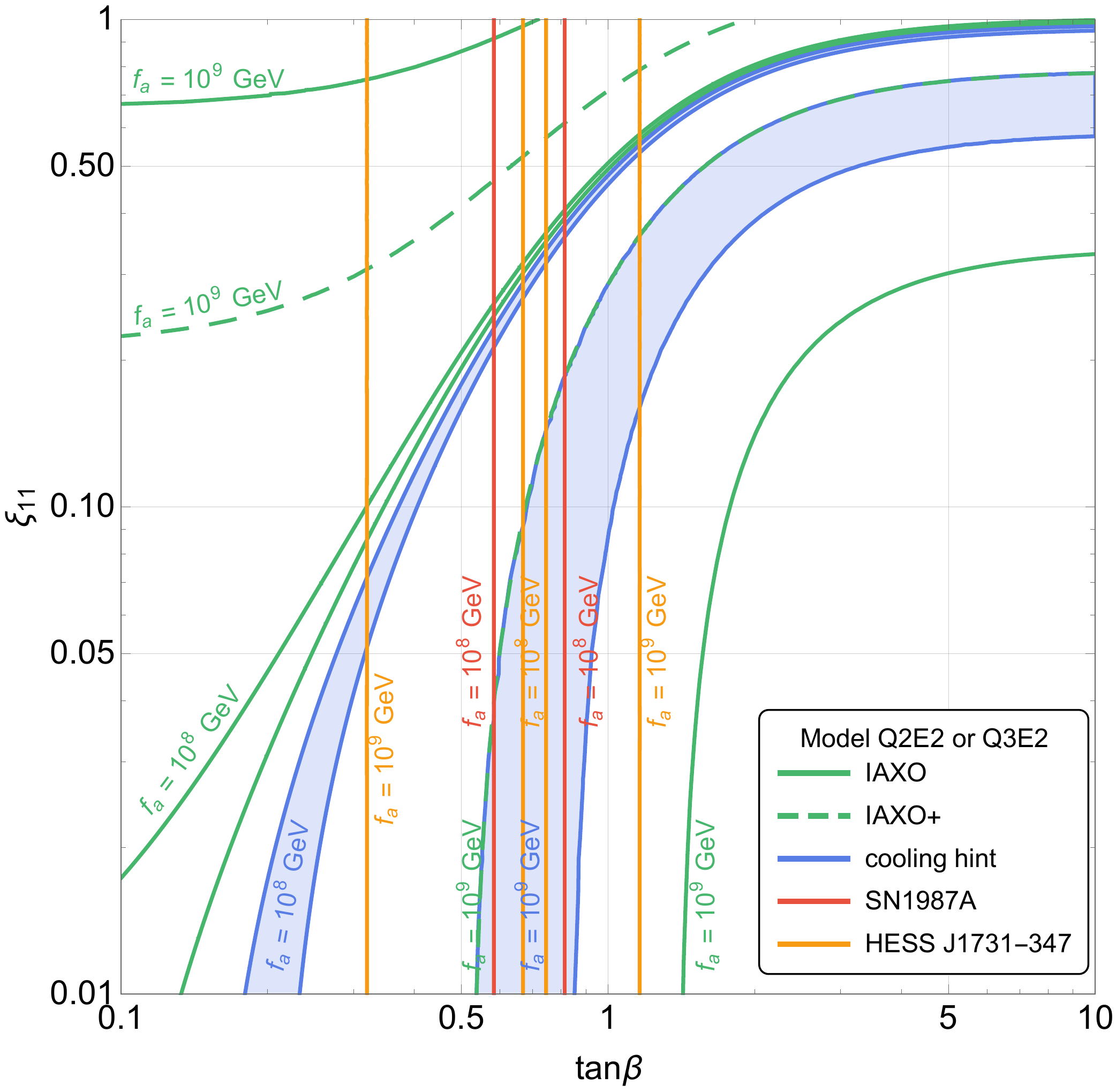}
	\includegraphics[width=0.4\textwidth]{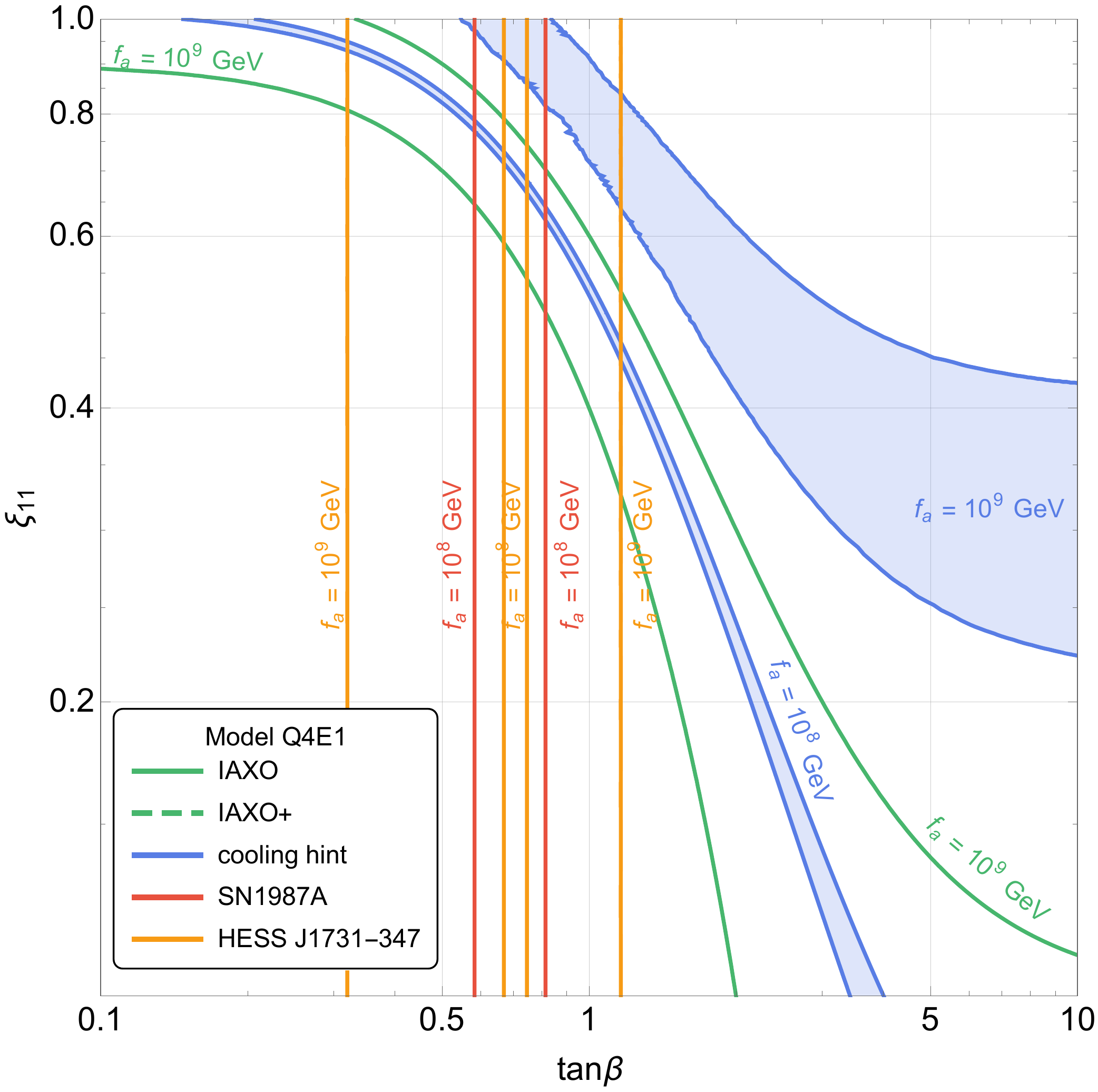}
	\includegraphics[width=0.4\textwidth]{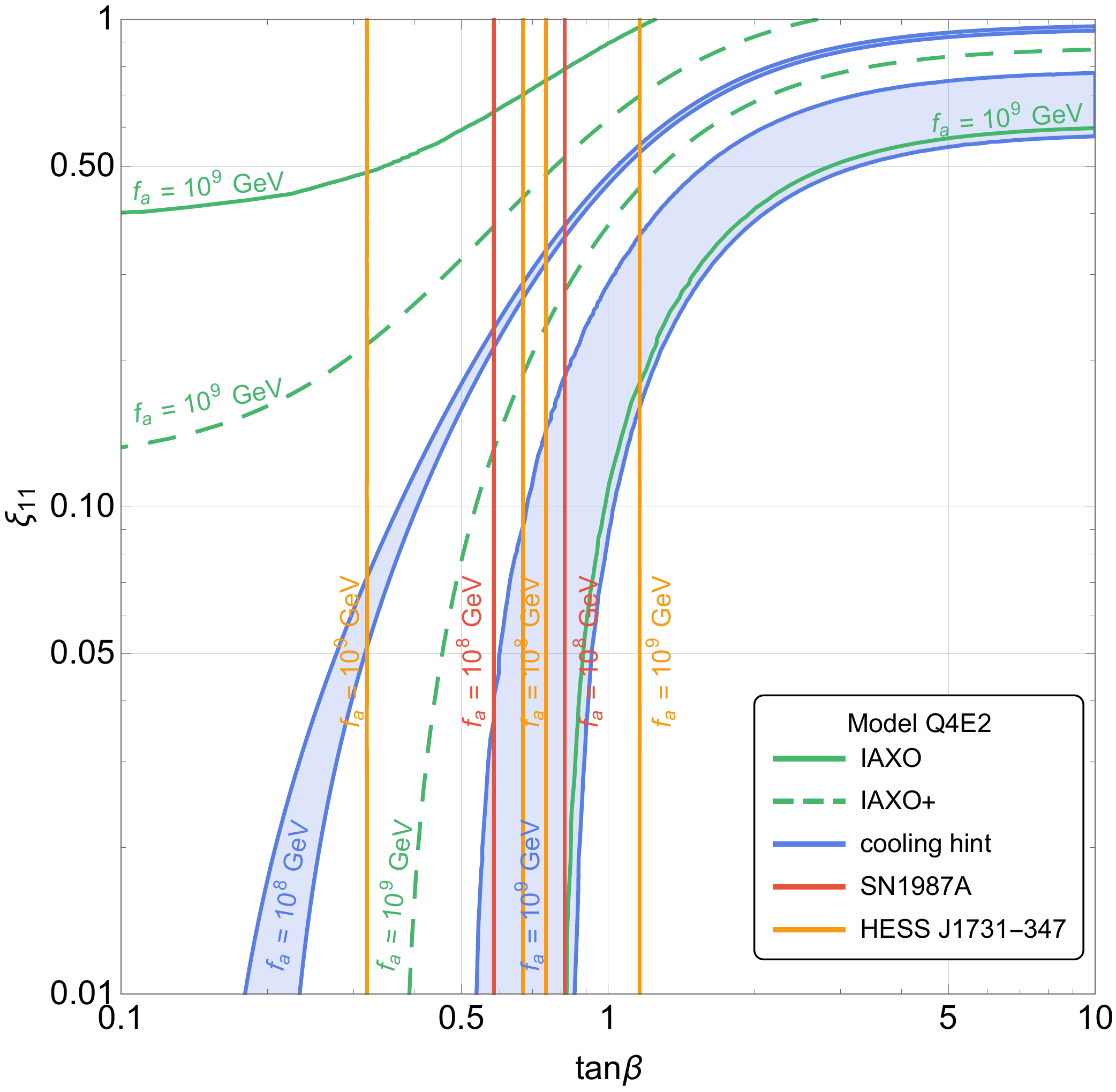}
\caption{
Projected potential of IAXO (green) and stellar cooling hint region (blue). For the indicated values of $f_a$ IAXO can probe the entire region between the solid green curves, while the dashed green contours show the region that can be tested by IAXO+ using a series of possible upgrades~\cite{IAXO3}. The blue regions are the $1\sigma$ cooling hint contours for a fixed value of $f_a$. The SN1987A and the neutron star bound from HESS J1731-347~\cite{Beznogov:2018fda} are given by the red and yellow curves, respectively.  \label{fig:IAXO}}
\end{center}
\end{figure}

We overlay the cooling hint region with the various constraints\footnote{We have also checked that additional bounds from the non-observation of X-rays from magnetic white dwarfs~\cite{Dessert:2021bkv} are always respected in the cooling hint region due to the rather large axion mass.} and the IAXO sensitivity in Fig. \ref{fig:IAXO}, which shows the $1 \sigma$ contours of the cooling hint, the reach of IAXO and IAXO+ as obtained in Ref.~\cite{CoolingAnom3}, and the bounds from supernovae and neutron stars for all the models, taking $\xi_{22}=0$. In particular, the solid green curves show the maximal reach of IAXO in its baseline configuration~\cite{IAXO2}. A more optimistic projection based on possible upgrades in the understanding of parameters of the magnet and the detectors is shown as a dashed line and it is labeled as IAXO+. On the other hand, the intermediate experimental stage called BabyIAXO~\cite{IAXO2} will have enough sensitivity to probe models with $E/N=-4/3$ (Q4E1) or $14/3$ (Q1E2) up to $f_a\sim 10^8$ GeV, assuming that the interaction with electrons is suppressed. The presence of a sizeable interaction with electrons may increase the actual sensitivity, due to a larger axion production rate in the sun.

The cooling hint regions are the same for all the E1 or E2 models. The IAXO reach, on the other hand, depends also on the value of $E/N$ and differs depending on both properties of the quark and lepton models, cf. Table \ref{Qmodels}-\ref{Lmodels}. Fig. \ref{fig:IAXO} shows that the area that satisfy both the cooling hint region and the different bounds dramatically depends on the type of the lepton model (E1 $vs$ E2). In particular, the models E1 prefer a region where $\xi_{11}\sim 1/\sqrt{2}$ ($\xi_{11}\gtrsim 0.6$) for $f_a=10^8$ ($10^9$) GeV, while the models E2 prefers a region with $\xi_{11}\sim 0.3$ ($\xi_{11}\lesssim 0.4$) for $f_a=10^8$ ($10^9$) GeV. Some of these regions could be probed by IAXO or IAXO$+$. The cooling regions for $f_a\sim 10^8$ GeV can all be probed by IAXO.\footnote{An absent curve for the IAXO reach in Fig.~\ref{fig:IAXO} implies that all of the corresponding parameter space will be probed by IAXO.} On the contrary, IAXO will not probe the cooling hint region for $f_a=10^9$ GeV for the models Q1E1 (upper left panel), and Q2E2 and Q3E2 (middle right panel), it will partially probe the region for the models Q2E1 and Q3E1 (middle left panel) and Q4E2 (bottom right panel), and it will completely probe the models Q1E2 (upper right panel) and Q4E1 (bottom left panel). The models that will not be probed by IAXO will be challenged by its upgrade IAXO$+$.

\section{Higgs Phenomenology}
\label{sec3}

All models can be tested also with precision Higgs physics at the LHC, provided that the heavy Higgs bosons are not decoupled. In particular all models predict flavor-violating decays of the SM-like Higgs, $h\to e \tau$, $h\to \mu \tau$, $h\to \mu e$, as well as modifications to the SM Higgs decays $h \to \tau\tau$ and $h \to \mu\mu$. All these observables can have large deviations from the SM prediction when the mass scale of the additional Higgs doublet is not too large.  The magnitude of this deviation for these Higgs decays is then only limited by the Higgs coupling measurements at the LHC and flavor-violating charged lepton decays like $\mu \to e \gamma$. In the following we discuss these constraints in detail, before we combine Higgs and axion phenomenology in the next section.

\subsection{Higgs coupling measurements}

The effects of non-decoupled heavy Higgs bosons can be parametrized by $c_{\beta-\alpha} \propto v^2/m_H^2$, which vanishes in the decoupling limit $m_H \to \infty$. The value of $c_{\beta-\alpha}$ affects all the couplings of the Higgs boson to SM particles. This implies that possible deviations of $c_{\beta-\alpha}$ from zero are constrained by the Higgs coupling measurements at the LHC. In order to study the impact of those constraints on the models 
we perform fits using the results from the combination of the measurements of Higgs boson production and decay by the ATLAS collaboration Ref.~\cite{Aad:2019mbh}. The fits are performed in the so-called $\kappa-\lambda$ framework where the parameters of the fit can be described in terms of measurements of the Higgs boson reduced couplings $\kappa$ and $\lambda$ as
\be
\kappa_{gZ}=\frac{\kappa_g \kappa_Z}{\kappa_h}\qquad\qquad \lambda_{ij}=\frac{\kappa_i}{\kappa_j} \, ,
\ee
with $(i,j) = (Z,g),(t,g),(W,Z),(\gamma,Z),(\tau,Z),(b,Z)$ and $\kappa_i$ are the couplings of the SM-like Higgs normalized to their values in the SM, e.g. the couplings to fermions are 
\be
\kappa_f=\frac{\sqrt{2}v }{ m_f}c_f^h \, ,
\ee
where $c_f^h$ are the couplings of the Higgs to the fermion $f$.

In absence of exotic Higgs decays, and neglecting the Higgs decays to $u$, $d$ and $e$, we have~\cite{Aad:2019mbh}:
\bea
\kappa_h^2 &\simeq& 0.58 \kappa_b^2+0.22\kappa_W^2+0.08\kappa_g^2+0.06\kappa_\tau^2+0.03\kappa_Z^2+0.03\kappa_c^2+2.3\times10^{-3}\kappa_\gamma^2\nonumber\\
&+& 1.5\times10^{-3}\kappa_{Z\gamma}^2+4\times10^{-4}\kappa_s^2+2.2\times10^{-4}\kappa_\mu^2 \, .
\eea
In the absence of new physics contributing to the effective couplings of the Higgs to gluon, photon and $Z\gamma$, we have the following scalings for the Higgs to gauge bosons
couplings~\cite{Aad:2019mbh}:
\bea
\kappa_W=\kappa_Z&=&\sin(\beta-\alpha)\nonumber\\
\kappa_{Z\gamma}^2 &\simeq& 0.00348\kappa_t^2 + 1.121\kappa_W^2 - 0.1249\,\kappa_t\kappa_W \, ,\nonumber\\
\kappa_g^2 &\simeq& 1.04\kappa_t^2 + 0.002\kappa_b^2 - 0.04\,\kappa_b\kappa_t\, ,\nonumber\\
\kappa_\gamma^2 &\simeq& 1.59\kappa_W^2 + 0.07\kappa_t^2 - 0.67\;\kappa_W\kappa_t \, .
\eea

The relevant experimental results are summarized in Table~\ref{tab:HC}, which can be found also in Table 12 of Ref.~\cite{Aad:2019mbh} and Fig. 30 of Ref.~\cite{ATLAS}.
\begin{table}[t!]
\begin{center}
{\footnotesize
\begin{tabular}{|c|c|c|}
\hline
&\hspace{0.15in}Mean\hspace{0.15in}	&\hspace{0.15in}RMS$\;\;\;\;$\\
\hline
$\kappa_{gZ}$ 					&1.06       &0.07 	\\
$\lambda_{Zg}$ 				&1.12	&0.15	\\
$\lambda_{tg}$ 					&1.10	&0.15	\\
$\lambda_{WZ}$ 				&0.95	&0.08	\\
$\left|\lambda_{\gamma Z}\right|$ 	&0.94	&0.07	\\
$\left|\lambda_{\tau Z}\right|$ 		&0.95	&0.13	\\
$\left|\lambda_{bZ}\right|$ 		&0.93	&0.15	\\
\hline
\end{tabular}
\qquad
\begin{tabular}{|c|ccccccc|}
\hline
& $\kappa_{gZ}$	&$\lambda_{Zg}$	&$\lambda_{tg}$	&$\lambda_{WZ}$	&$\left|\lambda_{\gamma Z}\right|$	&$\left|\lambda_{\tau Z}\right|$	&$\left|\lambda_{bZ}\right|$ \\
\hline
$\kappa_{gZ}$ 					&\;1.00   &-0.12  &-0.18     &-0.46    &-0.55     &-0.26     &-0.27\\
$\lambda_{Zg}$				&-0.12   &\;1.00  &\;0.44    &-0.56    &-0.33     &-0.32     &-0.66\\
$\lambda_{tg}$ 					&-0.18   &\;0.44  &\;1.00    &-0.21    &-0.16     &-0.21     &-0.32\\
$\lambda_{WZ}$				&-0.46   &-0.56   &-0.21    &\;1.00    &\;0.47    &\;0.27    &\;0.5\\
$\left|\lambda_{\gamma Z}\right|$	&-0.55   &-0.33   &-0.16    &\;0.47    &\;1.00    &\;0.38    &\;0.44\\
$\left|\lambda_{\tau Z}\right|$ 		&-0.26   &-0.32   &-0.21    &\;0.27    &\;0.38    &\;1.00    &\;0.34\\
$\left|\lambda_{bZ}\right|$ 		&-0.27   &-0.66   &-0.32    &\;0.5    &\;0.44    &\;0.34    &\;1.00\\
\hline
\end{tabular}
}
\caption{Higgs effective couplings in the $\kappa - \lambda$ framework from Ref.~\cite{Aad:2019mbh, ATLAS}. The root mean square (RMS) values are symmetrized in our fit procedure and are given in the table on the left (we opt for the conservative choice). The table on the right contains the correlation matrix amongst the seven free parameters defined in the text.\label{tab:HC}}
\end{center}

\end{table}%

The relevant $\kappa_f$ parameters are  
\begin{align}
\kappa_t &=  s_{\beta-\alpha} + \frac{c_{\beta-\alpha}}{ t_\beta}-\frac{ v }{ m_t} \frac{c_{\beta-\alpha} }{ s_\beta}\epsilon_{33}^u \, , &  \kappa_c &= s_{\beta-\alpha} + \frac{c_{\beta-\alpha}}{ t_\beta} -\frac{v }{ m_c} \frac{c_{\beta-\alpha} }{ s_\beta}\epsilon_{22}^u \,  ,  \nonumber\\
\kappa_b & = s_{\beta-\alpha} -  t_\beta  c_{\beta-\alpha} + \frac{  v }{ m_b}\frac{c_{\beta-\alpha}}{ c_\beta} \epsilon_{33}^d \, ,    &  \kappa_s & = s_{\beta-\alpha} -  t_\beta  c_{\beta-\alpha} + \frac{v }{ m_s}\frac{c_{\beta-\alpha}}{ c_\beta}\epsilon_{22}^d \, , \nonumber \\
 \kappa_\tau & = s_{\beta-\alpha} -  t_\beta  c_{\beta-\alpha}  + \frac{v }{ m_\tau}\frac{c_{\beta-\alpha}}{ c_\beta}\epsilon_{33}^e \, ,    & \kappa_\mu &= s_{\beta-\alpha} -  t_\beta  c_{\beta-\alpha} +\frac{v }{ m_\mu} \frac{c_{\beta-\alpha} }{ c_\beta}\epsilon_{22}^e \,.
\end{align}
\begin{table}[t!]
\begin{center}
\begin{tabular}{|c|c|c|c|c|}
\hline
&\hspace{0.15in}Q1\hspace{0.15in}	&\hspace{0.15in}Q2\hspace{0.15in}	&\hspace{0.15in}Q3\hspace{0.15in}	&\hspace{0.15in}Q4\;\;\;\;\\
\hline
\hline
$\kappa_t$ 					& $c_\alpha/s_\beta$       & $-s_\alpha/c_\beta$ & $-s_\alpha/c_\beta$ & $-s_\alpha/c_\beta$ 	\\
\hline
$\kappa_c$ 				         & $c_\alpha/s_\beta$	& $c_\alpha/s_\beta$ & $c_\alpha/s_\beta$ & $-s_\alpha/c_\beta$	\\
\hline
$\kappa_b$ 					& $c_\alpha/s_\beta$	& $c_\alpha/s_\beta$ & $c_\alpha/s_\beta$ & $-s_\alpha/c_\beta$	\\
\hline
$\kappa_s$ 				        & $c_\alpha/s_\beta$	& $-s_\alpha/c_\beta$ & $-s_\alpha/c_\beta$ & $-s_\alpha/c_\beta$	\\
\hline
\end{tabular}
\caption{Approximate $\kappa_f$ parameters for the models Q1 - Q4. \label{tab:ks}}
\end{center}
\end{table}
From Eqs.~\eqref{epsQ1} - \eqref{epsQ4} it is clear that the values of $\kappa_q$ depend on $\alpha$, $\beta$ and the model Q1-Q4, see Table~\ref{tab:ks}, while the couplings to leptons depend on the model class E1/E2 and  the chosen rotations  in the charged lepton sector (cf. Eqs.~\eqref{epsE1_xi} and \eqref{epsE2_xi}) 
\begin{align}
(\eps^e_{E1})_{ij} & =  \frac{m_{e_i} \delta_{ij} }{s_\beta v}   -  \frac{m_\tau}{s_\beta v} \sqrt{\xi^{e_L}_{ii} \xi^{e_R}_{jj}}   \, , &
(\eps^e_{E2})_{ij} & =  \frac{m_\tau}{s_\beta v}\sqrt{\xi^{e_L}_{ii} \xi^{e_R}_{jj}}   \, . 
\label{epsE}
\end{align}
%
\subsection{Constraints from flavor-violating Higgs decays}

The magnitude of the flavor-violating decays of the SM-like Higgs is controlled by the off-diagonal couplings $y_{\ell_i \ell_j}$ which are given by
\begin{equation}
y_{\ell_i \ell_j} = -\frac{1}{\sqrt{2}} \frac{c_{\alpha-\beta}}{c_\beta} \epsilon^e_{ij} \, , 
\label{yoffdiag}
\end{equation}
so that from Eq.~\eqref{epsE} we find the following prediction for the branching ratios of the SM-like Higgs decaying to a pair of leptons: 
\begin{equation}
\mathrm{BR}(h\to \ell_i \ell_j) = \frac{m_h}{16 \pi\,\Gamma_h}  \frac{c_{\alpha-\beta}^2}{c_\beta^2 s_\beta^2 }  \frac{m_\tau^2}{ v^2} \left( \xi_{ii}^{e_L} \xi_{jj}^{e_R} +  \xi_{ii}^{e_R} \xi_{jj}^{e_L} \right) \,,
\label{hBR}
\end{equation}
where we neglect tiny phase space effects. The total Higgs width is defined as~\cite{Aad:2019mbh}
\begin{equation}
\Gamma_h \simeq \frac{k_h^2}{1-B_{\mathrm{BSM}}} \Gamma_{\mathrm{SM}},
\end{equation}
where $\Gamma_{\mathrm{SM}}=4.1$ MeV~\cite{deFlorian:2016spz}. Here the branching ratio $B_{\mathrm{BSM}}$ denotes all the decay channels that are not present in the SM.

The constraints on the decays $h\to \ell_i \ell_j$ are as follows:
\begin{eqnarray}
\mathrm{BR}(h\to e\mu) &<& 6.1\times 10^{-5}\,\,  \mathrm{at\,\,95\%\,\,C.L.}\text{~\cite{Aad:2019ojw}} \, ,\nonumber\\
\mathrm{BR}(h\to e\tau)  &<& 2.2\times 10^{-3}\,\, \mathrm{at\,\,95\%\,\,C.L.}\text{~\cite{Sirunyan:2021ovv}} \, ,\nonumber\\
\mathrm{BR}(h\to \mu\tau)  &<& 1.5\times 10^{-3}\,\, \mathrm{at\,\,95\%\,\,C.L.}\text{~\cite{Sirunyan:2021ovv}} \, .
\end{eqnarray}
We note that both ATLAS~\cite{Aad:2019ugc} and CMS~\cite{Sirunyan:2021ovv} have observed a slight excess in the search for $h\to e\tau$ decays. We find that the weighted mean of the best-fit values reported by ATLAS and CMS is $\mathrm{BR}(h\to e\tau)^{\rm exp}\approx(0.09\pm0.07)\%$. Even though the excess  is just of the order of $1\sigma$, it is intriguing that both experiments have seen it, and an update of these analyses in the future will be interesting for the present scenario, where large effects in this channel are possible and actually expected as we are going to show later.

\subsection{Constraints from flavor-violating charged lepton decays}

Flavor-violating Higgs couplings are also constrained by the flavor-violating leptonic decays $\mu \to e \gamma$, $\tau \to e \gamma$ and $\tau \to \mu\gamma$~\cite{Harnik:2012pb,Crivellin:2013wna}.  
This class of decays are induced by one-loop penguin diagrams, with internal neutral or charged Higgs bosons, and by two-loop diagrams with top, $W$ or $Z$ running in the loop attached to the Higgs that induce the flavour violation~\cite{Chang:1993kw}.
Neglecting contributions for the heavy neutral and charged Higgs bosons, the branching ratio for the decay $\ell_i \to \ell_f \gamma$ is 
\begin{equation}
\mathrm{BR}(\ell_i \to \ell_f \gamma) = \frac{m_{\ell_i}^5}{4 \pi\,\Gamma_{\ell_i}}   \left( |c_{L,ij}|^2 + |c_{R,ij}|^2\right) \,,
\label{eq:ellBR}
\end{equation}
where $\Gamma_{\mu}$ is the initial state particle decay width,  
\begin{align}
c_{L,ij} &= \frac{e}{192 \pi^2\,m_h^2} \sum_{\ell=e,\mu,\tau} \biggl[   y_{\ell_f \ell} \,(y_{\ell_i\ell}^* + \delta_{\ell_i\ell} \Delta_{\ell_i}^{\mathrm{2-loop}})+\frac{m_{\ell_f}}{m_{\ell_i}}\,y_{\ell \ell_f}^* \,y_{\ell \ell_i}  \nonumber\\
&- \frac{m_\ell}{m_{\ell_i}} \,y_{\ell_f\ell} \,y_{\ell\ell_i}\,(9+6 \log(m_\ell^2/m_h^2))       \biggr] \,, \, 
\label{eq:ell2BR}
\end{align}
and $c_{R,ij}$ can be obtained from $c_{L,ij}$ upon replacing $y_{ij}$ by $y_{ji}^*$. Here $\Delta_{\ell_i}^{\mathrm{2-loop}}$ denotes the Barr-Zee two-loop contribution, which for $m_h=125$ GeV is given by $\Delta_{\ell_i}^{\mathrm{2-loop}} = -1.32 m_\tau/m_{\ell_i}$~\cite{Harnik:2012pb}. 
The two-loop and one-loop contributions are comparable for the decay $\tau \to e \gamma$. On the contrary, the two-loop diagrams are the dominant contributions for the $\mu \to e \gamma$ decay. 
The experimental constraints are given by~\cite{Aubert:2009ag,TheMEG:2016wtm}
\begin{eqnarray}
\mathrm{BR}(\tau \to \mu\gamma) &<& 4.4\times 10^{-8}\,\, \mathrm{at\,\,90\%\,\,C.L.} \, ,\nonumber\\
\mathrm{BR}(\tau \to e\gamma)  &<& 3.3\times 10^{-8}\,\, \mathrm{at\,\,90\%\,\,C.L.} \, ,\nonumber\\
\mathrm{BR}(\mu \to e\gamma)  &<& 4.2\times 10^{-13}\,\, \mathrm{at\,\,90\%\,\,C.L.} \, .
\end{eqnarray}
The bounds on $\tau \to \mu\gamma$ and $\tau \to e\gamma$ translate in rather weak bounds on 
\begin{equation}
\sqrt{|y_{\tau \mu}|^2+|y_{\mu \tau}|^2} < 1.6\times 10^{-2} \quad \mathrm{and}\quad \sqrt{|y_{\tau e}|^2+|y_{e \tau}|^2} < 1.4\times 10^{-2} \, ,
\end{equation}
respectively, assuming SM values for $y_{\tau\tau}$ and $y_{tt}$~\cite{Harnik:2012pb}. On the other hand, the latest experimental bound on $\mu \to e \gamma$ translates in 
\begin{equation}
\sqrt{|y_{e \mu}|^2+|y_{\mu e}|^2}~ < ~1.5\times 10^{-6} \, .
\end{equation}
This dramatically constrains the branching ratio for the $h\to \mu e$ decay to be\footnote{For this result we have neglected new physics contributions in the total Higgs decay width.} BR$(h \to \mu e)\lesssim 3 \times 10^{-9}$. %
Assuming instead that $y_{e \mu}$ and $y_{\mu e}$ are zero, one can obtain a bound on 
\begin{equation}
(|y_{\tau \mu}y_{e \tau}|^2+|y_{\mu \tau} y_{e\tau}|^2)^{1/4} <2.2 \times 10^{-4} \, .
\end{equation}
If $|y_{\tau \mu}|\sim |y_{e\tau}|$ the experimental bound on these couplings from $\mu \to e \gamma$ is much stronger than that from $\tau \to \mu\gamma$ and $\tau \to e\gamma$ and results in BR$(h \to \tau e)$ and BR$(h \to \tau \mu)$ below $\mathcal{O}(10^{-4})$. 
Therefore, only one of the branching ratios BR$(h \to \tau e)$ or BR$(h \to \tau \mu)$ can be large and close to the current LHC limits. 

\subsection{Summary of constraints from Higgs physics}

In order to simplify the discussion we introduce the following notation: $\xi_{ii}$ denotes the rotation of the chirality that defines the model (i.e. $\xi_{ii}= \xi_{ii}^{e_L}$ for E1L and E2L), while $\tilde{\xi}_{ii}$ denotes the rotation of opposite chirality (i.e. $\tilde{\xi}_{ii} = \xi_{ii}^{e_R}$ for E1L and E2L). In contrast to the axion phenomenology, the Higgs phenomenology depends on the rotation angles for both chiralities, i.e. both $\xi_{ii}$ and $\tilde{\xi}_{ii}$. In the following we will consider only the case where $\tilde{\xi}_{ii}=\xi_{ii}$. This choice does not have a strong impact on the resulting phenomenology, since $\tilde{\xi}_{ii}$  has only a subleading effect on Higgs phenomenology, as long as it is not larger than the specified one (i.e. as long as $\tilde{\xi}_{ii} \le \xi_{ii}$ ).

\begin{figure}[thbp]
\begin{center}
	\includegraphics[width=0.49\textwidth]{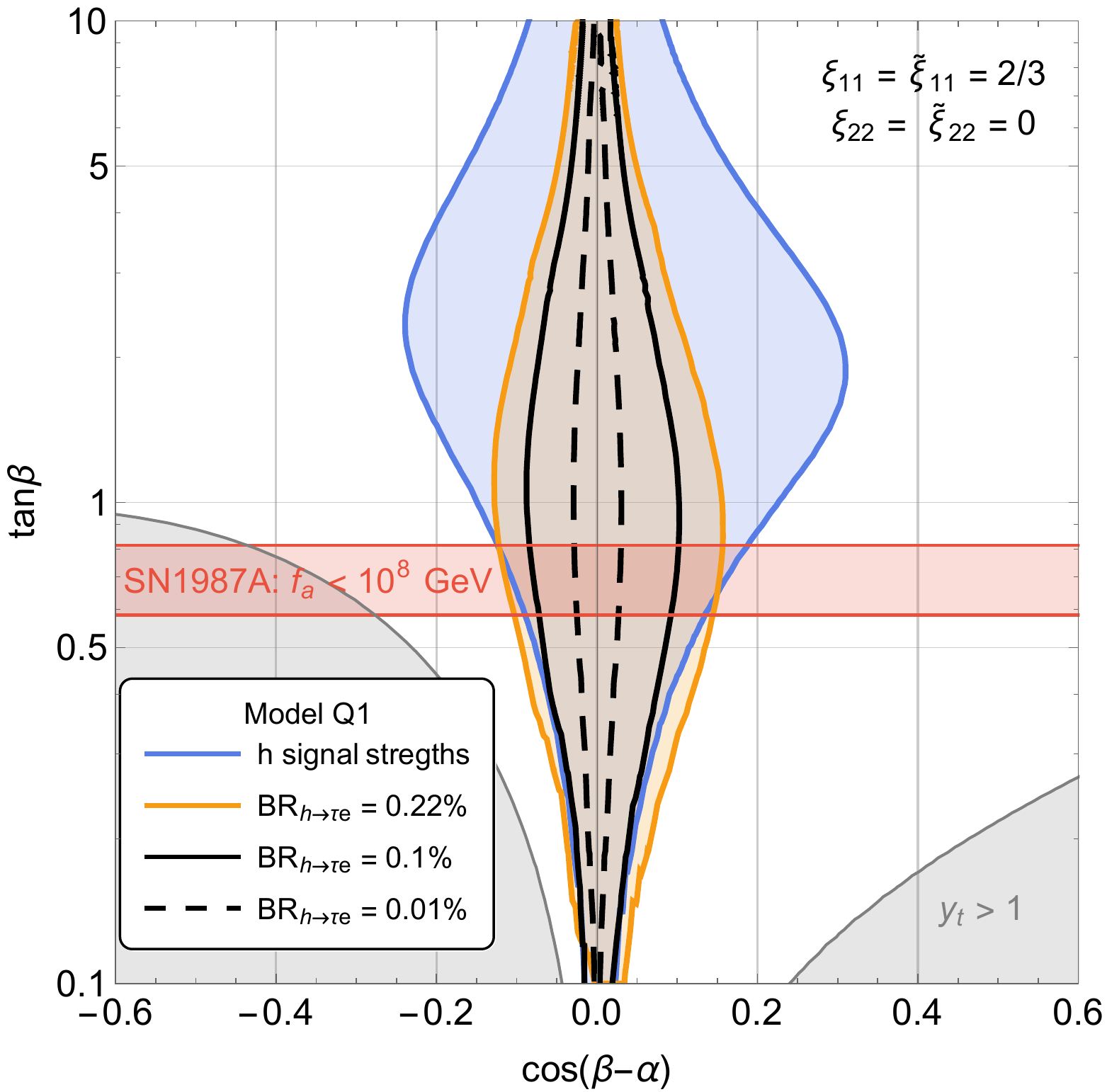}
	\includegraphics[width=0.49\textwidth]{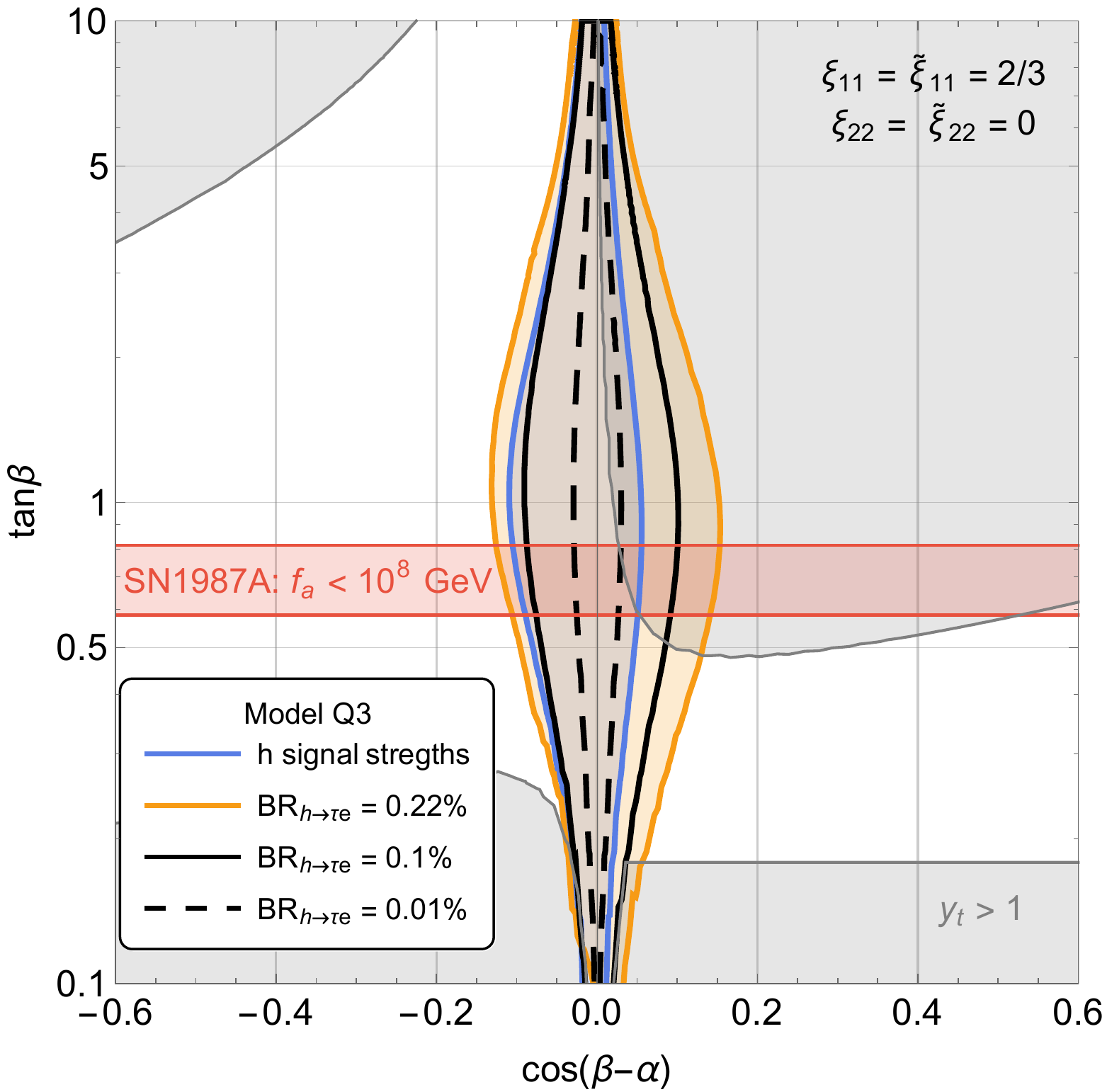}
	\includegraphics[width=0.49\textwidth]{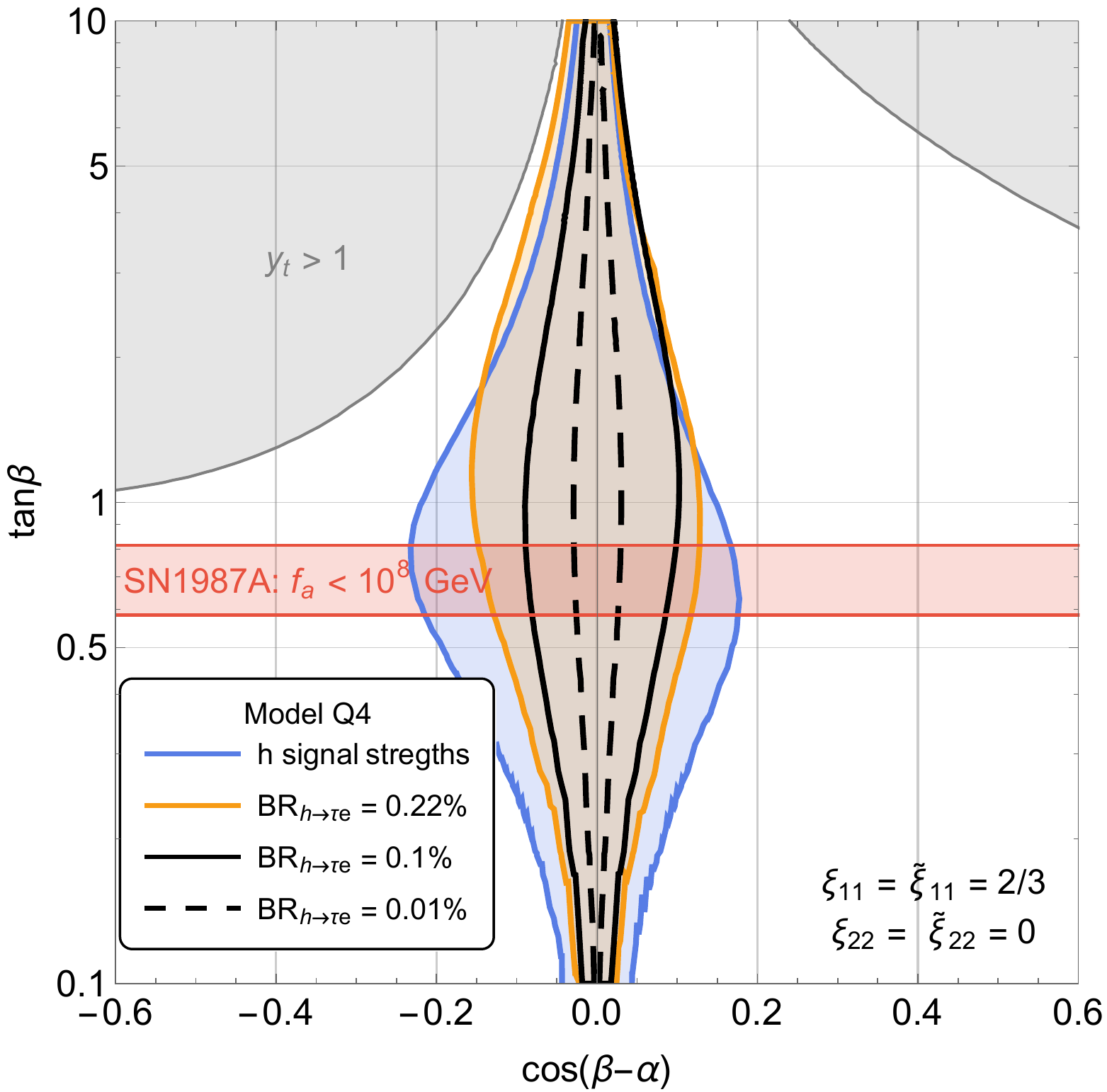}
\caption{Parameter space allowed by BR($h\to\tau e$) (orange region) and Higgs signal strengths (blue region)  for various quark models. 
The results for the Q2 model are the same as those for the Q3 model. We furthermore show the sensitivity for experiments that could prove branching ratios of $0.1\%$ and $0.01\%$. The gray areas denotes regions where Higgs couplings $y_t>1$, indicating the potential loss of perturbativity. Also shown is the region where the SN1987A constraint is satisfied for $f_a = 10^8 \GeV$. \label{fig:BR_etau}}
\end{center}
\end{figure}

The couplings $y_{ij}$ of the SM-like Higgs boson depend on $c_{\beta-\alpha}$, $t_\beta$ and the rotations $\xi_{ii},\tilde{\xi}_{ii}$.  Apart from the decoupling limit $c_{\beta-\alpha}\to 0$, which we are not interested in, one can avoid strong bounds on $y_{e \mu}$  and $y_{\mu e}$ from $\mu \to e \gamma$ by setting $\xi_{11}$ or $\xi_{22}$ to zero, as evident from Eqs.~\eqref{eq:ellBR}, \eqref{eq:ell2BR} and \eqref{yoffdiag}. As discussed in the axion phenomenology section, $\xi_{22}$ must be small to avoid the $\mu \to e a$ constraint, while sizeable $\xi_{11}$ is preferred to explain the cooling hints. This leads to a prediction of sizeable BR$(h \to \tau e)$, while BR$(h \to \mu e)$ and BR$(h \to \tau e)$ are strongly suppressed to avoid the $\mu \to e \gamma$ constraint, with a decay width proportional to 
\begin{align}
\Gamma (h \to \tau e) \propto \frac{c_{\alpha-\beta}^2}{c_\beta^2 s_\beta^2} \xi_{11} (1-\xi_{11}) \, . 
\end{align}
Thus the magnitude of ${\rm BR} (h\to \tau e)$ is controlled not only by $\xi_{11}$ but also  by $c_{\beta-\alpha}$, which cannot be arbitrarily large due to the constraints from the LHC Higgs coupling measurements.
In Fig.~\ref{fig:BR_etau} we show the $95\%$ allowed region for the ${\rm BR} (h\to \tau e)$ (yellow) and from the Higgs coupling measurement fit (blue) as a function of $c_{\beta-\alpha}$ and $\tan\beta$, with $\xi_{11}=2/3$ (1/3) and $\xi_{22}=0$ for models E1 (E2). Note that the different lepton models do not affect the Higgs coupling measurements for the indicated choice of $\xi_{ii}$. Furthermore, we show the future sensitivity of the branching ratios of $0.1\%$, that can be reached already at the Run 3 of the LHC, and $0.01\%$ which is the goal of high-luminosity LHC (HL--LHC)~\cite{deBlas:2019rxi}. The red band shows the region not excluded by SN1987A for axion models with $f_a\lesssim 10^8$ GeV, highlighting the most interesting region for axion phenomenology. Finally, the gray areas show regions where $y_t>1$, which indicates the potential loss of perturbativity. It is clear from this figure that perturbativity does not impose relevant constraints on the model parameter space consistent with Higgs couplings measurements (except for a small region for model Q3). 

The result for ${\rm BR} (h\to \tau e)$ is the same for all the lepton models, cf. Equation~\eqref{hBR}. On the other hand, the bounds from the Higgs coupling measurements strongly depend on the different quark models. In particular, the allowed region for models Q2 and Q3 is much smaller than the one for models Q1 and Q4. $|c_{\beta-\alpha}|$ for models Q2 and Q3 must be always below 0.1 while it can be about 0.2 for model Q4 or even 0.3 for model Q1 in agreement with the Higgs coupling measurement. This difference is due to the fact that, as seen from Table~\ref{tab:ks}, in models Q1 and Q4 $\kappa_b$ and $\kappa_t$ are approximately equal to each other while in models Q2 and Q3 $\kappa_b$ and $\kappa_t$ are anti-correlated i.e. when one is enhanced (suppressed) the other one is suppressed (enhanced). Note that $\kappa_t$ controls the magnitude of the Higgs production cross-section in the gluon fusion while $\kappa_b$ dominates the Higgs total width. This implies that for a given value of $c_{\beta-\alpha}$ deviations from the SM of the signal strengths for the Higgs decaying into electroweak gauge bosons (which are the best measured channels) are bigger in models Q2 and Q3 than in models Q1 and Q4 where the enhancement (suppression) of the Higgs production cross-section is partially compensated by the suppression (enhancement) of the Higgs branching ratios into gauge bosons (which is a consequence of the enhancement (suppression) of the Higgs total width). 

For this reason, at present the strongest constraints on $c_{\beta-\alpha}$ for the models Q1 and Q4 come from the bound on ${\rm BR} (h\to \tau e)$, and are almost independent of $\tan\beta$. On the other hand, although the constraints from the Higgs coupling measurements for the models Q2 and Q3 are quite stringent, it is still possible to have ${\rm BR} (h\to \tau e)\gtrsim 0.1\%$.   
In the remaining part of the paper we focus on the models Q1 and Q4 since they are able to predict larger rates for flavor-violating Higgs decays while being consistent with the Higgs coupling measurements.
 
 Finally, let us comment on the fact that as long as only Higgs phenomenology is considered, it is possible to choose $\xi_{11}=0$ to satisfy the $\mu\to e\gamma$ constraint. In such a case $\xi_{22}$ can be non-zero and the contours of ${\rm BR} (h\to \tau e)$ in Fig.~\ref{fig:BR_etau} would correspond to contours of ${\rm BR} (h\to \tau \mu)$ for $\xi_{22}=2/3$ and $\xi_{11}=0$. However, as we have discussed in the previous section, the stellar cooling anomalies together with the constraints from $\mu \to e a$ suggest that $\xi_{22}=0$ with $\xi_{11}$ free to vary.

\section{Interplay of Axion and Higgs Phenomenology} 
\label{sec4}

In this section we finally study the implications for Higgs physics in the parameter space where the axion can explain the stellar cooling hints.  In this way we fix the axion decay constant $f_a$ and study the maximal possible deviations for the Higgs decays $h \to \tau e$, $h \to \tau \tau$ and $h \to \mu \mu$, which can be obtained for a suitable value of $c_{\beta-\alpha}$ (while respecting all present constraints from precision Higgs physics).    

We will show results only for the models Q1E1L and Q4E1L, since they are less constrained by the Higgs coupling measurements. A different choice of lepton model would affect only the values of $\kappa_\mu$, $\kappa_\tau$ and the axion phenomenology, as discussed in the previous sections.
In order to avoid the bounds from $\mu\to e a$ we first fix $\xi_{22}=0$. As a consequence of this choice, we are left with three free parameters: $\xi_{11}$, $\tan\beta$ and $c_{\beta-\alpha}$. For given $\xi_{11}$ and $\tan\beta$ we fix $c_{\beta-\alpha}$ as the maximal value allowed by the Higgs coupling measurements, leading to a positive and a negative solution. We will show results only for the positive solution $c_{\beta-\alpha} > 0$ and comment on the difference with respect to the negative one. 
 
\begin{figure}[t!]
\begin{center}
	\includegraphics[width=0.49\textwidth]{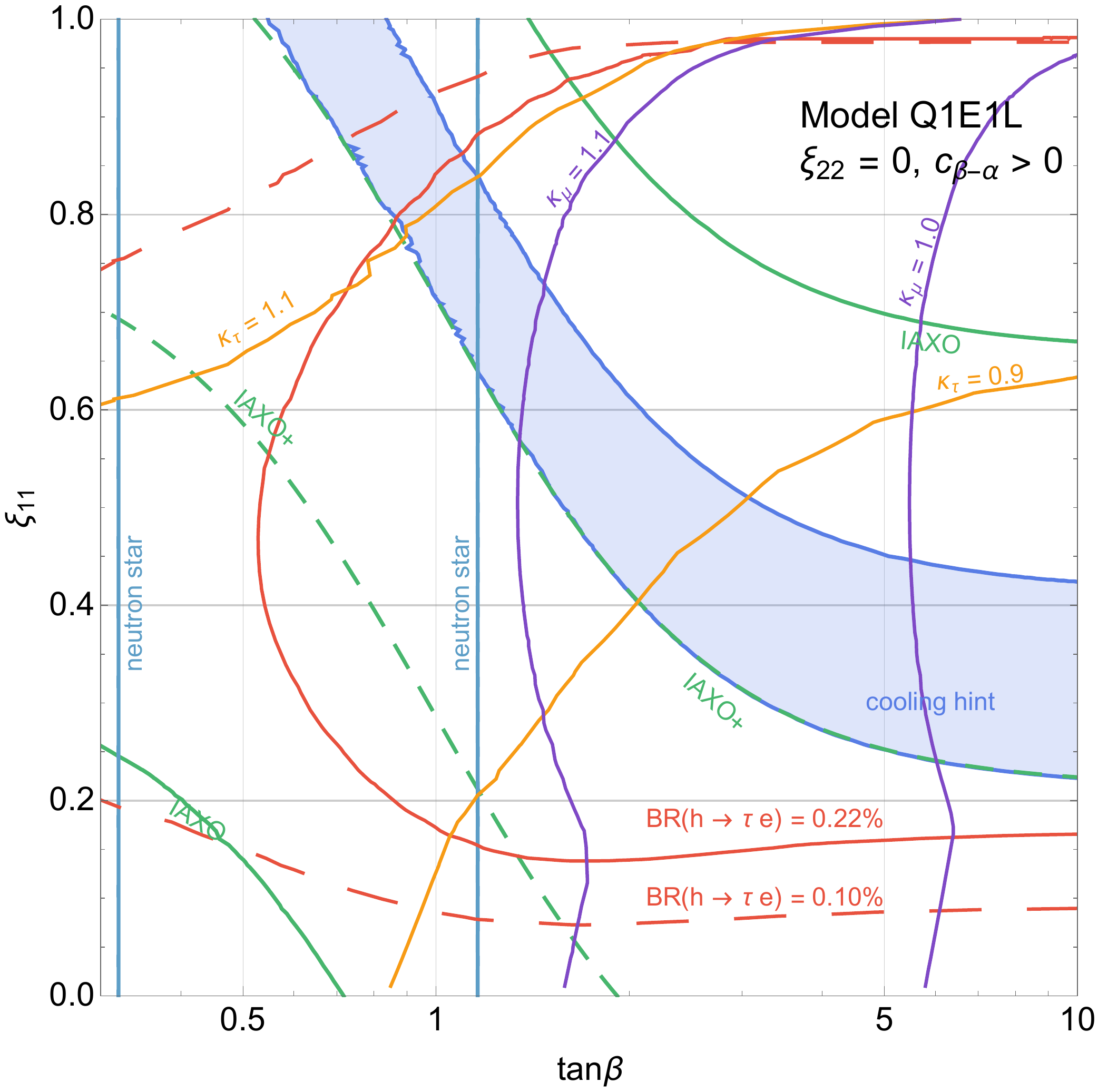}
	\includegraphics[width=0.49\textwidth]{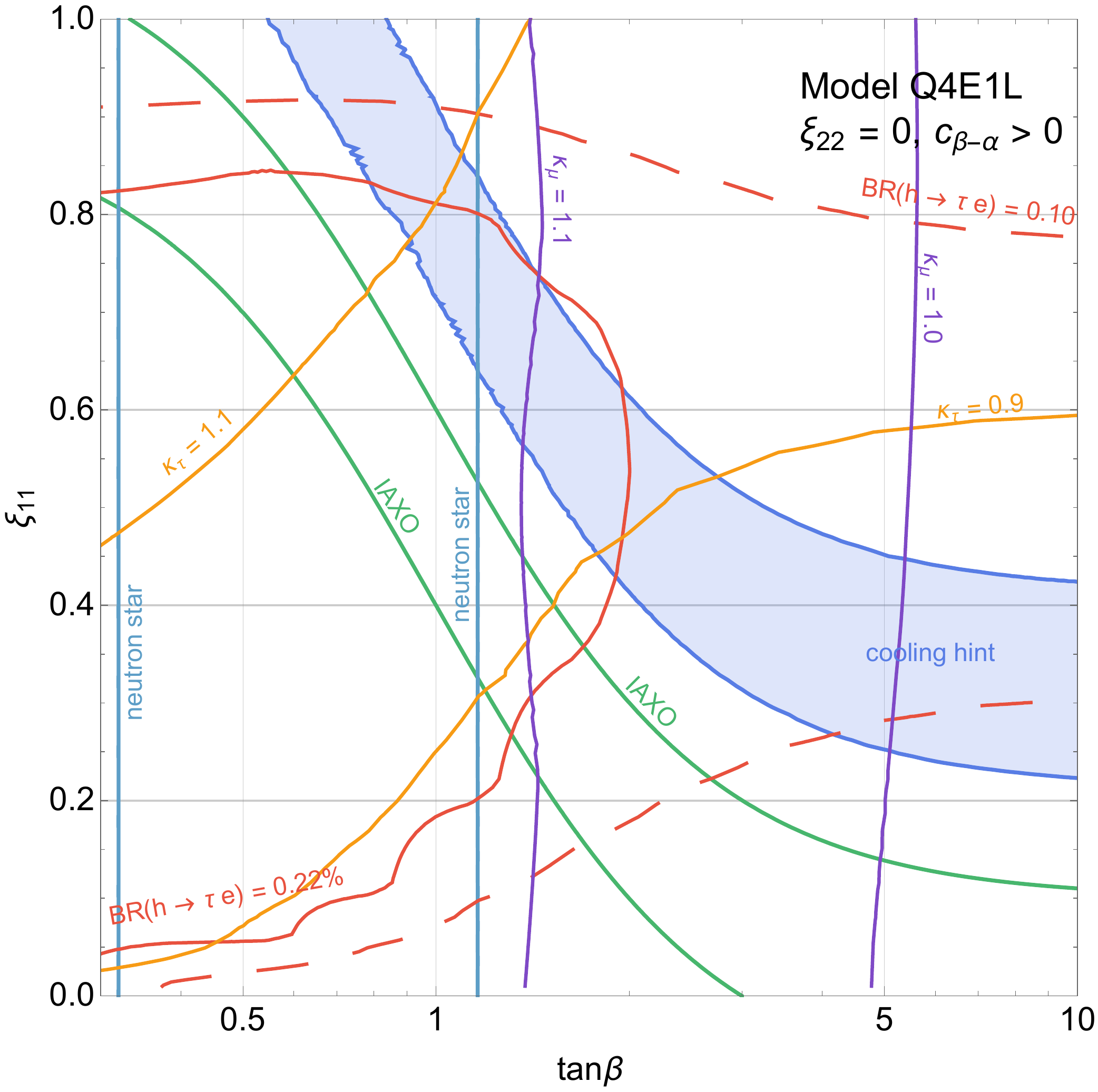}
\caption{ Contours of BR$(h\to\tau e)$ (solid, dashed and dotted red), $\kappa_\tau$ (orange), $\kappa_\mu$ (purple) obtained by taking the maximal value of $c_{\beta-\alpha}>0$ allowed by  LHC Higgs coupling measurements. The blue region is preferred at 1$\sigma$ by stellar cooling hints, the future helioscopes IAXO (IAXO+) will probe the whole region except the one between the green (dashed green) curves, and the region between light blue curves satisfies the neutron star bound for $f_a=10^9$ GeV.
\label{fig:BRcooling10t9}}
\end{center}
\end{figure}

In Fig.~\ref{fig:BRcooling10t9} we show contours of the maximal possible value ${\rm BR} (h\to \tau e)$ for the present (solid red) and future experimental reach (dashed red) consistent with the Higgs coupling measurements in the plane $\xi_{11}$ vs $\tan\beta$ for the models Q1E1L (left) and Q4E1L (right). These contours are overlaid with the 1$\sigma$ region explaining the cooling hint (blue), and various existing and future constraints on axions for $f_a=10^9$~GeV. In particular, we show the bound from neutron stars (light blue) and the future reach of helioscopes with IAXO (solid green) and IAXO+ (dashed green). Notice that for $f_a\simeq10^9$ GeV there is no bound from SN1987A.
In the region explaining the cooling hint, the branching ratio ${\rm BR} (h\to \tau e)$ can be as large as the current upper bound from CMS in both models. In the Q1E1L model, ${\rm BR} (h\to \tau e)$ can be maximal in the cooling hint region for any $\tan\beta$ above $\sim 0.8$, while the corresponding region with maximal ${\rm BR} (h\to \tau e)$ in the Q4E1L model is characterised by $\tan\beta\lesssim2$. A branching ratio of ${\rm BR} (h\to \tau e)\sim0.1$ could be obtained in most of the cooling hint region, just decreasing the value of $c_{\beta-\alpha}$.

We furthermore show contours for the reduced Higgs couplings $\kappa_\tau$ and $\kappa_\mu$. The deviations from the SM (which predicts $\kappa_\tau=\kappa_\mu=1$) can be $\mathcal{O}(10)\%$, within reach of the HL--LHC~\cite{ATLAS:2018jlh}. Interestingly, in the cooling hint region $\kappa_\tau$ and $\kappa_\mu$ are very different from each other. In particular, the cooling hint region cannot easily accommodate a SM like value for $\kappa_\tau$ and $\kappa_\mu$ simultaneously. An uncertainty below $10\%$ in the measurements of the $\kappa$ parameters would be enough to probe these models. It is worth to notice that taking into account the constraints from neutron stars and the cooling hint region, the Higgs couplings to muons and taus could deviate from the SM up to $\gtrsim 10\%$. This means that axion physics prefers the region $\tan\beta\lesssim1$ and $\xi_{11}\gtrsim0.6$, compatible with deviations in the $\mu$ and $\tau$ Yukawa couplings that may be observed already at the HL--LHC~\cite{ATLAS:2018jlh}.  The main difference between the Q1E1L and Q4E1L models is the future reach of IAXO and IAXO+. The IAXO helioscope, in its base configuration, will be able to probe all the cooling hint region for the model Q4E1L, independently of the value of $\tan\beta$ or $\xi_{11}$. On the other hand, the cooling hint region for the model Q1E1L could be partially probed by an advanced configuration of IAXO. Therefore, an interplay between axion and collider searches is needed in order to fully rule out this model.

The results for $c_{\beta-\alpha}<0$ are similar to the one in Fig.~\ref{fig:BRcooling10t9}. There are two differences that affect the results. On one hand, the change in sign affects  the observables that are linear in $c_{\beta-\alpha}$, such as $\kappa_\mu$ and $\kappa_\tau$. In particular, in Fig.~\ref{fig:BRcooling10t9}, the contour lines for $\kappa_{\mu/\tau} = 0.9$ become contour lines for $\kappa_{\mu/\tau} \simeq 1.1$. On the other hand, the difference in absolute value, due to the fact that the Higgs coupling measurements allowed region is not symmetric in $c_{\beta-\alpha}$ (see Fig.~\ref{fig:BR_etau}), influences both the observables that are linear or quadratic in $c_{\beta-\alpha}$, such as the ${\rm BR} (h\to \tau e)$.

\begin{figure}[t!]
\begin{center}
	\includegraphics[width=0.49\textwidth]{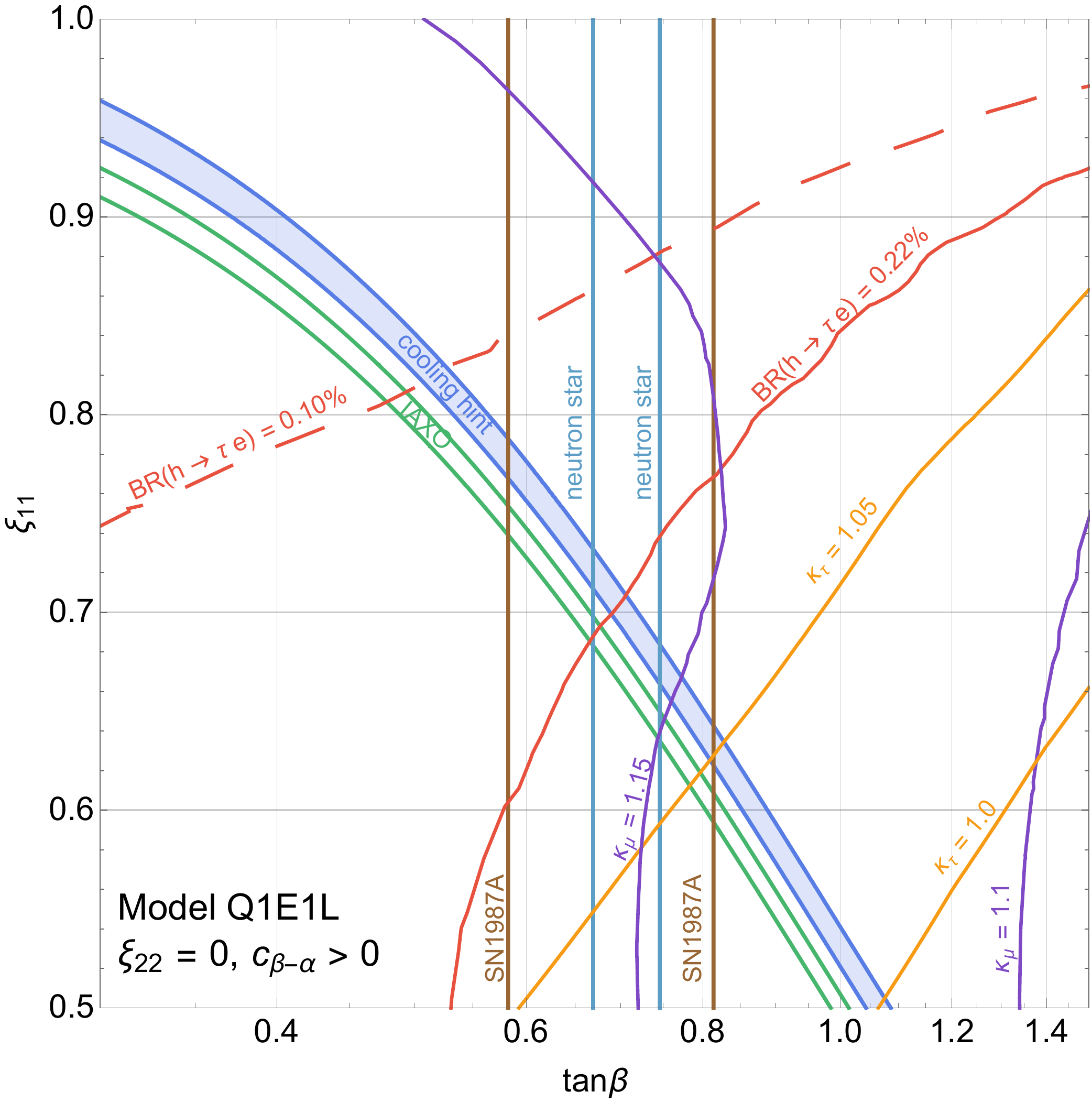}
	\includegraphics[width=0.49\textwidth]{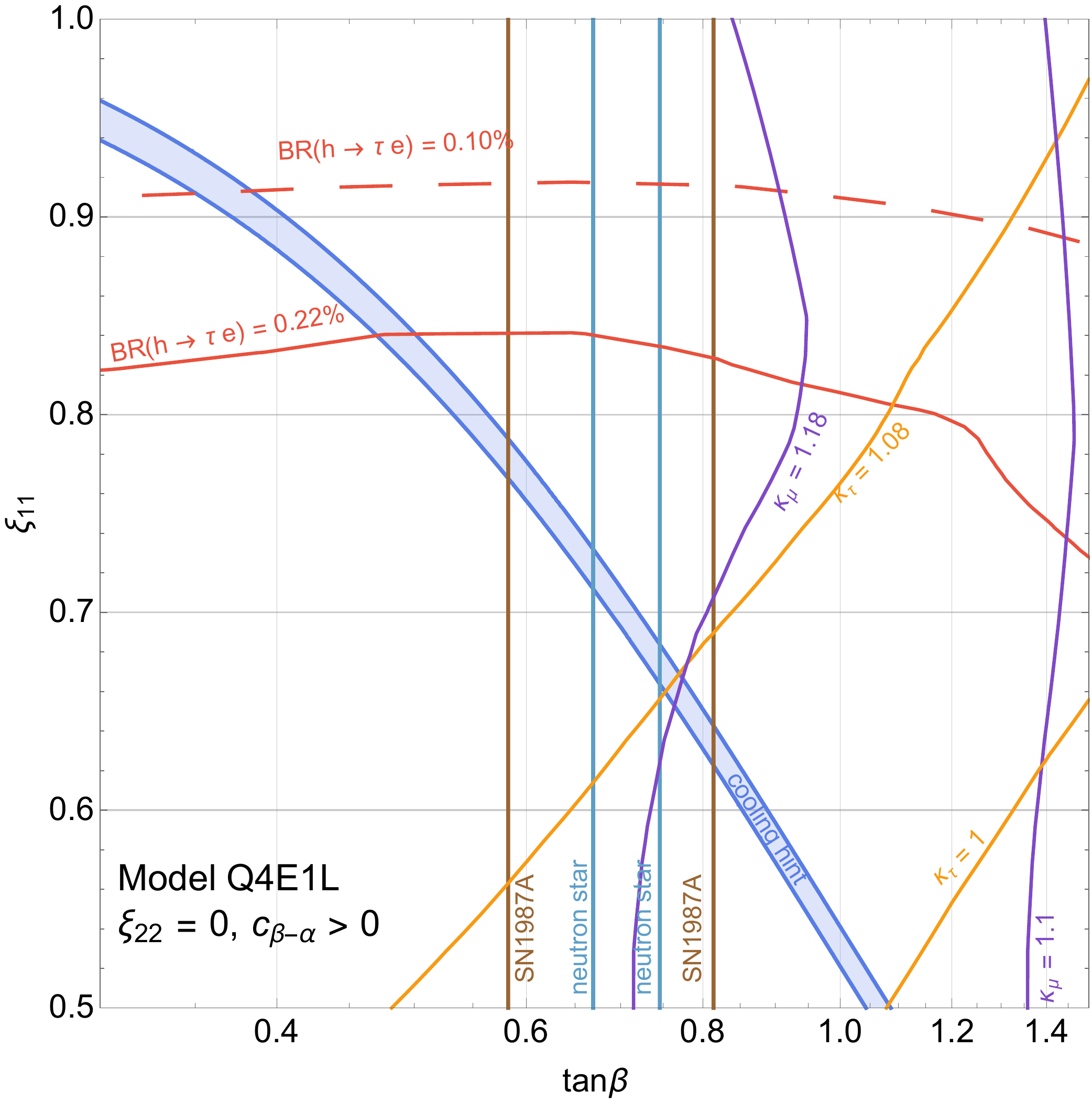}
	\caption{ The same as in Figure~\ref{fig:BRcooling10t9} but for $f_a=10^8$ GeV. 
	\label{fig:BRcooling10t8}}
\end{center}
\end{figure}

The results for $f_a=10^8$~GeV  are shown in Fig.~\ref{fig:BRcooling10t8} for the models Q1E1L (left) and Q4E1L (right). The contours of the Higgs observables are the same as in Fig.~\ref{fig:BRcooling10t9} while the cooling hint region and the other constraints on the axion are modified. In this case the constraint from SN1987A (brown curve) enforces $\tan\beta$ to be in a small range between about 0.6 and 0.8. Although the cooling hint region is much smaller in this case it is still possible to obtain ${\rm BR} (h\to \tau e)$ as large as $0.22\%$, i.e. the current upper bound from CMS~\cite{Sirunyan:2021ovv}. Interestingly, in the cooling hint region for $f_a=10^8$~GeV consistent with the constraint from SN1987A the maximal deviation of $\kappa_\mu$ from the SM  always exceeds $15\%$, while the deviation of $\kappa_\tau$ can be up to $10\%$. 
Therefore, the $f_a=10^8$~GeV case gives very sharp prediction for the pattern of the Higgs couplings to muons and taus. The projected sensitivity of ATLAS at the HL--LHC is around 7\% for $\kappa_\mu$ and 3\% for $\kappa_\tau$~\cite{ATLAS:2018jlh} which should be enough to test these models for $f_a=10^8$~GeV. Furthermore, this scenario will be easily probed by IAXO in its default configuration.\footnote{For the model Q4E1L the IAXO curve is not shown as IAXO will probe the whole parameter space.}

Let us emphasize that predictions for the Higgs couplings in our models substantially differ from Type-I and Type-II 2HDMs in which the Higgs couplings (normalized to the SM) to all down-type quarks and charged leptons are the same. In particular, in the Q4E1L model with $c_{\beta-\alpha}>0$, $\kappa_b$ is smaller than $\kappa_\mu$ ($\kappa_\tau$) by about 30\% (20\%) in the cooling hint region consistent with the neutron star bound for $f_a=10^8$~GeV. Therefore, it will be easy to experimentally distinguish this model from Type-I and Type-II 2HDMs.\footnote{In the Q1E1L model $\kappa_b$ is approximately equal to $\kappa_\mu$ but could differ from $\kappa_\tau$ by almost 10\%, so it may also be possible to distinguish this model from Type-I and Type-II 2HDMs. } We also note that in this region ${\rm BR} (h \to \mu \mu)$ could be up to $\sim60$\% larger than the SM prediction, while for $f_a=10^9$~GeV one can have deviations of order $\sim70$\%. This is a combined effect from simultaneously enhancing $\kappa_\mu$ and suppressing $\kappa_b$. In this case a significant excess in the muon decay channel may be observed already in the Run 3 of the LHC.

\section{Conclusions}
\label{sec5}
In this article we have explored the correlation of axion and Higgs phenomenology in variant axion models with a light second Higgs doublet. We restricted to a particular class of ``nucleophobic'' DFSZ-like models that allow to avoid the stringent constraints from SN1987A and neutron star cooling by suppressed couplings to nucleons, while couplings to electrons can be sizable and allow to address various stellar cooling anomalies. All axion couplings are fixed in terms of three relevant parameters, the axion decay constant $f_a$, the Higgs vacuum angle $\tan \beta$ and a free angular parameter $\xi_{11}$ that controls lepton flavor-violation. A compact region in this parameter space is selected by the stellar cooling hints, while imposing the astrophysical bounds on nucleon couplings and perturbativity, see Fig.~\ref{fig:IAXO}. This in particular restricts the values of the axion decay constant to values below about $4 \times 10^9 \GeV$, which corresponds to axion masses of the order of few meV. Large parts of this parameter space will be probed by  next-generation helioscope experiments such as IAXO.

Such heavy axions can still account for the observed Dark Matter abundance when produced by the decay of strings and domain walls in scenarios when PQ is broken after inflation, up to roughly $m_a \lesssim 4 \, {\rm meV}$~\cite{IAXO3}, although the abundance suffers from significant uncertainties~\cite{Gorghetto:2018myk}. Another production mechanism that also works for heavier axions is  parametric resonance from oscillations of the Peccei-Quinn symmetry breaking field~\cite{Co:2017mop}, which can yield the correct DM relic abundance up to axion masses of roughly $60 \, {\rm  meV}$. In both scenarios it is crucial to avoid  stable domain walls by having a trivial domain wall number~\cite{Vilenkin:1982ks}, which indeed is realized in the class of DFSZ models considered here (see also Ref.~\cite{Saikawa:2019lng}). 

Also the Higgs sector depends on the Higgs vacuum angle $\beta$ and the free angular parameter $\xi_{11}$ controlling lepton flavor-violation, in addition to the mass of the second Higgs doublet that enters Higgs couplings through the angle $\cos ({\beta - \alpha})$. While previous studies of similar models always decoupled the additional Higgs doublet, corresponding to the alignment limit when $\cos({\beta - \alpha}) \to 0$, here we analyzed in detail the resulting Higgs phenomenology when the deviation from alignment is as large as allowed by present experimental Higgs data. 

In this way we can correlate axion and Higgs phenomenology, since the cooling hints essentially fix the Higgs couplings to leptons as a function of $\cos ({\beta - \alpha})$. Of particular relevance  are the Higgs couplings to muons and tau leptons and the LFV Higgs decay $h \to \tau e$, which will be probed with upcoming LHC data for precision Higgs measurements. Maximal values of these observables can be predicted by taking the heavy Higgs doublet as light as possible consistent with present data, or equivalently maximizing $\cos ({\beta - \alpha})$.  Our results are summarized in Figs.~\ref{fig:BRcooling10t9} (for $f_a = 10^9 \GeV $) and Figs.~\ref{fig:BRcooling10t8} (for $f_a = 10^8 \GeV $), which show that lepton flavor-violating Higgs decays $h \to \tau e$ can saturate the current experimental bound ${\rm BR} (h \to \tau e) = 0.22 \%$, while deviations from the SM prediction for ${\rm BR} (h \to \mu \mu)$ can be as large as 70\% in the parameter region where the axion can explain the stellar cooling hints. 

The QCD axion models that we considered in this article to address the stellar cooling anomalies might therefore be testable not only by future helioscopes but also by precision Higgs data, highlighting the interplay of dedicated axion searches with IAXO and precision tests of the SM Higgs sector at the LHC.

\section*{Acknowledgements}

We would like to thank Kiwoon Choi for useful discussions. MB and RZ acknowledge the GGI Institute for Theoretical Physics for its hospitality and partial support where this work was initiated. MB has been partially supported by the National Science Centre, Poland, under research grants no. 2017/26/D/ST2/00225 and 2020/38/E/ST2/00243.
GGdC is supported by the INFN Iniziativa Specifica Theoretical
Astroparticle Physics (TAsP) and by the Frascati National 
Laboratories (LNF) through a Cabibbo Fellowship call 2019. This  work  is  partially  supported  by project C3b of the DFG-funded Collaborative Research Center TRR257, ``Particle Physics Phenomenology after the Higgs Discovery".

\appendix
\numberwithin{equation}{section}
\section{Generalized DFSZ Models}
\label{genDFSZ}

To the SM fermion fields we add two  Higgs doublets $h_i$ with hypercharge $Y=-1/2$ and a SM singlet $\phi$. The Lagrangian is taken to be invariant under a $U(1)_{\rm PQ}$ symmetry, with the most general charge assignment consistent with a $2+1$ flavor structure, as shown in Table~\ref{PQ}. Note that without loss of generality we can set the charges of the left-handed quark and lepton fields of the third generation (i.e. the flavor singlets) to zero, by redefining $U(1)_{\rm PQ}$ with the anomaly-free symmetries $B-L$ and $Y$. 
\begin{table}[t]
\centering
\begin{tabular}{|c|cccccccccccc|}
\hline
 &  $q_{L3}$ & $q_{L1,L2}$ & $u_{R3}$ & $u_{R1,R2}$ & $d_{R3}$ & $d_{R1,R2}$ & $l_{L3}$ & $l_{L1,L2}$ & $e_{R3}$ & $e_{R1,R2}$ & $h_i$ &  $\phi$ \\
 \hline
 $U(1)_{\rm PQ}$ & 0 & $X_{q}$ & $X_{u_3}$ & $X_{u}$ & $X_{d_3}$ & $X_{d}$ & 0 & $X_{l}$ & $X_{e_3}$ & $X_{e}$ & $X_{i}$  & $1$  \\
 \hline
 \end{tabular}
 \caption{\label{PQ} PQ charge assignment}
\end{table}
\noindent The Yukawa Lagrangian reads
\begin{align}
{\cal L} & = - y^{u}_{33} \overline{q}_{L3} u_{R3} h_{A_1} - y^{u}_{3a}  \overline{q}_{L3} u_{Ra} h_{A_2} - y^{u}_{a3} \overline{q}_{La} u_{R3} h_{A_3} - y^{u}_{ab}  \overline{q}_{La} u_{Rb} h_{A_4} \nonumber \\
&  + y^{d}_{33} \overline{q}_{L3} d_{R3} \tilde{h}_{A_5} + y^{d}_{3a}  \overline{q}_{L3} d_{Ra} \tilde{h}_{A_6} + y^{d}_{a3} \overline{q}_{La} d_{R3} \tilde{h}_{A_7} + y^{d}_{ab} \overline{q}_{La} d_{Rb} \tilde{h}_{A_8} \nonumber \\ 
& + y^{e}_{33} \overline{l}_{L3} e_{R3} \tilde{h}_{A_9} + y^{e}_{3a} \overline{l}_{L3} e_{Ra} \tilde{h}_{A_{10}} + y^{e}_{a3} \overline{l}_{La} e_{R3} \tilde{h}_{A_{11}} + y^{e}_{ab} \overline{l}_{La} e_{Rb} \tilde{h}_{A_{12}} 
+ {\rm h.c.}   
\label{eq:L1_app}
\end{align}
where $\tilde{h}_i = i \sigma^2 h_i^*$, $a,b = 1,2$ and $A_i \in \{ 1,2 \}$  is a free parameter that selects to which Higgs field the respective fermions couple to. This gives twelve constraints, which determines all fermion charges in terms of Higgs charges
\begin{align}
X_{u_3} & = - X_{A_1} \, , &  X_{u} & = - X_{A_2}  \, , &    X_{A_4} & = - X_{A_1}  + X_{A_2} + X_{A_3}  \, ,   \nonumber \\
X_{d_3} & =  X_{A_5} \, , &  X_{d} & =  X_{A_6}  \, , &  X_{A_7} & =  X_{A_1}  - X_{A_3} + X_{A_5}  \, ,   \nonumber \\
X_{e_3} & =  X_{A_9} \, , &  X_{e} & = X_{A_{10}}  \, , &   X_{A_8} & =  X_{A_1}  - X_{A_3} + X_{A_6}   \, ,  \nonumber \\
X_{q} & =  - X_{A_1} + X_{A_3} \, , & X_{l} & =  X_{A_9} - X_{A_{11}} \, ,  & X_{A_{12}} & = - X_{A_9}  + X_{A_{10}} + X_{A_{11}}   \, .
\end{align}
Compared to the SM, the above Yukawa Lagrangian has an extra $U(1)_h^2 \times U(1)_\phi$ global symmetry that needs to be broken to a single $U(1)_{\rm PQ}$ factor by adding two couplings in the scalar sector. Since  $U(1)_{\rm PQ} \ne U(1)_\phi$, we need at least one coupling of $\phi$. On the renormalizable level we can couple $h_2^\dagger h_1$ to an operator ${\cal O}_1 \in \{ \phi, \phi^*, \phi^2, \phi^{*2}\}$. This constrains the charges of the Higgs fields $h_2$ in terms of the $h_1$ charge $X_1$ and a free parameters $B$  that can take only discrete values
\begin{align}
X_{2} & = X_1 + B \, , 
\label{X12fix}
\end{align}
with the possible values
\begin{align}
B & \in \{ \pm 1, \pm 2\} \,   .
\end{align}
 The scalar potential is assumed to induce the Higgs and singlet vevs 
\begin{align}
\langle h_i \rangle & = v_i/\sqrt2 \, , &  \langle \phi \rangle & = v_\phi/\sqrt2 \, ,
\end{align} 
and one is free to make a field rotation such that only one linear combination of Higgs doublets takes a vev $v = 246 \GeV$
\begin{align}
h_v & \equiv \sum_{i} O_{1i} h_i \, , & O^T O & = 1 \, , & \langle h_v \rangle & = \frac{v}{\sqrt 2} \, , & v = \sum_{i} O_{1i} v_i   \, .
\end{align}
The Goldstone boson eaten up by the $Z$-boson resides in this linear combination, and with
\begin{align}
h_i & = \frac{v_i}{\sqrt2} \begin{pmatrix} 
1  \\ 0 \end{pmatrix} e^{ i a_i/v_i} + \hdots \, , & \phi & = \frac{v_\phi}{\sqrt2} \, e^{ i a_\phi/v_\phi} + \hdots \, , 
\end{align}
it is given by
\begin{align}
\phi_Z = \sum_i O_{1i} a_i   \, .
\end{align}
The anomalous $U(1)_{\rm PQ}$ current reads
\begin{align}
j_\mu^{\rm PQ} & 
= -i \left( \phi^\dagger \overset{\leftrightarrow}{\partial_\mu} \phi + \sum_{i=1}^N X_{i}  h_i^\dagger \overset{\leftrightarrow}{\partial_\mu} h_i + \hdots \right)
=  \partial_\mu \left( v_\phi a_\phi + \sum_{i} X_{i} v_i a_{i}  \right) + \hdots
\end{align}
where we omitted the fermionic part. This current creates the axion according to 
\begin{align}
j_\mu^{\rm PQ} = v_{\rm PQ} \partial_\mu a + \hdots \, , 
\end{align}
which defines the axion as the linear combination
\begin{align}
a = \frac{v_\phi}{v_{\rm PQ}} a_\phi + \sum_{i} X_{i} \frac{ v_i}{v_{\rm PQ}} a_{i}  \, , 
\label{adef}
\end{align}
with the PQ breaking scale 
\begin{align}
v_{\rm PQ}^2 =  v_\phi^2  + \sum_{i} X_{i}^2 v_i^2 \, .
\end{align}
We need to impose that the axion is orthogonal to the Goldstone eaten up by the $Z$, which gives the condition 
\begin{align}
0 = \sum_{i} O_{1i} X_{i} v_i \, . 
\end{align}
The rotation matrix $O$ can be constructed simply by taking $\vec{v}$ as first row and then flipping pairwise two entries with minus signs to be orthogonal. In particular one has
\begin{align}
O_{1i} & = v_i/v \, , & \sum_i v_i^2 &=  v^2 \, , 
\end{align}
which is the only input in the orthogonality condition
\begin{align}
0 =  \sum_{i} X_{i} v_{i}^2    = \sum_{i} X_{i} \frac{v_{i}^2}{v^2}  \, . 
\end{align}
Parameterizing the Higgs vevs with a single vacuum angle $\tan \beta = v_2/v_1$
\begin{align}
v_1 & = c v \, , & v_2 & = s  v \, ,
\end{align}
the orthogonality condition becomes
\begin{align}
0 =  X_{1} c^2 + X_{2} s^2     \, . 
\end{align}
Together with Eq.~\eqref{X12fix} this condition fixes the charge of $h_{1,2}$ (and therefore also the fermions) in terms of the vacuum angle $\tan \beta$ and the parameter $B$
\begin{align}
X_{1} & = - s^2 B \, , & X_{2} & = c^2 B   \, .
\label{X1fix}
\end{align}
Using the axion definition in Eq.~(\ref{adef}), it is easy to verify that mass terms and axion-fermion couplings arise from replacing the neutral Higgs field components by
\begin{align}
h_{A_i}^0 & \to \frac{v_{A_i}}{\sqrt 2} \, e^{i X_{A_i} a/v_{\rm PQ}} \, , & \tilde{h}_{A_i}^0 & \to - \frac{v_{A_i}}{\sqrt 2} \, e^{- i X_{A_i} a/v_{\rm PQ}}  \, . 
 \end{align}
Therefore axion-fermion couplings can be removed from the Yukawa Lagrangian by the flavor-diagonal fermion field redefinition (a local PQ transformation acting only on fermions) 
 \begin{align}
 f \to f \, e^{i X_f a/v_{\rm PQ}}   \, .
 \end{align}
Since this transformation is anomalous, it generates axion couplings to gauge field strengths, and since it is local it modifies the fermion kinetic terms. The anomalous couplings are given by

\begin{align}
{\cal L}_{\rm anom} = N \frac{a}{v_{\rm PQ}} \frac{\alpha_s}{4 \pi} G_{\mu \nu} \tilde{G}^{\mu \nu} + E \frac{a}{v_{\rm PQ}} \frac{\alpha_{\rm em}}{4 \pi} F_{\mu \nu} \tilde{F}^{\mu \nu} \, , 
\end{align}
with the dual field strength $\tilde{F}_{\mu \nu} = \frac{1}{2} \eps_{\mu \nu \rho \sigma} F^{\rho \sigma}$, $\eps^{0123} = -1$ and the color and electromagnetic anomaly coefficients
\begin{align}
N & = \frac{1}{2} \left( 4 X_q -  X_{u_3} - 2 X_u - X_{d_3} - 2 X_d \right) \nonumber \\
& = \frac{1}{2} \left(  - 3 X_{A_1} + 2 X_{A_2} + 4 X_{A_3} - X_{A_5}  - 2 X_{A_6} \right)  \, , \\
\nonumber \\
  E & = \frac{5}{3} \left(   2 X_q \right) - \frac{4}{3}  \left( X_{u_3} + 2 X_u \right) -  \frac{1}{3} \left(  X_{d_3} + 2 X_d \right) 
  + \left(  2 X_l \right) - X_{e_3}  - 2 X_e    \nonumber \\
  & = - 2 X_{A_1} + \frac{8}{3} X_{A_2} + \frac{10}{3} X_{A_3} - \frac{1}{3} X_{A_5} - \frac{2}{3} X_{A_6} + X_{A_9} - 2 X_{A_{10}} - 2 X_{A_{11}} \, .
 \end{align}
The kinetic terms give the following axion-fermion couplings in the flavor interaction basis
\begin{align}
{\cal L} &=  \frac{\partial_\mu a}{v_{\rm PQ}} \biggl[ \overline{u}_{i} \gamma^\mu \left( \tilde{C}^q_{ij} P_L + \tilde{C}^u_{ij} P_R \right)  u_{j} + \overline{d}_{i} \gamma^\mu \left( \tilde{C}^q_{ij} P_L + \tilde{C}^d_{ij} P_R \right)  d_{j}   \nonumber\\
&+ \overline{e}_{i} \gamma^\mu \left( \tilde{C}^l_{ij} P_L + \tilde{C}^e_{ij} P_R \right)  e_{j}   \biggr] \, , 
\end{align}
with
\begin{align}
\tilde{C}^q_{ij} & = \left(  X_{A_1} - X_{A_3} \right) \delta_{ij} + {\rm diag} (0, 0, X_{A_3} - X_{A_1})  \, , \nonumber \\
\tilde{C}^u_{ij} & =  X_{A_2} \delta_{ij} + {\rm diag} (0, 0,  X_{A_1} - X_{A_2})  \, , \nonumber \\
  \tilde{C}^d_{ij} & = - X_{A_6} \delta_{ij} +  {\rm diag} (0, 0, X_{A_6} - X_{A_5})  \, ,  \nonumber \\
 \tilde{C}^l_{ij} & = \left( X_{A_{11}} - X_{A_{9}} \right) \delta_{ij} + {\rm diag} (0, 0, X_{A_{9}} - X_{A_{11}}) \, ,  \nonumber \\
   \tilde{C}^e_{ij} & =  - X_{A_{10}} \delta_{ij} +  {\rm diag} (0, 0, X_{A_{10}} - X_{A_{9}})   \, .  
\end{align}
Finally we go to the mass basis.  The fermion mass terms are given by 
\begin{align}
- {\cal L}_{\rm mass} & = \overline{u}_{Li} M_{u,ij} u_{Rj} + \overline{d}_{Li} M_{d,ij} d_{Rj} + \overline{e}_{Li} M_{e,ij} e_{Rj} + {\rm h.c.} 
\end{align} 
They are diagonalized with bi-unitary transformations $f_{L,R} \to V_{fL,R} f_{L,R} $ such that
\begin{align}\label{eq:mass basis}
V_{UL}^\dagger M_u V_{UR} & = m_u^{\rm diag} \, , & 
V_{DL}^\dagger M_d V_{DR} & = m_d^{\rm diag} \, , &
V_{EL}^\dagger M_e V_{ER} & = m_e^{\rm diag} \, ,
\end{align}
and 
\begin{align}\label{eq:CKM}
V_{\rm CKM} & = V_{UL}^\dagger V_{DL} \, .
\end{align}
In this basis the axion-fermion couplings are given by 
\begin{align}
{\cal L} & = \frac{\partial_\mu a}{v_{\rm PQ}} \left[ \overline{u}  \gamma^\mu \left( g^q P_L + g^u P_R \right)  u + \overline{d}  \gamma^\mu \left( V_{\rm CKM}^\dagger g^q V_{\rm CKM} P_L + g^d P_R \right)  d  \right] \nonumber \\
& + \frac{\partial_\mu a}{v_{\rm PQ}} \left[  \overline{e}  \gamma^\mu \left( g^l P_L + g^e P_R \right)  e  +  \overline{\nu}_{L} V_{\rm PMNS}^\dagger g^l  V_{\rm PMNS}\gamma^\mu \nu_{L}  \right] \, , 
\end{align}
with
\begin{align}
C^q_{ij} & = \left(  X_{A_1} - X_{A_3} \right) \left[ \delta_{ij} -  (V_{UL})^*_{3i} (V_{UL})_{3j}  \right]   \, , \nonumber \\
C^u_{ij} & =  X_{A_2} \delta_{ij} + \left( X_{A_1} - X_{A_2} \right)  (V_{UR})^*_{3i} (V_{UR})_{3j}  \, , \nonumber \\
C^d_{ij} & =  - X_{A_6} \delta_{ij} - \left(  X_{A_5} - X_{A_6} \right) (V_{DR})^*_{3i} (V_{DR})_{3j}  \, ,  \nonumber \\
C^l_{ij} & = - \left( X_{A_9} - X_{A_{11}} \right) \left[ \delta_{ij} - (V_{EL})^*_{3i} (V_{EL})_{3j} \right] \, ,  \nonumber \\
 C^e_{ij} & = - X_{A_{10}} \delta_{ij} -  \left(X_{A_{9}} - X_{A_{10}} \right) (V_{ER})^*_{3i} (V_{ER})_{3j}  \, .  
 \label{Cij}
\end{align}
Now we adopt the standard convention for the axion decay constant $f_a =  v_{\rm PQ}/(2N)$, and write the Lagrangian as
\begin{align}
{\cal L} & = \frac{1}{2} (\partial_\mu a)^2 +  \frac{a}{f_a} \frac{\alpha_s}{8 \pi} G_{\mu \nu} \tilde{G}^{\mu \nu} + \frac{E}{N} \frac{a}{f_a} \frac{\alpha_{\rm em}}{8 \pi} F_{\mu \nu} \tilde{F}^{\mu \nu}  +\frac{\partial_\mu a}{2 f_a} \overline{f}_i \gamma^\mu \left[ C^V_{ij} + C^A_{ij} \gamma_5 \right] f_j \, ,
\end{align}
with
\begin{align}
C^V_{ij} & =   \frac{C^R_{ij} + C^L_{ij}}{2 N} \, , & C^A_{ij} & =  \frac{ C^R_{ij} - C^L_{ij}}{2 N}  \, ,
\label{CVCA}
\end{align}
where $L= q,l$ and $R=u,d,e$. The flavor structure is controlled by the matrices ($f = U,D,E; P=L,R$)
\begin{align}
\xi^{f_{P}}_{ij} \equiv (V_{fP})^*_{3i}  (V_{fP})_{3j} \, , 
\end{align}
 which satisfy
 \begin{align}
 0 & \le \xi^{f_P}_{ii} \le 1 \, , & \sum_i \xi^{f_P}_{ii}  & = 1 \, , &  |\xi^{f_P}_{ij}| & = \sqrt{\xi^{f_P}_{ii} \xi^{f_P}_{jj}} \, .
 \end{align}
The absolute values of these matrices depends only on two independent real parameters in each fermion sector. 
 Notice that the parameter $B$ only enters the domain wall number $N_{\rm DW} = 2 N$, since it is equivalent to the charge normalization of $\phi$ and thus drops out from all axion couplings which only depend on charge ratios. 
 
 In flavor-universal DFSZ models this setup reduces to 
 \begin{align}
 A_{1 \hdots 4} & = A_u \, , &
  A_{5 \hdots 8} & = A_d \, , &
   A_{9 \hdots 12} & = A_e \, . 
 \end{align}
 Without loss of generality one can choose 
 \begin{align}
 A_u & = 2 \, , & A_d & = 1  \, ,
&   A_e & = \begin{cases} A_u  =2 & {\rm (DFSZ-I)} \\   A_d = 1 &   {\rm (DFSZ-II)} \end{cases}
  \end{align}
   which gives
    \begin{align}
    2 N & = 3 B  \, ,  & \frac{E}{N} & =  \begin{cases} 2/3 & {\rm (DFSZ-I)} \\   8/3   &   {\rm (DFSZ-II)}   \end{cases}
    \end{align}
    and
    \begin{align}
   C^{A,V}_{u} & = \frac{c_\beta^2}{3} \, , & 
     C^{A,V}_{d} & = \frac{s_\beta^2}{3} \, , & 
       C^{A,V}_{e} & = \frac{1}{3} \begin{cases} - c_\beta^2  & {\rm (DFSZ-I)} \\   s_\beta^2  &   {\rm (DFSZ-II)} \end{cases} \, . 
    \end{align}

 \section{Higgs Mass Eigenstates}
 \label{AppB}
 Starting from Eq.~\eqref{eq:Yukawa couplings}, we parametrise the Higgs fields $h_i$ as
\begin{align}\label{eq:Higgs decomposition}
    h_i =
    \begin{pmatrix}
        h_i^0 \\
        - h_i^-    
    \end{pmatrix}
    =
    \begin{pmatrix}
        \frac{1}{\sqrt{2}}( v_i +  R_i - i J_i ) \\
        -h_i^-
    \end{pmatrix} \,,
\end{align}
and change into the Higgs basis $\Phi_i$ with
\begin{align}
    \begin{pmatrix}
        \Phi_1 \\
        \Phi_2
    \end{pmatrix}
    & =
    \begin{pmatrix}
    \ccb & \ssb  \\
    -\ssb & \ccb
    \end{pmatrix}
    \begin{pmatrix}
    -\tilde{h}_1 \\
    -\tilde{h}_2
    \end{pmatrix}  \,,
\end{align}
which yields the charged-Higgs boson $H^\pm$ and the pseudo scalar field $A$
\begin{align}\label{eq:Higgs basis}
    \Phi_1 & =
    \begin{pmatrix}
    \ccb \phi_1^+ + \ssb  \phi_2^+ \\
    \frac{\ccb}{\sqrt{2}} (v_1 +  R_1 + i J_1) + \frac{\ssb}{\sqrt{2}} (v_2 + R_2 + i J_2)
    \end{pmatrix} \notag\\
    &\equiv
    \begin{pmatrix}
    G^+ \\
    \frac{1}{\sqrt2} \left(v +  \ccb R_1 + \ssb R_2 + i G^0 \right)
    \end{pmatrix} \,, \nonumber \\
    \Phi_2 & = \begin{pmatrix}
    - \ssb \phi_1^+ + \ccb  \phi_2^+ \\
    - \frac{\ssb}{\sqrt{2}}  (v_1 + R_1 + i J_1) +  \frac{\ccb}{\sqrt{2}}  (v_2 + R_2 + i J_2)
    \end{pmatrix} \notag \\
    &\equiv
    \begin{pmatrix} H^+ \\
    \frac{1}{\sqrt2} \left( - \ssb R_1 + \ccb R_2 + i A \right)
    \end{pmatrix} \,.
\end{align}
Finally, the physical Higgs fields $h,H$ are related to the fields 
$R_1, R_2$ of the neutral Higgs decomposition in Eq.~\eqref{eq:Higgs decomposition}
via the orthogonal rotation
\begin{align}
    \begin{pmatrix}
        R_1 \\
        R_2
    \end{pmatrix}
    &=
    O
    \begin{pmatrix}
    H \\
    h
    \end{pmatrix}
    =
    \begin{pmatrix}
         \cca & \ssa \\
        -\ssa & \cca
    \end{pmatrix}
    \begin{pmatrix}
    H \\
    h
    \end{pmatrix} \,,
\end{align}
which is obtained by diagonalising the
$2\times 2$ mass block of the CP-even states $R_1, R_2$.
 
\bibliographystyle{JHEP} 
\bibliography{bib}

\providecommand{\href}[2]{#2}\begingroup\raggedright\begin{thebibliography}{10}

\bibitem{PQ1}
R.~D. Peccei and H.~R. Quinn, \emph{{CP Conservation in the Presence of
  Instantons}}, \href{https://doi.org/10.1103/PhysRevLett.38.1440}{\emph{Phys.
  Rev. Lett.} {\bfseries 38} (1977) 1440--1443}.

\bibitem{PQ2}
R.~D. Peccei and H.~R. Quinn, \emph{{Constraints Imposed by CP Conservation in
  the Presence of Instantons}},
  \href{https://doi.org/10.1103/PhysRevD.16.1791}{\emph{Phys. Rev.} {\bfseries
  D16} (1977) 1791--1797}.

\bibitem{WW1}
S.~Weinberg, \emph{{A New Light Boson?}},
  \href{https://doi.org/10.1103/PhysRevLett.40.223}{\emph{Phys. Rev. Lett.}
  {\bfseries 40} (1978) 223--226}.

\bibitem{WW2}
F.~Wilczek, \emph{{Problem of Strong p and t Invariance in the Presence of
  Instantons}}, \href{https://doi.org/10.1103/PhysRevLett.40.279}{\emph{Phys.
  Rev. Lett.} {\bfseries 40} (1978) 279--282}.

\bibitem{AxionDM1}
J.~Preskill, M.~B. Wise and F.~Wilczek, \emph{{Cosmology of the Invisible
  Axion}}, \href{https://doi.org/10.1016/0370-2693(83)90637-8}{\emph{Phys.
  Lett.} {\bfseries B120} (1983) 127--132}.

\bibitem{AxionDM2}
L.~F. Abbott and P.~Sikivie, \emph{{A Cosmological Bound on the Invisible
  Axion}}, \href{https://doi.org/10.1016/0370-2693(83)90638-X}{\emph{Phys.
  Lett.} {\bfseries B120} (1983) 133--136}.

\bibitem{AxionDM3}
M.~Dine and W.~Fischler, \emph{{The Not So Harmless Axion}},
  \href{https://doi.org/10.1016/0370-2693(83)90639-1}{\emph{Phys. Lett.}
  {\bfseries B120} (1983) 137--141}.

\bibitem{Raffelt:2011ft}
G.~G. Raffelt, J.~Redondo and N.~Viaux~Maira, \emph{{The meV mass frontier of
  axion physics}},
  \href{https://doi.org/10.1103/PhysRevD.84.103008}{\emph{Phys. Rev. D}
  {\bfseries 84} (2011) 103008},
  [\href{https://arxiv.org/abs/1110.6397}{{\ttfamily 1110.6397}}].

\bibitem{Ringwald:2015lqa}
A.~Ringwald, \emph{{The hunt for axions}},
  \href{https://doi.org/10.22323/1.244.0021}{\emph{PoS} {\bfseries NEUTEL2015}
  (2015) 021}, [\href{https://arxiv.org/abs/1506.04259}{{\ttfamily
  1506.04259}}].

\bibitem{Giannotti:2015dwa}
M.~Giannotti, \emph{{ALP hints from cooling anomalies}},  in \emph{{11th Patras
  Workshop on Axions, WIMPs and WISPs}}, 8, 2015,
  \href{https://arxiv.org/abs/1508.07576}{{\ttfamily 1508.07576}},
  \href{https://doi.org/10.3204/DESY-PROC-2015-02/giannotti_maurizio}{DOI}.

\bibitem{Giannotti:2015kwo}
M.~Giannotti, I.~Irastorza, J.~Redondo and A.~Ringwald, \emph{{Cool WISPs for
  stellar cooling excesses}},
  \href{https://doi.org/10.1088/1475-7516/2016/05/057}{\emph{JCAP} {\bfseries
  05} (2016) 057}, [\href{https://arxiv.org/abs/1512.08108}{{\ttfamily
  1512.08108}}].

\bibitem{Giannotti:2016hnk}
M.~Giannotti, \emph{{Hints of new physics from stars}},
  \href{https://doi.org/10.22323/1.282.0076}{\emph{PoS} {\bfseries ICHEP2016}
  (2016) 076}, [\href{https://arxiv.org/abs/1611.04651}{{\ttfamily
  1611.04651}}].

\bibitem{CoolingAnom3}
M.~Giannotti, I.~G. Irastorza, J.~Redondo, A.~Ringwald and K.~Saikawa,
  \emph{{Stellar Recipes for Axion Hunters}},
  \href{https://doi.org/10.1088/1475-7516/2017/10/010}{\emph{JCAP} {\bfseries
  1710} (2017) 010}, [\href{https://arxiv.org/abs/1708.02111}{{\ttfamily
  1708.02111}}].

\bibitem{SNbound2}
P.~Carenza, T.~Fischer, M.~Giannotti, G.~Guo, G.~Mart\'\i{}nez-Pinedo and
  A.~Mirizzi, \emph{{Improved axion emissivity from a supernova via
  nucleon-nucleon bremsstrahlung}},
  \href{https://doi.org/10.1088/1475-7516/2019/10/016}{\emph{JCAP} {\bfseries
  10} (2019) 016}, [\href{https://arxiv.org/abs/1906.11844}{{\ttfamily
  1906.11844}}].

\bibitem{Keller:2012yr}
J.~Keller and A.~Sedrakian, \emph{{Axions from cooling compact stars}},
  \href{https://doi.org/10.1016/j.nuclphysa.2012.11.004}{\emph{Nucl. Phys. A}
  {\bfseries 897} (2013) 62--69},
  [\href{https://arxiv.org/abs/1205.6940}{{\ttfamily 1205.6940}}].

\bibitem{Sedrakian:2015krq}
A.~Sedrakian, \emph{{Axion cooling of neutron stars}},
  \href{https://doi.org/10.1103/PhysRevD.93.065044}{\emph{Phys. Rev. D}
  {\bfseries 93} (2016) 065044},
  [\href{https://arxiv.org/abs/1512.07828}{{\ttfamily 1512.07828}}].

\bibitem{Hamaguchi:2018oqw}
K.~Hamaguchi, N.~Nagata, K.~Yanagi and J.~Zheng, \emph{{Limit on the Axion
  Decay Constant from the Cooling Neutron Star in Cassiopeia A}},
  \href{https://doi.org/10.1103/PhysRevD.98.103015}{\emph{Phys. Rev. D}
  {\bfseries 98} (2018) 103015},
  [\href{https://arxiv.org/abs/1806.07151}{{\ttfamily 1806.07151}}].

\bibitem{Beznogov:2018fda}
M.~V. Beznogov, E.~Rrapaj, D.~Page and S.~Reddy, \emph{{Constraints on
  Axion-like Particles and Nucleon Pairing in Dense Matter from the Hot Neutron
  Star in HESS J1731-347}},
  \href{https://doi.org/10.1103/PhysRevC.98.035802}{\emph{Phys. Rev. C}
  {\bfseries 98} (2018) 035802},
  [\href{https://arxiv.org/abs/1806.07991}{{\ttfamily 1806.07991}}].

\bibitem{Sedrakian:2018kdm}
A.~Sedrakian, \emph{{Axion cooling of neutron stars. II. Beyond hadronic
  axions}}, \href{https://doi.org/10.1103/PhysRevD.99.043011}{\emph{Phys. Rev.
  D} {\bfseries 99} (2019) 043011},
  [\href{https://arxiv.org/abs/1810.00190}{{\ttfamily 1810.00190}}].

\bibitem{DFSZ1}
M.~Dine, W.~Fischler and M.~Srednicki, \emph{{A Simple Solution to the Strong
  CP Problem with a Harmless Axion}},
  \href{https://doi.org/10.1016/0370-2693(81)90590-6}{\emph{Phys. Lett.}
  {\bfseries B104} (1981) 199--202}.

\bibitem{DFSZ2}
A.~R. Zhitnitsky, \emph{{On Possible Suppression of the Axion Hadron
  Interactions. (In Russian)}}, {\emph{Sov. J. Nucl. Phys.} {\bfseries 31}
  (1980) 260}.

\bibitem{Peccei:1986pn}
R.~D. Peccei, T.~T. Wu and T.~Yanagida, \emph{{A VIABLE AXION MODEL}},
  \href{https://doi.org/10.1016/0370-2693(86)90284-4}{\emph{Phys. Lett. B}
  {\bfseries 172} (1986) 435--440}.

\bibitem{Krauss:1986wx}
L.~M. Krauss and F.~Wilczek, \emph{{A SHORTLIVED AXION VARIANT}},
  \href{https://doi.org/10.1016/0370-2693(86)90244-3}{\emph{Phys. Lett. B}
  {\bfseries 173} (1986) 189--192}.

\bibitem{DiLuzio:2017ogq}
L.~Di~Luzio, F.~Mescia, E.~Nardi, P.~Panci and R.~Ziegler, \emph{{Astrophobic
  Axions}}, \href{https://doi.org/10.1103/PhysRevLett.120.261803}{\emph{Phys.
  Rev. Lett.} {\bfseries 120} (2018) 261803},
  [\href{https://arxiv.org/abs/1712.04940}{{\ttfamily 1712.04940}}].

\bibitem{Bjorkeroth:2019jtx}
F.~Bj\"orkeroth, L.~Di~Luzio, F.~Mescia, E.~Nardi, P.~Panci and R.~Ziegler,
  \emph{{Axion-electron decoupling in nucleophobic axion models}},
  \href{https://doi.org/10.1103/PhysRevD.101.035027}{\emph{Phys. Rev. D}
  {\bfseries 101} (2020) 035027},
  [\href{https://arxiv.org/abs/1907.06575}{{\ttfamily 1907.06575}}].

\bibitem{Saikawa:2019lng}
K.~Saikawa and T.~T. Yanagida, \emph{{Stellar cooling anomalies and variant
  axion models}},
  \href{https://doi.org/10.1088/1475-7516/2020/03/007}{\emph{JCAP} {\bfseries
  03} (2020) 007}, [\href{https://arxiv.org/abs/1907.07662}{{\ttfamily
  1907.07662}}].

\bibitem{Vilenkin:1982ks}
A.~Vilenkin and A.~E. Everett, \emph{{Cosmic Strings and Domain Walls in Models
  with Goldstone and PseudoGoldstone Bosons}},
  \href{https://doi.org/10.1103/PhysRevLett.48.1867}{\emph{Phys. Rev. Lett.}
  {\bfseries 48} (1982) 1867--1870}.

\bibitem{IAXO2}
E.~Armengaud et~al., \emph{{Conceptual Design of the International Axion
  Observatory (IAXO)}},
  \href{https://doi.org/10.1088/1748-0221/9/05/T05002}{\emph{JINST} {\bfseries
  9} (2014) T05002}, [\href{https://arxiv.org/abs/1401.3233}{{\ttfamily
  1401.3233}}].

\bibitem{Chiang:2015cba}
C.-W. Chiang, H.~Fukuda, M.~Takeuchi and T.~T. Yanagida, \emph{{Flavor-Changing
  Neutral-Current Decays in Top-Specific Variant Axion Model}},
  \href{https://doi.org/10.1007/JHEP11(2015)057}{\emph{JHEP} {\bfseries 11}
  (2015) 057}, [\href{https://arxiv.org/abs/1507.04354}{{\ttfamily
  1507.04354}}].

\bibitem{Chiang:2017fjr}
C.-W. Chiang, H.~Fukuda, M.~Takeuchi and T.~T. Yanagida, \emph{{Current Status
  of Top-Specific Variant Axion Model}},
  \href{https://doi.org/10.1103/PhysRevD.97.035015}{\emph{Phys. Rev. D}
  {\bfseries 97} (2018) 035015},
  [\href{https://arxiv.org/abs/1711.02993}{{\ttfamily 1711.02993}}].

\bibitem{Chiang:2018bnu}
C.-W. Chiang, M.~Takeuchi, P.-Y. Tseng and T.~T. Yanagida, \emph{{Muon $g-2$
  and rare top decays in up-type specific variant axion models}},
  \href{https://doi.org/10.1103/PhysRevD.98.095020}{\emph{Phys. Rev. D}
  {\bfseries 98} (2018) 095020},
  [\href{https://arxiv.org/abs/1807.00593}{{\ttfamily 1807.00593}}].

\bibitem{Villadoro2}
M.~Gorghetto and G.~Villadoro, \emph{{Topological Susceptibility and QCD Axion
  Mass: QED and NNLO corrections}},
  \href{https://doi.org/10.1007/JHEP03(2019)033}{\emph{JHEP} {\bfseries 03}
  (2019) 033}, [\href{https://arxiv.org/abs/1812.01008}{{\ttfamily
  1812.01008}}].

\bibitem{diCortona:2015ldu}
G.~Grilli~di Cortona, E.~Hardy, J.~Pardo~Vega and G.~Villadoro, \emph{{The QCD
  axion, precisely}},
  \href{https://doi.org/10.1007/JHEP01(2016)034}{\emph{JHEP} {\bfseries 01}
  (2016) 034}, [\href{https://arxiv.org/abs/1511.02867}{{\ttfamily
  1511.02867}}].

\bibitem{astrophobic}
L.~Di~Luzio, F.~Mescia, E.~Nardi, P.~Panci and R.~Ziegler, \emph{{Astrophobic
  Axions}}, \href{https://doi.org/10.1103/PhysRevLett.120.261803}{\emph{Phys.
  Rev. Lett.} {\bfseries 120} (2018) 261803},
  [\href{https://arxiv.org/abs/1712.04940}{{\ttfamily 1712.04940}}].

\bibitem{Krauss:1987ud}
L.~M. Krauss and D.~J. Nash, \emph{{A VIABLE WEAK INTERACTION AXION?}},
  \href{https://doi.org/10.1016/0370-2693(88)91864-3}{\emph{Phys. Lett. B}
  {\bfseries 202} (1988) 560--567}.

\bibitem{Hindmarsh:1997ac}
M.~Hindmarsh and P.~Moulatsiotis, \emph{{Constraints on variant axion models}},
  \href{https://doi.org/10.1103/PhysRevD.56.8074}{\emph{Phys. Rev. D}
  {\bfseries 56} (1997) 8074--8081},
  [\href{https://arxiv.org/abs/hep-ph/9708281}{{\ttfamily hep-ph/9708281}}].

\bibitem{Choi:2017gpf}
K.~Choi, S.~H. Im, C.~B. Park and S.~Yun, \emph{{Minimal Flavor Violation with
  Axion-like Particles}},
  \href{https://doi.org/10.1007/JHEP11(2017)070}{\emph{JHEP} {\bfseries 11}
  (2017) 070}, [\href{https://arxiv.org/abs/1708.00021}{{\ttfamily
  1708.00021}}].

\bibitem{MartinCamalich:2020dfe}
J.~Martin~Camalich, M.~Pospelov, P.~N.~H. Vuong, R.~Ziegler and J.~Zupan,
  \emph{{Quark Flavor Phenomenology of the QCD Axion}},
  \href{https://doi.org/10.1103/PhysRevD.102.015023}{\emph{Phys. Rev. D}
  {\bfseries 102} (2020) 015023},
  [\href{https://arxiv.org/abs/2002.04623}{{\ttfamily 2002.04623}}].

\bibitem{Heiles:2020plj}
M.~Heiles, M.~K\"onig and M.~Neubert, \emph{{Effective Field Theory for Heavy
  Vector Resonances Coupled to the Standard Model}},
  \href{https://arxiv.org/abs/2011.08205}{{\ttfamily 2011.08205}}.

\bibitem{Chala:2020wvs}
M.~Chala, G.~Guedes, M.~Ramos and J.~Santiago, \emph{{Running in the ALPs}},
  \href{https://doi.org/10.1140/epjc/s10052-021-08968-2}{\emph{Eur. Phys. J. C}
  {\bfseries 81} (2021) 181},
  [\href{https://arxiv.org/abs/2012.09017}{{\ttfamily 2012.09017}}].

\bibitem{Bauer:2020jbp}
M.~Bauer, M.~Neubert, S.~Renner, M.~Schnubel and A.~Thamm, \emph{{The
  Low-Energy Effective Theory of Axions and ALPs}},
  \href{https://doi.org/10.1007/JHEP04(2021)063}{\emph{JHEP} {\bfseries 04}
  (2021) 063}, [\href{https://arxiv.org/abs/2012.12272}{{\ttfamily
  2012.12272}}].

\bibitem{Choi:2021kuy}
K.~Choi, S.~H. Im, H.~J. Kim and H.~Seong, \emph{{Precision axion physics with
  running axion couplings}},
  \href{https://arxiv.org/abs/2106.05816}{{\ttfamily 2106.05816}}.

\bibitem{Anastassopoulos:2017ftl}
{\scshape CAST} collaboration, V.~Anastassopoulos et~al., \emph{{New CAST Limit
  on the Axion-Photon Interaction}},
  \href{https://doi.org/10.1038/nphys4109}{\emph{Nature Phys.} {\bfseries 13}
  (2017) 584--590}, [\href{https://arxiv.org/abs/1705.02290}{{\ttfamily
  1705.02290}}].

\bibitem{Ayala:2014pea}
A.~Ayala, I.~Dom\'\i{}nguez, M.~Giannotti, A.~Mirizzi and O.~Straniero,
  \emph{{Revisiting the bound on axion-photon coupling from Globular
  Clusters}}, \href{https://doi.org/10.1103/PhysRevLett.113.191302}{\emph{Phys.
  Rev. Lett.} {\bfseries 113} (2014) 191302},
  [\href{https://arxiv.org/abs/1406.6053}{{\ttfamily 1406.6053}}].

\bibitem{Armengaud:2014gea}
E.~Armengaud et~al., \emph{{Conceptual Design of the International Axion
  Observatory (IAXO)}},
  \href{https://doi.org/10.1088/1748-0221/9/05/T05002}{\emph{JINST} {\bfseries
  9} (2014) T05002}, [\href{https://arxiv.org/abs/1401.3233}{{\ttfamily
  1401.3233}}].

\bibitem{Carenza:2020cis}
P.~Carenza, B.~Fore, M.~Giannotti, A.~Mirizzi and S.~Reddy, \emph{{Enhanced
  Supernova Axion Emission and its Implications}},
  \href{https://doi.org/10.1103/PhysRevLett.126.071102}{\emph{Phys. Rev. Lett.}
  {\bfseries 126} (2021) 071102},
  [\href{https://arxiv.org/abs/2010.02943}{{\ttfamily 2010.02943}}].

\bibitem{WDbound}
M.~M. Miller~Bertolami, B.~E. Melendez, L.~G. Althaus and J.~Isern,
  \emph{{Revisiting the axion bounds from the Galactic white dwarf luminosity
  function}}, \href{https://doi.org/10.1088/1475-7516/2014/10/069}{\emph{JCAP}
  {\bfseries 1410} (2014) 069},
  [\href{https://arxiv.org/abs/1406.7712}{{\ttfamily 1406.7712}}].

\bibitem{Bollig:2020xdr}
R.~Bollig, W.~DeRocco, P.~W. Graham and H.-T. Janka, \emph{{Muons in
  supernovae: implications for the axion-muon coupling}},
  \href{https://doi.org/10.1103/PhysRevLett.125.051104}{\emph{Phys. Rev. Lett.}
  {\bfseries 125} (2020) 051104},
  [\href{https://arxiv.org/abs/2005.07141}{{\ttfamily 2005.07141}}].

\bibitem{Croon:2020lrf}
D.~Croon, G.~Elor, R.~K. Leane and S.~D. McDermott, \emph{{Supernova Muons: New
  Constraints on $Z$' Bosons, Axions and ALPs}},
  \href{https://doi.org/10.1007/JHEP01(2021)107}{\emph{JHEP} {\bfseries 01}
  (2021) 107}, [\href{https://arxiv.org/abs/2006.13942}{{\ttfamily
  2006.13942}}].

\bibitem{muea}
A.~Jodidio et~al., \emph{{Search for Right-Handed Currents in Muon Decay}},
  \href{https://doi.org/10.1103/PhysRevD.34.1967,
  10.1103/PhysRevD.37.237}{\emph{Phys. Rev.} {\bfseries D34} (1986) 1967}.

\bibitem{Bayes:2014lxz}
{\scshape TWIST} collaboration, R.~Bayes et~al., \emph{{Search for two body
  muon decay signals}},
  \href{https://doi.org/10.1103/PhysRevD.91.052020}{\emph{Phys. Rev. D}
  {\bfseries 91} (2015) 052020},
  [\href{https://arxiv.org/abs/1409.0638}{{\ttfamily 1409.0638}}].

\bibitem{Calibbi:2020jvd}
L.~Calibbi, D.~Redigolo, R.~Ziegler and J.~Zupan, \emph{{Looking forward to
  Lepton-flavor-violating ALPs}},
  \href{https://arxiv.org/abs/2006.04795}{{\ttfamily 2006.04795}}.

\bibitem{CoolingAnom2}
M.~Giannotti, I.~Irastorza, J.~Redondo and A.~Ringwald, \emph{{Cool WISPs for
  stellar cooling excesses}},
  \href{https://doi.org/10.1088/1475-7516/2016/05/057}{\emph{JCAP} {\bfseries
  1605} (2016) 057}, [\href{https://arxiv.org/abs/1512.08108}{{\ttfamily
  1512.08108}}].

\bibitem{Dessert:2021bkv}
C.~Dessert, A.~J. Long and B.~R. Safdi, \emph{{No evidence for axions from
  Chandra observation of magnetic white dwarf}},
  \href{https://arxiv.org/abs/2104.12772}{{\ttfamily 2104.12772}}.

\bibitem{IAXO3}
{\scshape IAXO} collaboration, E.~Armengaud et~al., \emph{{Physics potential of
  the International Axion Observatory (IAXO)}},
  \href{https://arxiv.org/abs/1904.09155}{{\ttfamily 1904.09155}}.

\bibitem{Aad:2019mbh}
{\scshape ATLAS} collaboration, G.~Aad et~al., \emph{{Combined measurements of
  Higgs boson production and decay using up to $80$ fb$^{-1}$ of proton-proton
  collision data at $\sqrt{s}=$ 13 TeV collected with the ATLAS experiment}},
  \href{https://doi.org/10.1103/PhysRevD.101.012002}{\emph{Phys. Rev. D}
  {\bfseries 101} (2020) 012002},
  [\href{https://arxiv.org/abs/1909.02845}{{\ttfamily 1909.02845}}].

\bibitem{ATLAS}
{\scshape ATLAS} collaboration, G.~Aad et~al., ``{Combined measurements of
  Higgs boson production and decay using up to $80$ fb$^{-1}$ of proton-proton
  collision data at $\sqrt{s}=$ 13 TeV collected with the ATLAS experiment}.''
  \url{https://atlas.web.cern.ch/Atlas/GROUPS/PHYSICS/PAPERS/HIGG-2018-57/}.

\bibitem{deFlorian:2016spz}
{\scshape LHC Higgs Cross Section Working Group} collaboration, D.~de~Florian
  et~al., \emph{{Handbook of LHC Higgs Cross Sections: 4. Deciphering the
  Nature of the Higgs Sector}},
  \href{https://arxiv.org/abs/1610.07922}{{\ttfamily 1610.07922}}.

\bibitem{Aad:2019ojw}
{\scshape ATLAS} collaboration, G.~Aad et~al., \emph{{Search for the Higgs
  boson decays $H \to ee$ and $H \to e\mu$ in $pp$ collisions at $\sqrt{s} =
  13$ TeV with the ATLAS detector}},
  \href{https://doi.org/10.1016/j.physletb.2019.135148}{\emph{Phys. Lett. B}
  {\bfseries 801} (2020) 135148},
  [\href{https://arxiv.org/abs/1909.10235}{{\ttfamily 1909.10235}}].

\bibitem{Sirunyan:2021ovv}
{\scshape CMS} collaboration, A.~M. Sirunyan et~al., \emph{{Search for
  lepton-flavor violating decays of the Higgs boson in the $\mu\tau$ and
  e$\tau$ final states in proton-proton collisions at $\sqrt{s}$ = 13 TeV}},
  \href{https://arxiv.org/abs/2105.03007}{{\ttfamily 2105.03007}}.

\bibitem{Aad:2019ugc}
{\scshape ATLAS} collaboration, G.~Aad et~al., \emph{{Searches for
  lepton-flavour-violating decays of the Higgs boson in $\sqrt{s}=13$ TeV pp
  collisions with the ATLAS detector}},
  \href{https://doi.org/10.1016/j.physletb.2019.135069}{\emph{Phys. Lett. B}
  {\bfseries 800} (2020) 135069},
  [\href{https://arxiv.org/abs/1907.06131}{{\ttfamily 1907.06131}}].

\bibitem{Harnik:2012pb}
R.~Harnik, J.~Kopp and J.~Zupan, \emph{{Flavor Violating Higgs Decays}},
  \href{https://doi.org/10.1007/JHEP03(2013)026}{\emph{JHEP} {\bfseries 03}
  (2013) 026}, [\href{https://arxiv.org/abs/1209.1397}{{\ttfamily 1209.1397}}].

\bibitem{Crivellin:2013wna}
A.~Crivellin, A.~Kokulu and C.~Greub, \emph{{Flavor-phenomenology of
  two-Higgs-doublet models with generic Yukawa structure}},
  \href{https://doi.org/10.1103/PhysRevD.87.094031}{\emph{Phys. Rev.}
  {\bfseries D87} (2013) 094031},
  [\href{https://arxiv.org/abs/1303.5877}{{\ttfamily 1303.5877}}].

\bibitem{Chang:1993kw}
D.~Chang, W.~S. Hou and W.-Y. Keung, \emph{{Two loop contributions of flavor
  changing neutral Higgs bosons to mu ---\ensuremath{>} e gamma}},
  \href{https://doi.org/10.1103/PhysRevD.48.217}{\emph{Phys. Rev. D} {\bfseries
  48} (1993) 217--224}, [\href{https://arxiv.org/abs/hep-ph/9302267}{{\ttfamily
  hep-ph/9302267}}].

\bibitem{Aubert:2009ag}
{\scshape BaBar} collaboration, B.~Aubert et~al., \emph{{Searches for Lepton
  Flavor Violation in the Decays tau+- ---\ensuremath{>} e+- gamma and tau+-
  ---\ensuremath{>} mu+- gamma}},
  \href{https://doi.org/10.1103/PhysRevLett.104.021802}{\emph{Phys. Rev. Lett.}
  {\bfseries 104} (2010) 021802},
  [\href{https://arxiv.org/abs/0908.2381}{{\ttfamily 0908.2381}}].

\bibitem{TheMEG:2016wtm}
{\scshape MEG} collaboration, A.~M. Baldini et~al., \emph{{Search for the
  lepton flavour violating decay $\mu ^+ \rightarrow \mathrm {e}^+ \gamma $
  with the full dataset of the MEG experiment}},
  \href{https://doi.org/10.1140/epjc/s10052-016-4271-x}{\emph{Eur. Phys. J. C}
  {\bfseries 76} (2016) 434},
  [\href{https://arxiv.org/abs/1605.05081}{{\ttfamily 1605.05081}}].

\bibitem{deBlas:2019rxi}
J.~de~Blas et~al., \emph{{Higgs Boson Studies at Future Particle Colliders}},
  \href{https://doi.org/10.1007/JHEP01(2020)139}{\emph{JHEP} {\bfseries 01}
  (2020) 139}, [\href{https://arxiv.org/abs/1905.03764}{{\ttfamily
  1905.03764}}].

\bibitem{ATLAS:2018jlh}
{\scshape ATLAS} collaboration, G.~Aad et~al., \emph{{Projections for
  measurements of Higgs boson cross sections, branching ratios, coupling
  parameters and mass with the ATLAS detector at the HL-LHC}}, .

\bibitem{Gorghetto:2018myk}
M.~Gorghetto, E.~Hardy and G.~Villadoro, \emph{{Axions from Strings: the
  Attractive Solution}},
  \href{https://doi.org/10.1007/JHEP07(2018)151}{\emph{JHEP} {\bfseries 07}
  (2018) 151}, [\href{https://arxiv.org/abs/1806.04677}{{\ttfamily
  1806.04677}}].

\bibitem{Co:2017mop}
R.~T. Co, L.~J. Hall and K.~Harigaya, \emph{{QCD Axion Dark Matter with a Small
  Decay Constant}},
  \href{https://doi.org/10.1103/PhysRevLett.120.211602}{\emph{Phys. Rev. Lett.}
  {\bfseries 120} (2018) 211602},
  [\href{https://arxiv.org/abs/1711.10486}{{\ttfamily 1711.10486}}].

\end{thebibliography}\endgroup



\providecommand{\href}[2]{#2}\begingroup\raggedright\endgroup

\end{document}